\documentclass{article}%
\usepackage{amsmath}
\usepackage{amsfonts}
\usepackage{amssymb}
\usepackage{graphicx}%
\pdfoutput=1
\newtheorem{theorem}{Theorem}
\newtheorem{acknowledgement}[theorem]{Acknowledgement}

\newtheorem{definition}[theorem]{Definition}

\newtheorem{exercise}[theorem]{Exercise}

\newtheorem{remark}[theorem]{Remark}

\newcommand{\bpartial}{\mathop{\partial\kern -4pt\raisebox{.8pt}{$|$}}}
\newcommand{\bra}{\mathopen{[\kern-1.6pt[}}
\newcommand{\ket}{\mathclose{]\kern-1.5pt]}}
\newcommand{\bbra}{\mathopen{[\kern-2.2pt[\kern-2.3pt[}}
\newcommand{\bket}{\mathclose{]\kern-2.1pt]\kern-2.3pt]}}
\newcommand{\slg}{\mbox{\bfseries\slshape g}}
\newcommand{\slx}{\mbox{\bfseries\slshape x}}
\newcommand{\slT}{\mbox{\bfseries\slshape T}}

\makeindex
\begin{document}

\title{Differential Forms on Riemannian (Lorentzian) and Riemann-Cartan Structures
and Some Applications to Physics\thanks{published in: \textit{Ann. Fond. L. de
Broglie} \textbf{32 }(special issue dedicate to torsion), 424-478 (2008).}}
\author{Waldyr Alves Rodrigues Jr.\\{\small Institute of Mathematics Statistics and Scientific Computation}\\{\small IMECC-UNICAMP CP 6065}\\{\small 13083760 Campinas SP Brazil}\\{\small e-mail: walrod@ime.unicamp.br or walrod@mpc.com.br}}
\maketitle

\begin{abstract}
In this paper after recalling some essential tools concerning the theory of
differential forms in the Cartan, Hodge and Clifford bundles over a Riemannian
or Riemann-Cartan space or a Lorentzian or Riemann-Cartan spacetime we solve
with details several exercises involving different grades of difficult. One of
the problems is to show that a recent formula given in \cite{evans101} for the
exterior covariant derivative of the Hodge dual of the torsion $2$-forms is
simply wrong. We believe that the paper will be useful for students (and
eventually for some experts) on applications of differential geometry to some
physical problems. A detailed account of the issues discussed in the paper
appears in the table of contents.

\end{abstract}
\tableofcontents

\section{Introduction}

In this paper we first recall some essential tools concerning the theory of
differential forms in the \textit{Cartan}, \textit{Hodge} and
\textit{Clifford} bundles over a $n$-dimensional manifold \ $M$ equipped with
a metric tensor $%
\slg
\in\sec T_{2}^{0}M$ of arbitrary signature $(p,q)$, $p+q=n$ \ and also
equipped with metric compatible connections, the Levi-Civita ($\mathring{D}$)
and a general Riemann-Cartan ($D$) one\footnote{A spacetime is a special
structure where the manifold is $4$-dimensional, the metric has signature
$(1,3)$ and which is equipped with a Levi-Civita or a a Riemann-Cartan
connection, orietability and time orientation. See below and, e.g.,
\cite{rodoliv2007,sawu} for more details, if needed.}. After that we solved
with details some exercises involving different grades of difficult, ranging
depending on the readers knowledge from kindergarten, intermediate to advanced
levels. In particular we show how to express the \textit{derivative (}$d$) and
\textit{coderivative }($\delta$) operators as functions of operators related
to the Levi-Civita or a Riemann-Cartan connection defined on a manifold,
namely the \textit{standard} Dirac operator ($%
\bpartial
$) and general Dirac operator ($\mbox{\boldmath$\partial$}$) . Those
operators\ are then used to express Maxwell equations in both a Lorentzian and
a Riemann-Cartan spacetime. We recall also important formulas (not well known
as they deserve to be) for the square of the general Dirac and standard Dirac
operators showing their relation with the \textit{Hodge} \textit{D'Alembertian
}($\Diamond$), the \textit{covariant D' Alembertian }($\mathring{\square}$)
and the \textit{Ricci \ operators }($\mathcal{\mathring{R}}^{\mathbf{a}}$,
$\mathcal{R}^{\mathbf{a}}$) and \textit{Einstein} operator ($\mathring
{\blacksquare}$) and the use of these operators in the
\textit{Einstein-Hilbert} gravitational theory. Finally, we study the
\textit{Bianchi} identities. Recalling that the first Bianchi identity is
\ $D\mathcal{T}^{\mathbf{a}}=\mathcal{R}_{\mathbf{b}}^{\mathbf{a}}\wedge
\theta^{\mathbf{b}}$, where $\mathcal{T}^{\mathbf{a}}$ and $\mathcal{R}%
_{\mathbf{b}}^{\mathbf{a}}$ are respectively the torsion and the curvature
$2$-forms and $\{\theta^{\mathbf{b}}\}$ is a cotetrad we ask the question: Who
is $D\star\mathcal{T}^{\mathbf{a}}$? We find the correct answer (Eq.(\ref{du6}%
)) using the tools introduced in previous sections of the paper. Our result
shows explicitly that the formula for .\textquotedblleft$D\star\mathcal{T}%
^{\mathbf{a}}=\star\mathcal{R}_{\mathbf{b}}^{\mathbf{a}}\wedge\theta
^{\mathbf{b}}$\textquotedblright\ recently found in \cite{evans101} and
claimed to imply a contradiction in Einstein-Hilbert gravitational theory is
\textit{wrong}. Two very simple counterexamples contradicting the wrong
formula for $D\star\mathcal{T}^{\mathbf{a}}$ are presented. A detailed account
of the issues discussed in the paper appears in the table of
contents\footnote{More on the subject may be found in, e.g.,
\cite{rodoliv2007} and recent advanced material may be found in several papers
of the author posted on the arXiv.}. We call also the reader attention that in
the physical applications we use natural units for which the numerical values
of $c$,$h$ and the gravitational constant $k$ (appearing in Einstein
equations) are equal to $1$.

\section{Classification of Metric Compatible Structures $(M,%
\slg
,D)$}

Let $M$ denotes a $n$-dimensional manifold\footnote{We left the toplogy of $M$
unspecified for a while.}. We denote as usual by $T_{x}M$ and $T_{x}^{\ast}M$
\ respectively the tangent and the cotangent spaces at $x\in M$. By $TM=%
{\displaystyle\bigcup\nolimits_{x\in M}}
$ $T_{x}M$ and $T^{\ast}M=%
{\displaystyle\bigcup\nolimits_{x\in M}}
$ $T_{x}^{x}M$ respectively the tangent and cotangent bundles. By $T_{s}^{r}M$
we denote the bundle of $r$-contravariant and $s$-covariant tensors and by
$\mathcal{T}M=\bigoplus\nolimits_{r,s=0}^{\infty}T_{s}^{r}M$ the tensor
bundle. By $%
{\displaystyle\bigwedge^{r}}
TM$ and $%
{\displaystyle\bigwedge^{r}}
T^{\ast}M$ denote respectively the bundles of $r$-multivector fields and of
$r$-form fields. We call $%
{\displaystyle\bigwedge}
TM=%
{\displaystyle\bigoplus\nolimits_{r=0}^{r=n}}
{\displaystyle\bigwedge^{r}}
TM$ the bundle of (non homogeneous) multivector fields and call $%
{\displaystyle\bigwedge}
T^{\ast}M=%
{\displaystyle\bigoplus\nolimits_{r=0}^{r=n}}
{\displaystyle\bigwedge^{r}}
T^{\ast}M$ the exterior algebra (Cartan) bundle. Of course, it is the bundle
of (non homogeneous) form fields. Recall that the real vector spaces are such
that $\dim\bigwedge^{r}T_{x}M$ $=\dim\bigwedge^{r}T_{x}^{\ast}M$ $=\binom
{n}{r}$ and $\dim%
{\displaystyle\bigwedge}
T^{\ast}M=2^{n}$. Some \textit{additional} structures will be introduced or
mentioned below when needed. Let\footnote{We denote by$\ \sec(X(M))$ the space
of the sections of a bundle $X(M)$. Note that all functions and differential
forms are supposed smooth, unless we explicitly say the contrary.} $%
\slg
\in\sec T_{2}^{0}M$ a metric of signature $(p,q)$ and $D$ an arbitrary metric
compatible connection on $M$, i.e., $D%
\slg
=0$. We denote by $\mathbf{R}$ and $\mathbf{T}$ respectively the \ (Riemann)
curvature and torsion tensors\footnote{The precise definitions of those
objects will be recalled below.} of the connection $D$, and recall that in
general a given manifold given some additional conditions may admit many
different metrics and many different connections.

Given \ a triple $(M,%
\slg
,D)$:

(a) it is called a Riemann-\textit{Cartan space }if and only if
\begin{equation}
D%
\slg
=0\qquad\mathrm{and}\qquad\mathbf{T}\neq0.
\end{equation}

(b) it is called \textit{Weyl space if and only if}%
\begin{equation}
D%
\slg
\neq0\qquad\mathrm{and}\qquad\mathbf{T}=0.
\end{equation}

(c) it is called a \textit{Riemann space} if and only if
\begin{equation}
D%
\slg
=0\qquad\mathrm{and}\qquad\mathbf{T}=0, \label{888}%
\end{equation}
and in that case the pair $(D,%
\slg
)$ is called \textit{Riemannian structure}.

\emph{(}d) it is called \textit{Riemann-Cartan-Weyl space} if and only if
\begin{equation}
D%
\slg
\neq0\qquad\mathrm{and}\qquad\mathbf{T}\neq0.
\end{equation}

(e) it is called \emph{(}Riemann\emph{)} flat if and only if
\[
D%
\slg
=0\qquad\mathrm{and}\qquad\mathbf{R}=0,\text{ }%
\]

(f) it is called teleparallel if and only if%
\begin{equation}
D%
\slg
=0,\text{ }\mathbf{T}\neq0\text{ }\mathrm{and}\text{ }\mathbf{R=}0.
\end{equation}

\subsection{Levi-Civita and Riemann-Cartan Connections}

For each metric tensor defined on the manifold $M$ there exists one and only
one connection in the conditions of Eq.(\ref{888}). It is is called
\textit{Levi-Civita connection\/} of the metric considered, and is denoted in
what follows by $\mathring{D}$. A connection satisfying the properties in (a)
above is called a Riemann-Cartan connection. In general both connections may
be defined in a given manifold and they are related by well established
formulas recalled below. \ A connection defines a rule for the parallel
transport of vectors (more generally tensor fields) in a manifold, something
which is conventional \cite{poincare}, and so the question concerning which
one is more important is according to our view meaningless\footnote{Even if it
is the case, that a particular one may be more convenient than others for some
purposes. See the example of the Nunes connections in Section 15.}. The author
knows that this assertion may surprise some readers, but he is sure that they
will be convinced \ of its correctness after studying Section 15. More on the
subject in \cite{rodoliv2007}. For \ implementations of these ideas for the
theory of gravitation see \cite{notterod}

\subsection{Spacetime Structures}

\begin{remark}
When $\dim M=4$ and the metric $%
\slg
$ has signature $(1,3)$ we sometimes substitute Riemann by Lorentz in the
previous definitions \emph{(c)},\emph{(e)} and \emph{(f)}.
\end{remark}

\begin{remark}
In order to represent a spacetime structure \ a Lorentzian or a Riemann-Cartan
structure $(M,%
\slg
,D)$ need be such that $M$ is connected and paracompact \emph{\cite{hawellis}}
and equipped with an orientation defined by the volume element $\tau_{%
\slg
}\in\sec%
{\displaystyle\bigwedge\nolimits^{4}}
T^{\ast}M$ and a time orientation denoted by $\uparrow$. We omit here the
details and ask to the interested reader to consult, e.g.,
\emph{\cite{rodoliv2007}}. A general spacetime will be represented by a
pentuple $(M,%
\slg
,D,\tau_{%
\slg
},\uparrow).$
\end{remark}

\section{ Absolute Differential and Covariant Derivatives}

Given a differentiable manifold $M$, let $X,Y\in\sec TM$, any vector fields,
$\alpha\in\sec T^{\ast}M$ any covector field . Let $\mathcal{T}M=\bigoplus
\nolimits_{r,s=0}^{\infty}T_{s}^{r}M$ be the tensor bundle of $M$ and
\ $\mathbf{P}\in\sec\mathcal{T}M$\ any general tensor field.

We now describe the main properties of a general connection\ $D$ (also called
absolute differential operator). We have%
\begin{align}
D  &  :\sec TM\times\sec\mathcal{T}M\rightarrow\sec\mathcal{T}M,\nonumber\\
(X,\mathbf{P})  &  \mapsto D_{X}\mathbf{P}, \label{CO TENSOR}%
\end{align}
where $D_{X}$ the covariant derivative in the direction of the vector field
$X$ satisfy the following properties: Given, differentiable functions
$f,g:M\rightarrow\mathbb{R}$, vector fields $X,Y\in\sec TM$ and $\mathbf{P,Q}%
\in\sec\mathcal{T}M$ we have%

\begin{align}
D_{fX+gY}\mathbf{P}  &  =fD_{X}\mathbf{P+}gD_{Y}\mathbf{P},\nonumber\\
D_{X}(\mathbf{P+Q})  &  =D_{X}\mathbf{P}+D_{X}\mathbf{Q},\nonumber\\
D_{X}(f\mathbf{P)}  &  =fD_{X}(\mathbf{P)+}X(f\mathbf{)P},\nonumber\\
D_{X}(\mathbf{P\otimes Q})  &  =D_{X}\mathbf{P\otimes Q}+\mathbf{P\otimes
}D_{X}\mathbf{Q}.
\end{align}

Given $\mathbf{Q\in}\sec T_{s}^{r}M$ the relation between $D\mathbf{Q}$, the
\textit{absolute differential} of $\mathbf{Q}$\ and $D_{X}\mathbf{Q}$ the
covariant derivative of $\mathbf{Q}$ in the direction of the vector filed $X$
is given by
\begin{align}
D  &  \mathbf{:}\sec T_{s}^{r}M\rightarrow\sec T_{s+1}^{r}M,\nonumber\\
&  D\mathbf{Q(}X\mathbf{,}X_{1},...,X_{s},\alpha_{1},...,\alpha_{r}%
)\nonumber\\
&  =D_{X}\mathbf{Q(}X_{1},...,X_{s},\alpha_{1},...,\alpha_{r}),\nonumber\\
X_{1},...,X_{s}  &  \in\sec TM,\alpha_{1},...\alpha_{r}\in\sec T^{\ast}M.
\label{absolute diff}%
\end{align}

\ Let $U\subset M$ \ and consider a chart of the maximal atlas of $M$ covering
$U$ coordinate functions $\{%
\slx
^{\mu}\}$. Let \texttt{ }$%
\slg
\in\sec T_{2}^{0}M$ be a metric field for $M$. Let
$\{{\mbox{\boldmath$\partial$}}_{\mu}\}$ be a basis for $TU$, $U\subset M$ and
let $\{\theta^{\mu}=dx^{\mu}\}$ be the dual basis of
$\{{\mbox{\boldmath$\partial$}}_{\mu}\}$. The reciprocal basis of
$\{\theta^{\mu}\}$ is denoted $\{\theta_{\mu}\}$, and \texttt{g}$(\theta^{\mu
},\theta_{\nu}):=$ $\theta^{\mu}\underset{\mathtt{\ }}{\cdot}\theta_{\nu
}=\delta_{\nu}^{\mu}$. Introduce next a set of differentiable functions
$q_{\mu}^{\mathbf{a}},q_{\mathbf{b}}^{\nu}:U\rightarrow\mathbb{R}$ such that :%

\begin{equation}
q_{\mathbf{a}}^{\mu}q_{\mu}^{\mathbf{b}}=\mathbf{\delta}_{\mathbf{a}%
}^{\mathbf{b}}\text{, }\qquad q_{\mathbf{a}}^{\mu}q_{\nu}^{\mathbf{a}}%
=\delta_{\nu}^{\mu}\text{ .}%
\end{equation}
It is trivial to verify the formulas
\begin{align}
g_{\mu\nu}  &  =q_{\mu}^{\mathbf{a}}q_{\nu}^{\mathbf{b}}\eta_{\mathbf{ab}%
}\text{, }\qquad g^{\mu\nu}=q_{\mathbf{a}}^{\mu}q_{\mathbf{b}}^{\nu}%
\eta^{\mathbf{ab}},\nonumber\\
\eta_{\mathbf{ab}}  &  =q_{\mathbf{a}}^{\mu}q_{\mathbf{b}}^{\nu}g_{\mu\nu
},\qquad\eta^{\mathbf{ab}}=q_{\mu}^{\mathbf{a}}q_{\nu}^{\mathbf{b}}g^{\mu\nu},
\end{align}
with \
\begin{equation}%
\begin{array}
[c]{c}%
\eta_{\mathbf{ab}}=\mathrm{diag}(\underbrace{1,...,1}\underbrace{-1,...-1})\\
\text{ \ \ \ \ \ \ \ \ \ \ \ \ \ \ \ \ \ }p\text{ {\footnotesize times
\ \ \ \ \ \ }}q\text{ {\footnotesize times}}%
\end{array}
. \label{eta}%
\end{equation}

Moreover, defining
\[
\mathtt{\ }\mathbf{e}_{\mathbf{b}}=q_{\mathbf{b}}^{\nu}%
{\mbox{\boldmath$\partial$}}_{\nu}%
\]
the set $\{\mathbf{e}_{\mathbf{a}}\}$ with $\mathbf{e}_{\mathbf{a}}\in\sec TM$
is an orthonormal basis for $TU$. The dual basis of $TU$ is $\{\theta
^{\mathbf{a}}\}$, with $\theta^{\mathbf{a}}=q_{\mu}^{\mathbf{a}}dx^{\mu}$.
Also, $\{\theta_{\mathbf{b}}\}$ is the reciprocal basis of \ $\{\theta
^{\mathbf{a}}\}$, i.e. $\theta^{\mathbf{a}}\cdot\theta_{\mathbf{b}}%
=\delta_{\mathbf{b}}^{\mathbf{a}}$.

\begin{remark}
When $\dim M=4$ the basis $\{\mathbf{e}_{\mathbf{a}}\}$ of $TU$ is called a
\textit{tetrad} and the \emph{(}dual\emph{)} basis $\{\theta^{\mathbf{a}}\}$
of $T^{\ast}U$ is called a cotetrad. The names are appropriate ones if we
recall the Greek origin of the word.
\end{remark}

The connection coefficients associated to the respective covariant derivatives
in the respective basis will be denoted as:%

\begin{align}
D_{{\mbox{\boldmath$\partial$}}_{\mu}}{\mbox{\boldmath$\partial$}}_{\nu}  &
=\Gamma_{\mu\nu}^{\rho}{\mbox{\boldmath$\partial$}}_{\rho}%
,\;\;\;D_{{\mbox{\boldmath$\partial$}}_{\sigma}}{\mbox{\boldmath$\partial$}}%
^{\mu}=-\Gamma_{\sigma\alpha}^{\mu}{\mbox{\boldmath$\partial$}}^{\alpha},\\
D_{\mathtt{\ }\mathbf{e}_{\mathbf{a}}}\mathtt{\ }\mathbf{e}_{\mathbf{b}}  &
=\omega_{\mathbf{ab}}^{\mathbf{c}}\mathtt{\ }\mathbf{e}_{\mathbf{c}},\qquad
D_{\mathtt{\ }\mathbf{e}_{\mathbf{a}}}\mathtt{\ }\mathbf{e}^{\mathbf{b}%
}=-\omega_{\mathbf{ac}}^{\mathbf{b}}\mathtt{\ }\mathbf{e}^{\mathbf{c}%
},\;\;D_{{\mbox{\boldmath$\partial$}}_{\mu}}\mathtt{\ }\mathbf{e}_{\mathbf{b}%
}=\omega_{\mu\mathbf{b}}^{\mathbf{c}}\mathtt{\ }\mathbf{e}_{\mathbf{c}%
},\nonumber\\
\;D_{{\mbox{\boldmath$\partial$}}_{\mu}}dx^{\nu}  &  =-\Gamma_{\mu\alpha}%
^{\nu}dx^{\alpha},\;\;\;D_{{\mbox{\boldmath$\partial$}}_{\mu}}\theta_{\nu
}=\Gamma_{\mu\nu}^{\rho}\theta_{\rho},\label{14 esp}\\
D_{\mathtt{\ }\mathbf{e}_{\mathbf{a}}}\theta^{\mathbf{b}}  &  =-\omega
_{\mathbf{ac}}^{\mathbf{b}}\theta^{\mathbf{c}}%
,\;\;\;D_{{\mbox{\boldmath$\partial$}}_{\mu}}\theta^{\mathbf{b}}=-\omega
_{\mu\mathbf{a}}^{\mathbf{b}}\theta^{\mathbf{a}}\\
D_{\mathtt{\ }\mathbf{e}_{\mathbf{a}}}\theta^{\mathbf{b}}  &  =-\omega
_{\mathbf{cab}}\theta^{\mathbf{c}},\nonumber\\
\text{ }\omega_{\mathbf{abc}}  &  =\eta_{\mathbf{ad}}\omega_{\mathbf{bc}%
}^{\mathbf{d}}=-\omega_{\mathbf{cba}},\text{ }\omega_{\mathbf{a}}%
^{\mathbf{bc}}=\eta^{\mathbf{bk}}\omega_{\mathbf{kal}}\eta^{\mathbf{cl}%
},\text{ }\omega_{\mathbf{a}}^{\mathbf{bc}}=-\omega_{\mathbf{a}}^{\mathbf{cb}%
}\nonumber\\
&  etc...
\end{align}

\begin{remark}
The connection coefficients of the Levi-Civita Connection in a coordinate
basis are called Christoffel symbols. We write in what follows%
\begin{equation}
\mathring{D}_{{\mbox{\boldmath$\partial$}}_{\mu}}{\mbox{\boldmath$\partial$}}%
_{\nu}=\mathring{\Gamma}_{\mu\nu}^{\rho}{\mbox{\boldmath$\partial$}}_{\rho
},\mathring{D}_{{\mbox{\boldmath$\partial$}}_{\mu}}dx^{\nu}=-\mathring{\Gamma
}_{\mu\rho}^{\nu}dx^{\rho}\text{.} \label{levi-civita}%
\end{equation}

\end{remark}

To understood how $D$ works, consider its action, e.g., on the sections of
$T_{1}^{1}M=TM\otimes T^{\ast}M$.
\begin{equation}
D(X\otimes\alpha)=(DX)\otimes\alpha+X\otimes D\alpha. \label{COV D1}%
\end{equation}

For every vector field $V\in\sec TU$ and a covector field $C\in\sec T^{\ast}U$
we have%

\begin{equation}
D_{{\mbox{\boldmath$\partial$}}_{\mu}}V=D_{{\mbox{\boldmath$\partial$}}_{\mu}%
}(V^{\alpha}{\mbox{\boldmath$\partial$}}_{\alpha}),\quad\text{ }%
D_{{\mbox{\boldmath$\partial$}}_{\mu}}C=D_{{\mbox{\boldmath$\partial$}}_{\mu}%
}(C_{\alpha}\theta^{\alpha}) \label{15'}%
\end{equation}
and using the properties of a covariant derivative operator introduced above,
$D_{{\mbox{\boldmath$\partial$}}_{\mu}}V$ can be written as:%
\begin{align}
D_{{\mbox{\boldmath$\partial$}}_{\mu}}V  &  =D_{{\mbox{\boldmath$\partial$}}%
_{\mu}}(V^{\alpha}{\mbox{\boldmath$\partial$}}_{\alpha}%
)=(D_{{\mbox{\boldmath$\partial$}}_{\mu}}V)^{\alpha}%
{\mbox{\boldmath$\partial$}}_{\alpha}\nonumber\\
&  =({\mbox{\boldmath$\partial$}}_{\mu}V^{\alpha}){\mbox{\boldmath$\partial$}}%
_{\alpha}+V^{\alpha}D_{{\mbox{\boldmath$\partial$}}_{\mu}}%
{\mbox{\boldmath$\partial$}}_{\alpha}\nonumber\\
&  =\left(  \frac{\partial V^{\alpha}}{\partial x^{\mu}}+V^{\rho}\Gamma
_{\mu\rho}^{\alpha}\right)  {\mbox{\boldmath$\partial$}}_{\alpha}:=(D_{\mu
}^{+}V^{\alpha}){\mbox{\boldmath$\partial$}}_{\alpha}, \label{16'}%
\end{align}
where it is to be kept in mind that the symbol $D_{\mu}^{+}V^{\alpha}$ is a
short notation for
\begin{equation}
D_{\mu}^{+}V^{\alpha}:=(D_{{\mbox{\boldmath$\partial$}}_{\mu}}V)^{\alpha}
\label{16bis}%
\end{equation}

Also, we have
\begin{align}
D_{{\mbox{\boldmath$\partial$}}_{\mu}}C  &  =D_{{\mbox{\boldmath$\partial$}}%
_{\mu}}(C_{\alpha}\theta^{\alpha})=(D_{{\mbox{\boldmath$\partial$}}_{\mu}%
}C)_{\alpha}\theta^{\alpha}\nonumber\\
&  =\left(  \frac{\partial C_{\alpha}}{\partial x^{\mu}}-C_{\beta}\Gamma
_{\mu\alpha}^{\beta}\right)  \theta^{\alpha},\nonumber\\
&  :=(D_{\mu}^{-}C_{\alpha})\theta^{\alpha} \label{17'}%
\end{align}
where it is to be kept in mind that \footnote{Recall that other authors prefer
the notations $(\mathbf{\nabla}_{{\mbox{\boldmath$\partial$}}_{\mu}}%
V)^{\alpha}:=V_{:\mu}^{\alpha}$ and $(\mathbf{\nabla}%
_{{\mbox{\boldmath$\partial$}}_{\mu}}C)_{\alpha}:=C_{\alpha:\mu}$. What is
important is always to have in mind the meaning of the symbols.} that the
symbol $D_{\mu}^{-}C_{\alpha}$ is a short notation for
\begin{equation}
D_{\mu}^{-}C_{\alpha}:=(D_{{\mbox{\boldmath$\partial$}}_{\mu}}C)_{\alpha}.
\label{17new}%
\end{equation}

\begin{remark}
The necessity of precise notation becomes obvious when we calculate
\begin{align*}
D_{\mu}^{-}q_{\nu}^{\mathbf{a}}  &  :=(D_{{\mbox{\boldmath$\partial$}}_{\mu}%
}\theta^{\mathbf{a}})_{\nu}=(D_{{\mbox{\boldmath$\partial$}}_{\mu}}q_{\nu
}^{\mathbf{a}}dx^{\nu})_{\nu}={\partial}_{\mu}q_{\nu}^{\mathbf{a}}-\Gamma
_{\mu\nu}^{\rho}q_{\rho}^{\mathbf{a}}=\omega_{\mu\mathbf{b}}^{\mathbf{a}%
}q_{\nu}^{\mathbf{b}},\\
D_{\mu}^{+}q_{\nu}^{\mathbf{a}}  &  :=(D_{{\mbox{\boldmath$\partial$}}_{\mu}%
}q_{\nu}^{\mathbf{a}}\mathbf{e}_{\mathbf{a}})^{\mathbf{a}}=\partial_{\mu
}q_{\nu}^{\mathbf{a}}+\omega_{\mu\nu}^{\rho}q_{\rho}^{\mathbf{a}}=\Gamma
_{\mu\nu}^{\rho}q_{\rho}^{\mathbf{a}},
\end{align*}
\ thus verifying that $D_{\mu}^{-}q_{\nu}^{\mathbf{a}}\neq D_{\mu}^{+}q_{\nu
}^{\mathbf{a}}\neq0$ and that
\begin{equation}
\partial_{\mu}q_{\nu}^{\mathbf{a}}+\omega_{\mu\mathbf{b}}^{\mathbf{a}}q_{\nu
}^{\mathbf{b}}-\Gamma_{\mu\mathbf{b}}^{\mathbf{a}}q_{\nu}^{\mathbf{b}}=0.
\label{freshman}%
\end{equation}
Moreover, if we define the object
\begin{equation}
\mathtt{\ }\mathbf{q=}\mathtt{\ }\mathbf{e}_{\mathbf{a}}\otimes\theta
^{\mathbf{a}}=q_{\mu}^{\mathbf{a}}\mathtt{\ }\mathbf{e}_{\mathbf{a}}\otimes
dx^{\mu}\in\sec T_{1}^{1}U\subset\sec T_{1}^{1}M,
\end{equation}
which is clearly the identity endormorphism acting on sections of $TU$, we
find%
\begin{equation}
D_{\mu}q_{\nu}^{\mathbf{a}}:=(D_{{\mbox{\boldmath$\partial$}}_{\mu}%
}\mathbf{q)}_{\nu}^{\mathbf{a}}=\partial_{\mu}q_{\nu}^{\mathbf{a}}+\omega
_{\mu\mathbf{b}}^{\mathbf{a}}q_{\nu}^{\mathbf{b}}-\Gamma_{\mu\mathbf{b}%
}^{\mathbf{a}}q_{\nu}^{\mathbf{b}}=0. \label{TETRAD}%
\end{equation}

\end{remark}

\begin{remark}
Some authors call $\mathbf{q}\in\sec T_{1}^{1}U$ \emph{(}a single
object\emph{)} a tetrad, thus forgetting the Greek meaning of that word. We
shall avoid this nomenclature. Moreover, \emph{Eq.(\ref{TETRAD}) }is presented
in many textbooks \emph{(}see, e.g., \emph{\cite{carroll,gsw,rovelli}) }and
articles under the name\ `tetrad postulate' and it is said that the covariant
derivative of the "tetrad" vanish. It is obvious that \emph{Eq.(\ref{TETRAD})}
it is not a postulate, it is a trivial \emph{(}freshman\emph{)} identity. In
those books, since authors do not distinguish clearly the derivative operators
$D^{+},D^{-}$ and $D$, \emph{Eq.(\ref{TETRAD})} becomes sometimes
misunderstood as meaning $D_{\mu}^{-}q_{\nu}^{\mathbf{a}}$ or $D_{\mu}%
^{+}q_{\nu}^{\mathbf{a}}$, thus generating a big confusion and producing
errors \emph{(}see below\emph{)}.
\end{remark}

\section{Calculus on the Hodge Bundle $(%
{\displaystyle\bigwedge}
T^{\ast}M,\cdot,\tau_{%
\slg
})$}

We call in what follows Hodge bundle the quadruple $(%
{\displaystyle\bigwedge}
T^{\ast}M,\wedge,\cdot,\tau_{%
\slg
})$. We now recall the meaning of the above symbols.

\subsection{Exterior Product}

We suppose in what follows that any reader of this paper knows the meaning of
the exterior product of form fields and its main properties\footnote{We use
the conventions of \cite{rodoliv2007}.}. We simply recall\ here that if
$\mathcal{A}_{r}\in\sec\bigwedge^{r}T^{\ast}M$, $\mathcal{B}_{s}\in
\sec\bigwedge^{s}T^{\ast}M$ then
\begin{equation}
\mathcal{A}_{r}\wedge\mathcal{B}_{s}=(-1)^{rs}\mathcal{B}_{s}\wedge
\mathcal{A}_{r}. \label{exterior}%
\end{equation}

\subsection{Scalar Product and Contractions}

Let be $\mathcal{A}_{r}=a_{1}\wedge...\wedge a_{r}$ $\in\sec\bigwedge
^{r}T^{\ast}M$, $\mathcal{B}_{r}=b_{1}\wedge...\wedge b_{r}\in\sec
\bigwedge^{r}T^{\ast}M$ \ where $a_{i},b_{j}\in\sec\bigwedge^{1}T^{\ast}M$
$(i,j=1,2,...,r).$

(i) The scalar product $\mathcal{A}_{r}\cdot\mathcal{B}_{r}$ is defined by
\begin{align}
\mathcal{A}_{r}\cdot\mathcal{B}_{r}  &  =(a_{1}\wedge...\wedge a_{r}%
)\cdot(b_{1}\wedge...\wedge b_{r})\nonumber\\
&  =\left\vert
\begin{array}
[c]{lll}%
a_{1}\cdot b_{1} & .... & a_{1}\cdot b_{r}\\
.......... & .... & ..........\\
a_{r}\cdot b_{1} & .... & a_{r}\cdot b_{r}%
\end{array}
\right\vert . \label{scalarprod}%
\end{align}
where $a_{i}\cdot b_{j}:=\mathtt{g}(a_{i},b_{j})$.

We agree that if $r=s=0$, the scalar product is simple the ordinary product in
the real field.

Also, if $r\neq s$, then $\mathcal{A}_{r}\cdot\mathcal{B}_{s}=0$. Finally, the
scalar product is extended by linearity for all sections of $\mathcal{%
{\displaystyle\bigwedge}
}T^{\ast}M$.

For $r\leq s,$ $\mathcal{A}_{r}=a_{1}\wedge...\wedge a_{r},$ $\mathcal{B}%
_{s}=b_{1}\wedge...\wedge b_{s\text{ }}$we define the \textit{left
contraction} by
\begin{equation}
\lrcorner:(\mathcal{A}_{r},\mathcal{B}_{s})\mapsto\mathcal{A}_{r}%
\lrcorner\mathcal{B}_{s}=%
{\displaystyle\sum\limits_{i_{1}\,<...\,<i_{r}}}
\epsilon^{i_{1}....i_{s}}(a_{1}\wedge...\wedge a_{r})\cdot(b_{_{i_{1}}}%
\wedge...\wedge b_{i_{r}})^{\sim}b_{i_{r}+1}\wedge...\wedge b_{i_{s}}
\label{7}%
\end{equation}
where $\sim$ is the reverse mapping (\emph{reversion}) defined by
\begin{equation}
\sim:\sec%
{\displaystyle\bigwedge\nolimits^{p}}
T^{\ast}M\ni a_{1}\wedge...\wedge a_{p}\mapsto a_{p}\wedge...\wedge a_{1}
\label{8}%
\end{equation}
and extended by linearity to all sections of $%
{\displaystyle\bigwedge}
T^{\ast}M$. We agree that for $\alpha,\beta\in\sec\bigwedge^{0}T^{\ast}M$ the
contraction is the ordinary (pointwise) product in the real field and that if
$\alpha\in\sec\bigwedge^{0}T^{\ast}M$, $\mathcal{A}_{r}\in\sec\bigwedge
^{r}T^{\ast}M$, $\mathcal{B}_{s}\in\sec\bigwedge^{s}T^{\ast}M$ then
$(\alpha\mathcal{A}_{r})\lrcorner\mathcal{B}_{s}=\mathcal{A}_{r}%
\lrcorner(\alpha\mathcal{B}_{s})$. Left contraction is extended by linearity
to all pairs of elements of sections of $%
{\displaystyle\bigwedge}
T^{\ast}M$, i.e., for $\mathcal{A},\mathcal{B}\in\sec%
{\displaystyle\bigwedge}
T^{\ast}M$%

\begin{equation}
\mathcal{A\lrcorner B}=\sum_{r,s}\langle\mathcal{A}\rangle_{r}\lrcorner
\langle\mathcal{B}\rangle_{s},\text{ }r\leq s, \label{9bis}%
\end{equation}
where $\langle\mathcal{A}\rangle_{r}$ means the projection of $\mathcal{A}$ in
$%
{\displaystyle\bigwedge\nolimits^{r}}
T^{\ast}M$.

It is also necessary to introduce the operator of \emph{right contraction}
denoted by $\llcorner$. The definition is obtained from the one presenting the
left contraction with the imposition that $r\geq s$ and taking into account
that now if $\mathcal{A}_{r}\in\sec\bigwedge^{r}T^{\ast}M$, $\mathcal{B}%
_{s}\in\sec\bigwedge^{s}T^{\ast}M\ $then $\mathcal{B}_{s}\lrcorner
\mathcal{A}_{r}=(-1)^{s(r-s)}\mathcal{A}_{r}\llcorner\mathcal{B}_{s}$.

\subsection{Hodge Star Operator $\star$}

The Hodge star operator is the mapping
\[
\star:\sec\bigwedge\nolimits^{k}T^{\ast}M\rightarrow\sec\bigwedge
\nolimits^{n-k}T^{\ast}M,\text{ }\mathcal{A}_{k}\mapsto\star\mathcal{A}_{k}%
\]
where for $\mathcal{A}_{k}\in\sec\bigwedge^{k}T^{\ast}M$%
\begin{equation}
\lbrack\mathcal{B}_{k}\cdot\mathcal{A}_{k}]\tau_{%
\slg
}=\mathcal{B}_{k}\wedge\star\mathcal{A}_{k},\text{ }\forall\mathcal{B}_{k}%
\in\sec\bigwedge\nolimits^{k}T^{\ast}M \label{11a}%
\end{equation}
$\tau_{%
\slg
}\in\bigwedge^{n}T^{\ast}M$ is the \textit{metric volume element}. Of course,
the Hodge star operator is naturally extended to an isomorphism $\star
:\sec\bigwedge T^{\ast}M\rightarrow\sec\bigwedge T^{\ast}M$ by linearity. The
inverse $\star^{-1}:\sec\bigwedge^{n-r}T^{\ast}M\rightarrow\sec\bigwedge
^{r}T^{\ast}M$ of the Hodge star operator is given by:
\begin{equation}
\star^{-1}=(-1)^{r(n-r)}\mathrm{sgn}%
\slg
\star,
\end{equation}
where \textrm{sgn }$%
\slg
$ $=\det%
\slg
/|\det%
\slg
|$ denotes the sign of the determinant of the matrix $(g_{\alpha\beta}%
=$\texttt{ }$%
\slg
(e_{\alpha},e_{\beta}))$, where $\{e_{\alpha}\}$ is an \textit{arbitrary}
basis of $TU$.

We can show that (see, e.g., \cite{rodoliv2007}) that
\begin{equation}
\star\mathcal{A}_{k}=\widetilde{\mathcal{A}}_{k}\lrcorner\tau_{%
\slg
}, \label{11b}%
\end{equation}
where as noted before, in this paper $\widetilde{\mathcal{A}}_{k}$ denotes the
\textit{reverse} of $\mathcal{A}_{k}$.

Let $\{\vartheta^{\alpha}\}$ be the dual basis of $\{e_{\alpha}\}$ (i.e., it
is a basis for $T^{\ast}U\equiv\bigwedge\nolimits^{1}T^{\ast}U$) then
\texttt{g}$(\vartheta^{\alpha},\vartheta^{\beta})=g^{\alpha\beta}$, with
$g^{\alpha\beta}g_{\alpha\rho}=\delta_{\rho}^{\beta}$. Writing $\vartheta
^{\mu_{1}...\mu_{p}}=\vartheta^{\mu_{1}}\wedge...\wedge\vartheta^{\mu_{p}}$,
$\vartheta^{\nu_{p+1}...\nu_{n}}=\vartheta^{\nu_{p+1}}\wedge...\wedge
\vartheta^{\nu_{n}}$ we have from Eq.(\ref{11b})
\begin{equation}
{}\star\theta^{\mu_{1}...\mu_{p}}=\frac{1}{(n-p)!}\sqrt{\left\vert \det%
\slg
\right\vert }g^{\mu_{1}\nu_{1}}...g^{\mu_{p}\nu_{p}}\epsilon_{\nu_{1}%
...\nu_{n}}\vartheta^{\nu_{p+1}...\nu_{n}}. \label{hodge dual}%
\end{equation}
Some identities (used below) involving the Hodge star operator, the exterior
product and contractions are\footnote{See also the last formula in
Eq.(\ref{10}) which uses the Clifford product.}:%

\begin{equation}%
\begin{array}
[c]{l}%
A_{r}\wedge\star B_{s}=B_{s}\wedge\star A_{r};\quad r=s\\
A_{r}\cdot\star B_{s}=B_{s}\cdot\star A_{r};\quad r+s=n\\
A_{r}\wedge\star B_{s}=(-1)^{r(s-1)}\star(\tilde{A}_{r}\lrcorner B_{s});\quad
r\leq s\\
A_{r}\lrcorner\star B_{s}=(-1)^{rs}\star(\tilde{A}_{r}\wedge B_{s});\quad
r+s\leq n\\
\star\tau_{%
\slg
}=\mathrm{sign}\text{ }%
\slg
;\quad\star1=\tau_{%
\slg
}.
\end{array}
\label{440new}%
\end{equation}

\subsection{Exterior derivative $d$ and Hodge coderivative $\delta$}

The \textit{exterior derivative is a mapping\/}
\[
d:\sec\bigwedge T^{\ast}M\rightarrow\sec\bigwedge T^{\ast}M,
\]
satisfying:
\begin{equation}%
\begin{array}
[c]{ll}%
\text{(i)} & d(A+B)=dA+dB;\\
\text{(ii)} & d(A\wedge B)=dA\wedge B+\bar{A}\wedge dB;\\
\text{(iii)} & df(v)=v(f);\\
\text{(iv)} & d^{2}=0,
\end{array}
\label{ext deriv}%
\end{equation}
for every $A,B\in\sec\bigwedge T^{\ast}M$, $f\in\sec\bigwedge^{0}T^{\ast}M$
and $v\in\sec TM$.

The \textit{Hodge codifferential} operator in the Hodge bundle is the mapping
$\delta:\sec\bigwedge^{r}T^{\ast}M\rightarrow\sec\bigwedge^{r-1}T^{\ast}M$,
given for homogeneous multiforms, by:%
\begin{equation}
\delta=(-1)^{r}\star^{-1}d\star, \label{hodge}%
\end{equation}
where $\star$ is the Hodge star operator. The operator $\delta$ extends by
linearity to all $\bigwedge T^{\ast}M$

The \textit{Hodge Laplacian (or Hodge D'Alembertian)} operator is the mapping
\[
\Diamond:\sec\bigwedge T^{\ast}M\rightarrow\sec\bigwedge T^{\ast}M
\]
given by:
\begin{equation}
\Diamond=-(d\delta+\delta d). \label{705}%
\end{equation}

The exterior derivative, the Hodge codifferential and the Hodge D' Alembertian
satisfy the relations:
\begin{equation}%
\begin{array}
[c]{l}%
dd=\delta\delta=0;\quad\Diamond=(d-\delta)^{2}\\
d\Diamond=\Diamond d;\quad\delta\Diamond=\Diamond\delta\\
\delta\star=(-1)^{r+1}\star d;\quad\star\delta=(-1)^{r}d\star\\
d\delta\star=\star\delta d;\quad\star d\delta=\delta d\star;\quad\star
\Diamond=\Diamond\star{.}%
\end{array}
\label{545}%
\end{equation}

\section{Clifford Bundles}

Let $(M,%
\slg
,\nabla)$ be a Riemannian, Lorentzian or Riemann-Cartan
structure\footnote{$\nabla$ may be the Levi-Civita connection $\mathring{D}$
of $%
\slg
$ or an arbitrary Riemann-Cartan connection $D$.}. As before let
$\mathtt{g}\in\sec T_{0}^{2}M$ be the metric on the cotangent bundle
associated with $%
\slg
\in\sec T_{2}^{0}M$. Then $T_{x}^{\ast}M\simeq\mathbb{R}^{p,q}$, where
$\mathbb{R}^{p,q}$ is a vector space equipped with a scalar product
$\bullet\equiv\left.  \mathtt{g}\right\vert _{x}$ of signature $(p,q)$. The
Clifford bundle of differential forms $\mathcal{C}\!\ell(M,\mathtt{g})$ is the
bundle of algebras, i.e., $\mathcal{C}\ell(M,\mathtt{g})=\cup_{x\in
M}\mathcal{C}\ell(T_{x}^{\ast}M,\bullet)$, where $\forall x\in M$,
$\mathcal{C}\!\ell(T_{x}^{\ast}M,\mathbf{\bullet})=\mathbb{R}_{p,q}$, a real
Clifford algebra. When the structure $(M,%
\slg
,\nabla)$ is part of a Lorentzian or Riemann-Cartan spacetime $\mathcal{C}%
\ell(T_{x}^{\ast}M,\mathbf{\bullet})=\mathbb{R}_{1,3}$ the so called
\emph{spacetime} \emph{algebra}. Recall also that $\mathcal{C}\ell
(M,\mathtt{g})$ is a vector bundle associated with the \emph{\ }$\mathtt{g}%
$-\emph{orthonormal coframe bundle \ }$\mathbf{P}_{\mathrm{SO}_{(p,q)}^{e}%
}(M,\mathtt{g})$, i.e., $\mathcal{C}\ell(M,\mathtt{g})$ $=P_{\mathrm{SO}%
_{(p,q)}^{e}}(M,\mathtt{g})\times_{ad}\mathbb{R}_{1,3}$ (see more details in,
e.g., \cite{mosnawal,rodoliv2007}). For any $x\in M$, $\mathcal{C}\ell
(T_{x}^{\ast}M,\bullet)$ is a linear space over the real field $\mathbb{R}$.
Moreover, $\mathcal{C}\ell(T_{x}^{\ast}M)$ is isomorphic as a real vector
space to the Cartan algebra $\bigwedge T_{x}^{\ast}M$ of the cotangent space.
Then, sections of $\mathcal{C}\ell(M,\mathtt{g})$ can be represented as a
\textit{sum} of non homogeneous differential forms. Let now $\{\mathbf{e}%
_{\mathbf{a}}\}$ be an orthonormal basis for $TU$ and $\{\theta^{\mathbf{a}%
}\}$ its dual basis. Then, \texttt{g}$(\theta^{\mathbf{a}},\theta^{\mathbf{b}%
})=\eta^{\mathbf{ab}}.$

\subsection{Clifford Product}

The fundamental \emph{Clifford product} (in what follows to be denoted by
juxtaposition of symbols) is generated by%
\begin{equation}
\theta^{\mathbf{a}}\theta^{\mathbf{b}}+\theta^{\mathbf{b}}\theta^{\mathbf{a}%
}=2\eta^{\mathbf{ab}} \label{cp}%
\end{equation}
and if $\mathcal{C}\in\mathcal{C}\ell(M,\mathtt{g})$ we have%

\begin{equation}
\mathcal{C}=s+v_{\mathbf{a}}\theta^{\mathbf{a}}+\frac{1}{2!}b_{\mathbf{ab}%
}\theta^{\mathbf{a}}\theta^{\mathbf{b}}+\frac{1}{3!}a_{\mathbf{abc}}%
\theta^{\mathbf{a}}\theta^{\mathbf{b}}\theta^{\mathbf{c}}+p\theta^{n+1}\;,
\label{3}%
\end{equation}
where $\tau_{%
\slg
}:=\theta^{n+1}=\theta^{0}\theta^{1}\theta^{2}\theta^{3}...\theta^{n}$ is the
volume element and $s$, $v_{\mathbf{a}}$, $b_{\mathbf{ab}}$, $a_{\mathbf{abc}%
}$, $p\in\sec\bigwedge^{0}T^{\ast}M\hookrightarrow\sec\mathcal{C}%
\ell(M,\mathtt{g})$.

Let $\mathcal{A}_{r},\in\sec\bigwedge^{r}T^{\ast}M\hookrightarrow
\sec\mathcal{C}\!\ell(M,\mathtt{g}),\mathcal{B}_{s}\in\sec\bigwedge^{s}%
T^{\ast}M\hookrightarrow\sec\mathcal{C}\ell(M,\mathtt{g})$. For $r=s=1$, we
define the \emph{scalar product} as follows:

For $a,b\in\sec\bigwedge^{1}T^{\ast}M\hookrightarrow\sec\mathcal{C}%
\ell(M,\mathtt{g}),$%
\begin{equation}
a\cdot b=\frac{1}{2}(ab+ba)=\mathtt{g}(a,b). \label{4}%
\end{equation}
We identify the \emph{exterior product} ($\forall r,s=0,1,2,3)$ of homogeneous
forms (already introduced above) by
\begin{equation}
\mathcal{A}_{r}\wedge\mathcal{B}_{s}=\langle\mathcal{A}_{r}\mathcal{B}%
_{s}\rangle_{r+s}, \label{5}%
\end{equation}
where $\langle\rangle_{k}$ is the \textit{component} in $\bigwedge^{k}T^{\ast
}M$ \ (projection) of the Clifford field. The exterior product is extended by
linearity to all sections of $\mathcal{C}\ell(M,\mathtt{g})$.

The scalar product, the left and the right are defined for homogeneous form
fields that are sections of the Clifford bundle in exactly the same way as in
the Hodge bundle and they are extended by linearity for all sections of
$\mathcal{C}\ell(M,\mathtt{g})$.

In particular, for $\mathcal{A},\mathcal{B}\in\sec\mathcal{C}\ell
(M,\mathtt{g})$ we have%

\begin{equation}
\mathcal{A\lrcorner B}=\sum_{r,s}\langle\mathcal{A}\rangle_{r}\lrcorner
\langle\mathcal{B}\rangle_{s},\text{ }r\leq s. \label{9}%
\end{equation}

The main formulas used in the present paper can be obtained (details may be
found in \cite{rodoliv2007}) from the following ones (where $a\in\sec
\bigwedge^{1}T^{\ast}M\hookrightarrow\sec\mathcal{C}\!\ell(M,\mathtt{g})$):
\begin{align}
a\mathcal{B}_{s}  &  =a\lrcorner\mathcal{B}_{s}+a\wedge\mathcal{B}%
_{s},\;\;\mathcal{B}_{s}a=\mathcal{B}_{s}\llcorner a+\mathcal{B}_{s}\wedge
a,\nonumber\\
a\lrcorner\mathcal{B}_{s}  &  =\frac{1}{2}(a\mathcal{B}_{s}-(-1)^{s}%
\mathcal{B}_{s}a),\nonumber\\
\mathcal{A}_{r}\lrcorner\mathcal{B}_{s}  &  =(-1)^{r(s-r)}\mathcal{B}%
_{s}\llcorner\mathcal{A}_{r},\nonumber\\
a\wedge\mathcal{B}_{s}  &  =\frac{1}{2}(a\mathcal{B}_{s}+(-1)^{s}%
\mathcal{B}_{s}a),\nonumber\\
\mathcal{A}_{r}\mathcal{B}_{s}  &  =\langle\mathcal{A}_{r}\mathcal{B}%
_{s}\rangle_{|r-s|}+\langle\mathcal{A}_{r}\mathcal{B}_{s}\rangle
_{|r-s|+2}+...+\langle\mathcal{A}_{r}\mathcal{B}_{s}\rangle_{|r+s|}\nonumber\\
&  =\sum\limits_{k=0}^{m}\langle\mathcal{A}_{r}\mathcal{B}_{s}\rangle
_{|r-s|+2k}\text{ }\nonumber\\
\mathcal{A}_{r}\cdot\mathcal{B}_{r}  &  =\mathcal{B}_{r}\cdot\mathcal{A}%
_{r}=\widetilde{\mathcal{A}}_{r}\text{ }\lrcorner\mathcal{B}_{r}%
=\mathcal{A}_{r}\llcorner\widetilde{\mathcal{B}}_{r}=\langle\widetilde
{\mathcal{A}}_{r}\mathcal{B}_{r}\rangle_{0}=\langle\mathcal{A}_{r}%
\widetilde{\mathcal{B}}_{r}\rangle_{0},\nonumber\\
\star\mathcal{A}_{k}  &  =\widetilde{\mathcal{A}}_{k}\lrcorner\tau_{%
\slg
}=\widetilde{\mathcal{A}}_{k}\tau_{%
\slg
}. \label{10}%
\end{align}
Two other important identities to be used below are:%

\begin{equation}
a\lrcorner(\mathcal{X}\wedge\mathcal{Y})=(a\lrcorner\mathcal{X})\wedge
\mathcal{Y}+\mathcal{\hat{X}}\wedge(a\lrcorner\mathcal{Y}), \label{T54}%
\end{equation}
for any $a\in\sec%
{\displaystyle\bigwedge\nolimits^{1}}
T^{\ast}M$ and $\mathcal{X},\mathcal{Y}\in\sec%
{\displaystyle\bigwedge}
T^{\ast}M$, and
\begin{equation}
A\lrcorner(B\lrcorner C)=(A\wedge B)\lrcorner C, \label{T50}%
\end{equation}
for any $A,B,C\in\sec\bigwedge T^{\ast}M\hookrightarrow\mathcal{C}%
\ell(M,\mathtt{g})$

\subsection{Dirac Operators Acting on Sections of a Clifford Bundle
$\mathcal{C}\ell(M,%
\slg
)$}

\subsubsection{The Dirac Operator\ $\mbox{\boldmath$\partial$}$ Associated to
$D$}

The Dirac operator associated to a general Riemann-Cartan structure $(M,%
\slg
,D)$ acting on sections of $\mathcal{C}\!\ell(M,\mathtt{g})$ is the invariant
first order differential operator
\begin{equation}
\mbox{\boldmath$\partial$}=\theta^{\mathbf{a}}D_{\mathbf{e}_{\mathbf{a}}%
}=\vartheta^{\alpha}D_{e_{\alpha}}. \label{12}%
\end{equation}
For any $\mathcal{A}\in\sec\bigwedge T^{\ast}M\hookrightarrow\sec
\mathcal{C}\!\ell(M,\mathtt{g})$ we define%

\begin{align}
\mbox{\boldmath$\partial$}\mathcal{A}  &  =\mbox{\boldmath$\partial$}\wedge
\mathcal{A}+\mbox{\boldmath$\partial$}\lrcorner\mathcal{A}\nonumber\\
\mbox{\boldmath$\partial$}\wedge\mathcal{A}  &  =\theta^{\mathbf{a}}%
\wedge(D_{\mathbf{e}_{\mathbf{a}}}\mathcal{A}),\hspace{0.1in}%
\,\mbox{\boldmath$\partial$}\lrcorner\mathcal{A}=\theta^{\mathbf{a}}%
\lrcorner(D_{\mathbf{e}_{\mathbf{a}}}\mathcal{A}). \label{DIRAC}%
\end{align}

\subsubsection{Clifford Bundle Calculation of $D_{\mathbf{e}_{\mathbf{a}}%
}\mathcal{A}$}

Recall that the \textit{reciprocal }basis of $\{\theta^{\mathbf{b}}\}$ is
denoted $\{\theta_{\mathbf{a}}\}$ with $\theta_{\mathbf{a}}\cdot
\theta_{\mathbf{b}}=\mathbf{\eta}_{\mathbf{ab}}$ ($\mathbf{\eta}_{\mathbf{ab}%
}=\mathrm{diag}(1,...,1,-1,...,-1)$) and that
\begin{equation}
D_{\mathbf{e}_{\mathbf{a}}}\theta^{\mathbf{b}}=-\omega_{\mathbf{ac}%
}^{\mathbf{b}}\theta^{\mathbf{c}}=-\omega_{\mathbf{a}}^{\mathbf{bc}}%
\theta_{\mathbf{c}}, \label{12n}%
\end{equation}
with $\omega_{\mathbf{a}}^{\mathbf{bc}}=-\omega_{\mathbf{a}}^{\mathbf{cb}}$,
and $\omega_{\mathbf{a}}^{\mathbf{bc}}=\mathbf{\eta}^{\mathbf{bk}}%
\omega_{\mathbf{kal}}\mathbf{\eta}^{\mathbf{cl}},$ $\omega_{\mathbf{abc}%
}=\mathbf{\eta}_{\mathbf{ad}}\omega_{\mathbf{bc}}^{\mathbf{d}}=-\omega
_{\mathbf{cba}}$. Defining
\begin{equation}
\mathbf{\omega}_{\mathbf{a}}=\frac{1}{2}\omega_{\mathbf{a}}^{\mathbf{bc}%
}\theta_{\mathbf{b}}\wedge\theta_{\mathbf{c}}\in\sec%
{\displaystyle\bigwedge\nolimits^{2}}
T^{\ast}M\hookrightarrow\sec\mathcal{C}\ell(M,\mathtt{g}), \label{12nn}%
\end{equation}
we have (by linearity) that \cite{mosnawal} for any $\mathcal{A}\in
\sec\bigwedge T^{\ast}M\hookrightarrow\sec\mathcal{C}\ell(M,\mathtt{g})$
\begin{equation}
D_{\mathbf{e}_{\mathbf{a}}}\mathcal{A}=\partial_{\mathbf{e}_{\mathbf{a}}%
}\mathcal{A}+\frac{1}{2}[\mathbf{\omega}_{\mathbf{a}},\mathcal{A}],
\label{12nnn}%
\end{equation}
where $\partial_{\mathbf{e}_{\mathbf{a}}}$ is the Pfaff derivative, i.e., for
any $A=\frac{1}{p!}A_{\mathbf{i}_{1}...\mathbf{i}_{p}}\theta^{_{\mathbf{i}%
_{1}}}...\theta^{_{.\mathbf{i}_{p}\text{ \ }}}\in\sec\bigwedge^{p}T^{\ast
}M\hookrightarrow\sec\mathcal{C}\ell(M,\mathtt{g})$ it is:
\begin{equation}
\partial_{\mathbf{e}_{\mathbf{a}}}A=\frac{1}{p!}[\mathbf{e}_{\mathbf{a}%
}(A_{\mathbf{i}_{1}...\mathbf{i}_{p}})]\theta^{_{\mathbf{i}_{1}}}%
...\theta^{_{.\mathbf{i}_{p}\text{ }}}. \label{pfaff}%
\end{equation}

\subsubsection{The Dirac Operator $%
\protect\bpartial
$ Associated to $\mathring{D}$}

Using Eq.(\ref{12nnn}) \ we can show that for the case of a Riemannian or
Lorentzian structure $(M,%
\slg
,\mathring{D})$ the standard Dirac operator defined by:\hspace{-0.06in}%
\begin{align}%
\bpartial
&  =\theta^{\mathbf{a}}\mathring{D}_{\mathbf{e}_{\mathbf{a}}}=\vartheta
^{\alpha}\mathring{D}_{e_{\alpha}},\nonumber\\%
\bpartial
\mathcal{A}  &  =%
\bpartial
\wedge\mathcal{A}+%
\bpartial
\lrcorner\mathcal{A} \label{sd1}%
\end{align}
for any $\mathcal{A}\in\sec\bigwedge T^{\ast}M\hookrightarrow\sec
\mathcal{C}\ell(M,\mathtt{g})$ is such that%

\begin{equation}%
\bpartial
\wedge\mathcal{A}=d\mathcal{A}\text{, \ \ }%
\bpartial
\lrcorner\mathcal{A=-}\delta\mathcal{A} \label{sd2}%
\end{equation}
i.e.,%
\begin{equation}%
\bpartial
=d-\delta\label{sd3}%
\end{equation}

\section{Torsion, Curvature and Cartan Structure Equations}

As we said in the beginning of Section 1 a given structure $(M,%
\slg
)$ may admit many different metric compatible connections. Let then
$\mathring{D}$ be the Levi-Civita connection of $%
\slg
$ and $D$ a Riemann-Cartan connection acting on the tensor fields defined on
$M$.

Let $U\subset M$ and consider a chart of the maximal atlas of $M$ covering $U$
with arbitrary coordinates $\{x^{\mu}\}$. Let $\{{\mbox{\boldmath$\partial$}}%
_{\mu}\}$ be a basis for $TU$, $U\subset M$ and let $\{\theta^{\mu}=dx^{\mu
}\}$ be the dual basis of $\{{\mbox{\boldmath$\partial$}}_{\mu}\}$. The
reciprocal basis of $\{\theta^{\mu}\}$ is denoted $\{\theta^{\mu}\}$, and
\texttt{g}$(\theta^{\mu},\theta_{\nu}):=$ $\theta^{\mu}\cdot\theta_{\nu
}=\delta_{\nu}^{\mu}$.

Let also $\{\mathbf{e}_{\mathbf{a}}\}$ be an orthonormal basis for $TU\subset
TM$ \ with $\mathbf{e}_{\mathbf{b}}=q_{\mathbf{b}}^{\nu}%
{\mbox{\boldmath$\partial$}}_{\nu}$.$\mathtt{\ }$The dual basis of $TU$ is
$\{\theta^{\mathbf{a}}\}$, with $\theta^{\mathbf{a}}=q_{\mu}^{\mathbf{a}%
}dx^{\mu}$. Also, $\{\theta_{\mathbf{b}}\}$ is the reciprocal basis of
$\{\theta^{\mathbf{a}}\}$, i.e. $\theta^{\mathbf{a}}\cdot\theta_{\mathbf{b}%
}=\delta_{\mathbf{b}}^{\mathbf{a}}$. An arbitrary frame on $TU\subset TM$,
coordinate or orthonormal will be denote by $\{e_{\alpha}\}$. Its dual frame
will be denoted by $\{\vartheta^{\rho}\}$ (i.e., $\vartheta^{\rho}(e_{\alpha
})=\delta_{\alpha}^{\rho}$ ).

\subsection{Torsion and Curvature Operators}

\begin{definition}
The \textit{torsion and curvature operators }$\mathbf{\tau}$ and
$\mathbf{\rho}$ of a connection $D$, are respectively the mappings:%
\begin{align}
\mathbf{\tau}(\mathbf{u,v})  &  =D_{\mathbf{u}}\mathbf{v}-D_{\mathbf{v}%
}\mathbf{u}-[\mathbf{u,v}],\label{top}\\
\mathbf{\rho(u,v)}  &  =D_{\mathbf{u}}D_{\mathbf{v}}-D_{\mathbf{v}%
}D_{\mathbf{u}}-D_{[\mathbf{u,v}]}, \label{cop}%
\end{align}
for every $\mathbf{u,v}\in\sec TM$.
\end{definition}

\subsection{Torsion and Curvature Tensors}

\begin{definition}
The \textit{torsion and\/curvature }tensors of a connection $D$, are
respectively the mappings:%

\begin{align}
\mathbf{T}(\alpha,\mathbf{u,v})  &  =\alpha\left(  \mathbf{\tau}%
(\mathbf{u},\mathbf{v})\right)  ,\label{to op}\\
\mathbf{R}(\mathbf{w},\alpha,\mathbf{u,v})  &  =\alpha(\mathbf{\rho(u,v)w}),
\label{curv op}%
\end{align}
for every $\mathbf{u,v,w}\in\sec TM$ and $\alpha\in\sec\bigwedge^{1}T^{\ast}M$.
\end{definition}

We recall that for any differentiable functions $f,g$ and $h$ we have%
\begin{align}
\mathbf{\tau}(g\mathbf{u,}h\mathbf{v})  &  =gh\mathbf{\tau}(\mathbf{u,v}%
),\nonumber\\
\mathbf{\rho(}g\mathbf{u,}h\mathbf{v)}f\mathbf{w}  &  \mathbf{=}%
ghf\mathbf{\rho(u,v)w} \label{exerctocur}%
\end{align}

\subsubsection{Properties of the Riemann Tensor for a Metric Compatible
Connection}

Note that it is quite obvious that
\begin{equation}
\mathbf{R}(\mathbf{w},\alpha,\mathbf{u,v})=\mathbf{R}(\mathbf{w}%
,\alpha,\mathbf{v,u}).
\end{equation}
Define the tensor field $\mathbf{R}^{\prime}$ as the mapping such that for
every $\mathbf{a,u,v,w}\in\sec TM$ and $\alpha\in\sec\bigwedge^{1}T^{\ast}M$.
\begin{equation}
\mathbf{R}^{\prime}(\mathbf{w},\mathbf{a},\mathbf{u,v})=\mathbf{R}%
(\mathbf{w},\alpha,\mathbf{v,u}). \label{ri3}%
\end{equation}
It is quite ovious that
\begin{equation}
\mathbf{R}^{\prime}(\mathbf{w},\mathbf{a},\mathbf{u,v})=\mathbf{a\cdot
}(\mathbf{\rho(u,v)w}), \label{ri4}%
\end{equation}
where
\begin{equation}
\alpha=%
\slg
(\mathbf{a,\;)}\text{, }\mathbf{a}=\mathtt{g}(\alpha,\;) \label{ri5}%
\end{equation}
We now show that for any structure $(M,%
\slg
,D)$ such that $D%
\slg
=0$ we have for $\mathbf{c},\mathbf{u,v}\in\sec TM$,
\begin{equation}
\mathbf{R}^{\prime}(\mathbf{c},\mathbf{c},\mathbf{u,v})=\mathbf{c\cdot
}(\mathbf{\rho(u,v)c})=0. \label{ri6}%
\end{equation}

We start recalling that for every metric compatible connection it holds:
\begin{align}
\mathbf{u(v(c\cdot c)}  &  \mathbf{=u(}D_{\mathbf{v}}\mathbf{c\cdot c+c\cdot
}D_{\mathbf{v}}\mathbf{c)=}2\mathbf{u(}D_{\mathbf{v}}\mathbf{c\cdot
c)}\nonumber\\
&  =2\mathbf{(}D_{\mathbf{u}}D_{\mathbf{v}}\mathbf{c)\cdot c}+2\mathbf{(}%
D_{\mathbf{u}}\mathbf{c)\cdot D_{\mathbf{v}}c,} \label{ri7}%
\end{align}
Exachanging $\mathbf{u}\leftrightarrow\mathbf{v}$ in the last equation we get%
\begin{equation}
\mathbf{v(u(c\cdot c)=}2\mathbf{(}D_{\mathbf{v}}D_{\mathbf{u}}\mathbf{c)\cdot
c}+2\mathbf{(}D_{\mathbf{v}}\mathbf{c)\cdot D_{\mathbf{u}}c.} \label{ri8}%
\end{equation}
Subtracting Eq.(\ref{ri7}) from Eq.(\ref{ri8}) we have
\begin{equation}
\lbrack\mathbf{u,v](c\cdot c)=}2([D_{\mathbf{u}},D_{\mathbf{v}}%
]\mathbf{c)\cdot c} \label{ri9}%
\end{equation}
But since
\begin{equation}
\lbrack\mathbf{u,v](c\cdot c)=}D_{[\mathbf{u,v]}}(\mathbf{c\cdot
c})=2(D_{[\mathbf{u,v]}}\mathbf{c)\cdot c,} \label{ri10}%
\end{equation}
we have from Eq.(\ref{ri9}) that
\begin{equation}
([D_{\mathbf{u}},D_{\mathbf{v}}]\mathbf{c-}D_{[\mathbf{u,v]}}\mathbf{c)\cdot
c}=0\text{,} \label{ri11}%
\end{equation}
and it follows that $\mathbf{R}^{\prime}(\mathbf{c},\mathbf{c},\mathbf{u,v}%
)=0$ as we wanted to show.

\begin{exercise}
Prove that for any metric compatible connection,
\begin{equation}
\mathbf{R}^{\prime}(\mathbf{c},\mathbf{d},\mathbf{u,v})=\mathbf{R}^{\prime
}(\mathbf{d},\mathbf{c},\mathbf{v,u}). \label{ri12}%
\end{equation}

\end{exercise}

Given an arbitrary frame $\{e_{\alpha}\}$ on $TU\subset TM$, let
$\{\vartheta^{\rho}\}$ be the \textit{dual frame}. We write:
\begin{equation}%
\begin{array}
[c]{ccl}%
\lbrack e_{\alpha}\mathbf{,}e_{\beta}] & = & c_{\alpha\beta}^{\rho}e_{\rho}\\
D_{e_{\alpha}}e_{\beta} & = & \mathbf{L}_{\alpha\beta}^{\rho}e_{\rho},
\end{array}
\end{equation}
where $c_{\alpha\beta}^{\rho}$ are the \textit{structure coefficients\/}of the
frame $\{e_{\alpha}\}$ and $\mathbf{L}_{\alpha\beta}^{\rho}$ are the
\textit{connection coefficients\/}in this frame. Then, the components of the
torsion and curvature tensors are given, respectively, by:
\begin{equation}%
\begin{array}
[c]{c}%
\mathbf{T}(\vartheta^{\rho},e_{\alpha}\mathbf{,}e_{\beta})=T_{\alpha\beta
}^{\rho}=\mathbf{L}_{\alpha\beta}^{\rho}-\mathbf{L}_{\beta\alpha}^{\rho
}-c_{\alpha\beta}^{\rho}\\
\mathbf{R(}e_{\mu},\vartheta^{\rho},e_{\alpha}\mathbf{,}e_{\beta}%
\mathbf{)}=R_{\mu}{}^{\rho}{}_{\!\alpha\beta}=e_{\alpha}(\mathbf{L}_{\beta\mu
}^{\rho})-e_{\beta}(\mathbf{L}_{\alpha\mu}^{\rho})+\mathbf{L}_{\alpha\sigma
}^{\rho}\mathbf{L}_{\beta\mu}^{\sigma}-\mathbf{L}_{\beta\sigma}^{\rho
}\mathbf{L}_{\alpha\mu}^{\sigma}-c_{\alpha\beta}^{\sigma}\mathbf{L}_{\sigma
\mu}^{\rho}.
\end{array}
\label{585}%
\end{equation}

It is important for what follows to keep in mind the definition of the
(symmetric) Ricci tensor, here denoted $\mathbf{Ric}\in\sec T_{2}^{0}M$ and
which in an arbitrary basis is written as
\begin{equation}
\mathbf{Ric=}R\mathbf{_{\mu\nu}\vartheta^{\mu}\otimes\vartheta^{\nu}:=}R_{\mu
}{}^{\rho}{}_{\!\rho\nu}\vartheta^{\mu}\otimes\vartheta^{\nu} \label{Ricci}%
\end{equation}
It is crucial here to take into account the \textit{place} where the
contractions in the Riemann tensor takes place according to our conventions.

We also have:
\begin{equation}%
\begin{array}
[c]{l}%
d\vartheta^{\rho}=-\frac{1}{2}c_{\alpha\beta}^{\rho}\vartheta^{\alpha}%
\wedge\vartheta^{\beta}\\
D_{e_{\alpha}}\vartheta^{\rho}=-\mathbf{L}_{\alpha\beta}^{\rho}\vartheta
^{\beta}%
\end{array}
\label{608}%
\end{equation}
where $\omega_{\beta}^{\rho}\in\sec\bigwedge^{1}T^{\ast}M$ are the
\textit{connection 1-forms, }$\mathbf{L}_{\alpha\beta}^{\rho}$ \ are said to
be the connection coefficients in the given basis, and the $\mathcal{T}^{\rho
}\in\sec\bigwedge^{2}T^{\ast}M$ are the \textit{torsion 2-forms\/} and the
$\mathcal{R}_{\beta}^{\rho}\in\sec\bigwedge^{2}T^{\ast}M$\textbf{\ }are the
\textit{curvature 2-forms\/}, given by:
\begin{align}
&  \omega_{\beta}^{\rho}=\mathbf{L}_{\alpha\beta}^{\rho}\vartheta^{\alpha
},\nonumber\\
&  \mathcal{T}^{\rho}=\frac{1}{2}T_{\alpha\beta}^{\rho}\vartheta^{\alpha
}\wedge\theta^{\beta}\label{620}\\
&  \mathcal{R}_{\mu}^{\rho}=\frac{1}{2}R_{\mu}{}^{\rho}{}_{\!\alpha\beta
}\vartheta^{\alpha}\wedge\vartheta^{\beta}.\nonumber
\end{align}

Multiplying Eqs.(\ref{585}) by $\frac{1}{2}\vartheta^{\alpha}\wedge
\vartheta^{\beta}$ and using Eqs.(\ref{608}) and~(\ref{620}), we get:

\subsection{Cartan Structure Equations\/}%

\begin{equation}%
\begin{array}
[c]{l}%
d\vartheta^{\rho}+\omega_{\beta}^{\rho}\wedge\vartheta^{\beta}=\mathcal{T}%
^{\rho},\\
d\omega_{\mu}^{\rho}+\omega_{\beta}^{\rho}\wedge\omega_{\mu}^{\beta
}=\mathcal{R}_{\mu}^{\rho}.
\end{array}
\label{559}%
\end{equation}

We can show that the torsion and (Riemann) curvature tensors can be written as%
\begin{align}
\mathbf{T}  &  =e_{\alpha}\otimes\mathcal{T}^{\alpha},\\
\mathbf{R}  &  =e_{\rho}\otimes e^{\mu}\otimes\mathcal{R}_{\mu}^{\rho}.
\end{align}

\section{Exterior Covariant Derivative $\mathbf{D}$}

Sometimes, Eqs.(\ref{559}) are written by some authors \cite{thiwal} as:
\begin{align}
\mathbf{D}\vartheta^{\rho}  &  =\mathcal{T}^{\rho},\label{559a}\\
\text{\textquotedblleft\ }\mathbf{D}\omega_{\mu}^{\rho}  &  =\mathcal{R}_{\mu
}^{\rho}.\text{\textquotedblright} \label{559b}%
\end{align}
and $\mathbf{D}:\sec\bigwedge T^{\ast}M\rightarrow\sec\bigwedge T^{\ast}M$ is
said to be the \textit{exterior covariant derivative }\/related to the
connection $D$. Now, Eq.(\ref{559b}) has been printed with quotation marks due
to the fact that it is an \textit{incorrect} equation. Indeed, a\textit{
legitimate} exterior covariant derivative operator\footnote{Sometimes also
called exterior covariant differential.} is a concept that can be defined for
$(p+q)$-indexed $r$-form fields\footnote{Which is not the case of the
connection $1$-forms $\omega_{\beta}^{\alpha}$, despite the name. More
precisely, the $\omega_{\beta}^{\alpha}$ are not true indexed forms, i.e.,
there does not exist a tensor field $\mathbf{\omega}$ such that
$\mathbf{\omega(}e_{i},e_{\beta},\vartheta^{\alpha})=$ $\omega_{\beta}%
^{\alpha}(e_{i}).$} as follows. Suppose that $X\in\sec T_{p}^{r+q}M$ and let
\begin{equation}
X_{\nu_{1}....\nu_{q}}^{\mu_{1}....\mu_{p}}\in\sec\bigwedge\nolimits^{r}%
T^{\ast}M, \label{559new}%
\end{equation}
such that for $v_{i}\in\sec TM,$ $i=0,1,2,..,r$,
\begin{equation}
X_{\nu_{1}....\nu_{q}}^{\mu_{1}....\mu_{p}}(v_{1},...,v_{r})=X(v_{1}%
,...,v_{r},e_{\nu_{1}},...,e_{\nu_{q}},\vartheta^{\mu_{1}},...,\vartheta
^{\mu_{p}}). \label{559new1}%
\end{equation}

The exterior covariant differential $\mathbf{D}$\textbf{ }of $X_{\nu
_{1}....\nu_{q}}^{\mu_{1}....\mu_{p}}$ on a manifold with a general connection
$D$ is the mapping:%

\begin{equation}
\mathbf{D:}\sec\bigwedge\nolimits^{r}T^{\ast}M\rightarrow\sec\bigwedge
\nolimits^{r+1}T^{\ast}M\text{, }0\leq r\leq4, \label{559new2}%
\end{equation}
such that\footnote{As usual the inverted hat over a symbol (in
Eq.(\ref{559new3})) means that the corresponding symbol is missing in the
expression.}%
\begin{align}
&  (r+1)\mathbf{D}X_{\nu_{1}....\nu_{q}}^{\mu_{1}....\mu_{p}}(v_{0}%
,v_{1},...,v_{r})\nonumber\\
&  =\sum\limits_{\nu=0}^{r}(-1)^{\nu}D_{\mathbf{e}_{\nu}}X(v_{0}%
,v_{1},...,\check{v}_{\nu},...v_{r},e_{\nu_{1}},...,e_{\nu_{q}},\vartheta
^{\mu_{1}},...,\vartheta^{\mu_{p}})\nonumber\\
&  -\sum\limits_{0\leq\lambda,\varsigma\,\leq r}(-1)^{\nu+\varsigma
}X(\mathbf{T(}v_{\lambda},v_{\varsigma}),v_{0},v_{1},...,\check{v}_{\lambda
},...,\check{v}_{\varsigma},...,v_{r},e_{\nu_{1}},...,e_{\nu_{q}}%
,\vartheta^{\mu_{1}},...,\vartheta^{\mu_{p}}). \label{559new3}%
\end{align}

Then, we may verify that
\begin{align}
\mathbf{D}X_{\nu_{1}....\nu_{q}}^{\mu_{1}....\mu_{p}}  &  =dX_{\nu_{1}%
....\nu_{q}}^{\mu_{1}....\mu_{p}}+\omega_{\mu_{s}}^{\mu_{1}}\wedge X_{\nu
_{1}....\nu_{q}}^{\mu_{s}....\mu_{p}}+...+\text{ }\omega_{\mu_{s}}^{\mu_{1}%
}\wedge X_{\nu_{1}....\nu_{q}}^{\mu_{1}....\mu_{p}}\label{559new4}\\
&  -\omega_{\nu_{1}}^{\nu_{s}}\wedge X_{\nu_{s}....\nu_{q}}^{\mu_{1}%
....\mu_{p}}-...-\text{ }\omega_{\mu_{s}}^{\mu_{1}}\wedge X_{\nu_{1}%
....\nu_{s}}^{\mu_{1}....\mu_{p}}.\nonumber
\end{align}

\begin{remark}
Note that if \emph{Eq.(\ref{559new4})} is applied on any one of the connection
$1$-forms $\omega_{\nu}^{\mu}$ we would get $\mathbf{D}\omega_{\nu}^{\mu
}=d\omega_{\nu}^{\mu}+\omega_{\alpha}^{\mu}\wedge\omega_{\nu}^{\alpha}%
-\omega_{\nu}^{\alpha}\wedge\omega_{\alpha}^{\mu}$. So, we see that the symbol
$\mathbf{D}\omega_{\nu}^{\mu}$ in \emph{Eq.(\ref{559b})}, supposedly defining
the curvature $2$-forms is simply wrong despite this being an equation printed
in many Physics textbooks and many professional articles\emph{\footnote{The
authors of reference \cite{thiwal} knows exactly what they are doing and use
"$\mathbf{D}\omega_{\mu}^{\rho}=\mathcal{R}_{\mu}^{\rho}$" only as a short
notation. Unfortunately this is not the case for some other authors.}}! .
\end{remark}

\subsection{ Properties of $\mathbf{D}$}

The exterior covariant derivative $\mathbf{D}$ satisfy the following properties:

\textbf{(a)} For any $X^{J}$ $\in\sec\bigwedge\nolimits^{r}T^{\ast}M$ and
$Y^{K}$ $\in\sec\bigwedge\nolimits^{s}T^{\ast}M$ are sets of indexed
forms\footnote{Multi indices are here represented by $J$ and $K$.
\par
{}}, then%
\begin{equation}
\mathbf{D(}X^{J}\wedge Y^{K}\mathbf{)=D}X^{J}\wedge Y^{K}+(-1)^{rs}X^{J}%
\wedge\mathbf{D}Y^{K}. \label{559new5}%
\end{equation}

\textbf{(b)} For any $X^{\mu_{1}....\mu_{p}}\in\sec\bigwedge\nolimits^{r}%
T^{\ast}M$ then%
\begin{equation}
\mathbf{DD}X^{\mu_{1}....\mu_{p}}=dX^{\mu_{1}....\mu_{p}}+\mathcal{R}_{\mu
_{s}}^{\mu_{1}}\wedge X^{\mu_{s}....\mu_{p}}+...\mathcal{R}_{\mu_{s}}^{\mu
_{p}}\wedge X^{\mu_{1}....\mu_{s}}. \label{559new6}%
\end{equation}

\textbf{(c)} For any metric-compatible connection $D$ if $g=g_{\mu\nu
}\vartheta^{\mu}\otimes\vartheta^{\nu}$ then,%
\begin{equation}
\mathbf{D}g_{\mu\nu}=0. \label{559new7}%
\end{equation}

\subsection{Formula for Computation of the Connection $1$- Forms
$\omega_{\mathbf{b}}^{\mathbf{a}}$}

In an orthonormal cobasis $\{\theta^{\mathbf{a}}\}$ we have (see, e.g.,
\cite{rodoliv2007}) for the connection $1$-forms
\begin{equation}
\omega^{\mathbf{cd}}=\frac{1}{2}\left[  \theta^{\mathbf{d}}\lrcorner
d\theta^{\mathbf{c}}-\theta^{\mathbf{c}}\lrcorner d\theta^{\mathbf{d}}%
+\theta^{\mathbf{c}}\lrcorner(\theta^{\mathbf{d}}\lrcorner d\theta
_{\mathbf{a}})\theta^{\mathbf{a}}\right]  , \label{conn}%
\end{equation}
or taking into account that $d\theta^{\mathbf{a}}=-\frac{1}{2}c_{\mathbf{jk}%
}^{\mathbf{a}}\theta^{\mathbf{j}}\wedge\theta^{\mathbf{k}},$
\begin{equation}
\omega^{\mathbf{cd}}=\frac{1}{2}(-c_{\mathbf{jk}}^{\mathbf{c}}\eta
^{\mathbf{dj}}+c_{\mathbf{jk}}^{\mathbf{d}}\eta^{\mathbf{cj}}-\eta
^{\mathbf{ca}}\eta_{\mathbf{bk}}\eta^{\mathbf{dj}}c_{\mathbf{ja}}^{\mathbf{b}%
})\theta^{\mathbf{k}}. \label{conna}%
\end{equation}

\section{Relation Between the Connection $\mathring{D}$ and $D$}

As we said above a given structure $(M,%
\slg
)$ in general admits many different connections. Let then $\mathring{D}$ and
\ $D$ be the Levi-Civita connection of $%
\slg
$ on $M$ and $D$ and arbitrary Riemann-Cartan connection. Given an arbitrary
basis $\{e_{\alpha}\}$ on $TU\subset TM$, let $\{\vartheta^{\rho}\}$ be the
dual frame. We write for the connection coefficients of the Riemann-Cartan and
the Levi-Civita connections in the arbitrary bases $\{e_{\alpha}%
\}$,$\{\vartheta^{\rho}\}$:
\begin{align}
D_{e_{\alpha}}e_{\beta}  &  =\mathbf{L}_{\alpha\beta}^{\rho}e_{\rho},\text{
}D_{e_{\alpha}}\vartheta^{\rho}=-\mathbf{L}_{\alpha\beta}^{\rho}%
\vartheta^{\beta},\nonumber\\
\mathring{D}_{e_{\alpha}}e_{\beta}  &  =\mathbf{\mathring{L}}_{\alpha\beta
}^{\rho}e_{\rho},\text{ }\mathring{D}_{e_{\alpha}}\vartheta^{\rho
}=-\mathbf{\mathring{L}}_{\alpha\beta}^{\rho}\vartheta^{\beta}.
\label{L and R}%
\end{align}
Moreover, the structure coefficients of the arbitrary basis $\{e_{\alpha}\}$
are:
\begin{equation}
\lbrack e_{\alpha}\mathbf{,}e_{\beta}]=c_{\alpha\beta}^{\rho}e_{\rho}.
\label{LR}%
\end{equation}
Let moreover,%
\begin{equation}
b_{\alpha\beta}^{\rho}=-(\pounds _{e^{\rho}}%
\slg
)_{\alpha\beta}, \label{lie}%
\end{equation}
where $\pounds _{e^{\rho}}$ is the Lie derivative in the direction of the
vector field $e^{\rho}$. \ Then, we have the noticeable formula (for a proof,
see, e.g., \cite{rodoliv2007}):%
\begin{equation}
\mathbf{L}_{\alpha\beta}^{\rho}=\mathbf{\mathring{L}}_{\alpha\beta}^{\rho
}+\frac{1}{2}T_{\alpha\beta}^{\rho}+\frac{1}{2}S_{\alpha\beta}^{\rho},
\label{LR1}%
\end{equation}
where the tensor $S_{\alpha\beta}^{\rho}$ is called the strain tensor of the
connection and can be decomposed as:%

\begin{equation}
S_{\alpha\beta}^{\rho}=\breve{S}_{\alpha\beta}^{\rho}+\frac{2}{n}s^{\rho
}g_{\alpha\beta} \label{LR2}%
\end{equation}
where $\breve{S}_{\alpha\beta}^{\rho}$ is its traceless part, is called the
\textit{shear\/} of the connection, and
\begin{equation}
s^{\rho}=\frac{1}{2}g^{\mu\nu}S_{\mu\nu}^{\rho} \label{thes}%
\end{equation}
is its trace part, is called the \textit{dilation \/}of the connection. We
also have that connection coefficients of the Levi-Civita connection can be
written as:%
\begin{equation}
\mathbf{\mathring{L}}_{\alpha\beta}^{\rho}=\frac{1}{2}(b_{\alpha\beta}^{\rho
}+c_{\alpha\beta}^{\rho}). \label{LR3}%
\end{equation}

Moreover, we introduce the \textit{contorsion tensor} whose components in an
arbitrary basis are defined by%

\begin{equation}
K_{\alpha\beta}^{\!\rho}=\mathbf{L}_{\alpha\beta}^{\rho}-\mathbf{\mathring{L}%
}_{\alpha\beta}^{\rho}=\frac{1}{2}(T_{\alpha\beta}^{\rho}+S_{\alpha\beta
}^{\rho}), \label{1168}%
\end{equation}
and which can be written as
\begin{equation}
K_{\alpha\beta}^{\!\rho}=-\frac{1}{2}g^{\rho\sigma}(g_{\mu\alpha}%
T_{\sigma\beta}^{\mu}+g_{\mu\beta}T_{\sigma\alpha}^{\mu}-g_{\mu\sigma
}T_{\alpha\beta}^{\mu}). \label{1130}%
\end{equation}

We now present the relation between the Riemann curvature tensor $R_{\mu}%
{}^{\rho}{}_{\!\alpha\beta}$ associated with the Riemann-Cartan connection $D$
and the Riemann curvature tensor $\mathring{R}{}_{\mu}{}^{\rho}{}%
_{\!\alpha\beta}$ of the Levi-Civita connection $\mathring{D}$.
\begin{equation}
R_{\mu}{}^{\rho}{}_{\!\alpha\beta}=\mathring{R}_{\mu}{}^{\rho}{}%
_{\!\alpha\beta}+J_{\mu}{}^{\rho}{}_{\![\alpha\beta]}, \label{1070}%
\end{equation}
where:
\begin{equation}
J_{\!\mu}{}^{\rho}{}_{\!\alpha\beta}=\mathring{D}_{\!\alpha}K_{\beta\mu
}^{\!\rho}-K_{\beta\sigma}^{\!\rho}K_{\alpha\mu}^{\!\sigma}=D_{\alpha}%
K_{\beta\mu}^{\!\rho}-K_{\alpha\sigma}^{\!\rho}K_{\beta\mu}^{\!\sigma
}+K_{\alpha\beta}^{\!\sigma}K_{\sigma\mu}^{\!\rho}.
\end{equation}
Multiplying both sides of Eq.(\ref{1070}) by $\frac{1}{2}\theta^{\alpha}%
\wedge\theta^{\beta}$ we get:
\begin{equation}
\mathcal{R}_{\mu}^{\rho}=\mathcal{\mathring{R}}_{\mu}^{\rho}+\mathfrak{J}%
_{\mu}^{\rho}, \label{1208}%
\end{equation}
where
\begin{equation}
\mathfrak{J}_{\mu}^{\rho}=\frac{1}{2}J_{\!\mu}{}^{\rho}{}_{\![\alpha\beta
]}\theta^{\alpha}\wedge\theta^{\beta}.
\end{equation}

From Eq.(\ref{1070}) we also get the relation between the Ricci tensors of the
connections $D$ and $\mathring{D}$. We write for the \textit{Ricci tensor of
}$D$
\begin{align}
\mathbf{Ric}  &  =R_{\mu\alpha}dx^{\mu}\otimes dx^{\nu}\nonumber\\
R_{\mu\alpha}  &  :=R_{\mu}{}^{\rho}{}_{\!\alpha\rho} \label{ricci}%
\end{align}
Then, we have
\begin{equation}
R_{\mu\alpha}=\mathring{R}_{\mu\alpha}+J_{\mu\alpha}, \label{1174}%
\end{equation}
with
\begin{align}
J_{\mu\alpha}  &  =\mathring{D}_{\!\alpha}K_{\rho\mu}^{\!\rho}-\mathring
{D}_{\!\rho}K_{\alpha\mu}^{\!\rho}+K_{\alpha\sigma}^{\!\rho}K_{\rho\mu
}^{\!\sigma}-K_{\rho\sigma}^{\!\rho}K_{\alpha\mu}^{\!\sigma}\nonumber\\
&  =D_{\!\alpha}K_{\rho\mu}^{\!\rho}-D_{\!\rho}K_{\alpha\mu}^{\!\rho
}-K_{\sigma\alpha}^{\!\rho}K_{\rho\mu}^{\!\sigma}+K_{\rho\sigma}^{\!\rho
}K_{\alpha\mu}^{\!\sigma}.
\end{align}
Observe that since the connection $D$ is arbitrary, its Ricci tensor will be
\textit{not} be symmetric in general. Then, since the Ricci tensor
$\mathring{R}_{\mu\alpha}$ of $\mathring{D}$ is necessarily symmetric, we can
split Eq.(\ref{1174}) into:
\begin{equation}%
\begin{tabular}
[c]{c}%
$R_{[\mu\alpha]}=J_{[\mu\alpha]},$\\
\\
$R_{(\mu\alpha)}=\mathring{R}_{(\mu\alpha)}+J_{(\mu\alpha)}.$%
\end{tabular}
\ \ \ \ \ \label{1190}%
\end{equation}

\section{Expressions for $d$ and $\delta$ in Terms of Covariant Derivative
Operators $\mathring{D}$ and $D$}

We have the following noticeable formulas whose proof can be found in, e.g.,
\cite{rodoliv2007}. Let $\mathcal{Q}\in\sec%
{\displaystyle\bigwedge}
T^{\ast}M$. Then as we already know
\begin{align}
d\mathcal{Q}  &  =\vartheta^{\alpha}\wedge(\mathring{D}_{e_{\alpha}%
}\mathcal{Q})=%
\bpartial
\wedge\mathcal{Q},\nonumber\\
\delta\mathcal{Q}  &  =\mathcal{-}\vartheta^{\alpha}\lrcorner(\mathring
{D}_{e_{\alpha}}\mathcal{Q})=%
\bpartial
\lrcorner\mathcal{Q}. \label{d and delta}%
\end{align}
We have also the important formulas%

\begin{align}
d\mathcal{Q}  &  =\vartheta^{\alpha}\wedge(D_{e_{\alpha}}\mathcal{Q}%
)-\mathcal{T}^{\alpha}\wedge(\vartheta_{\alpha}\lrcorner\mathcal{Q)=}%
{\mbox{\boldmath$\partial$}}\wedge\mathcal{Q-T}^{\alpha}\wedge(\vartheta
_{\alpha}\lrcorner\mathcal{Q)},\nonumber\\
\delta\mathcal{Q}=\mathcal{-}  &  \vartheta^{\alpha}\lrcorner(D_{e_{\alpha}%
}\mathcal{Q})-\mathcal{T}^{\alpha}\lrcorner(\vartheta_{\alpha}\wedge
\mathcal{Q)}=-{\mbox{\boldmath$\partial$}}\lrcorner\mathcal{Q}-\mathcal{T}%
^{\alpha}\lrcorner(\vartheta_{\alpha}\wedge\mathcal{Q).} \label{d delta bis}%
\end{align}

\section{Square of Dirac Operators and D' Alembertian, Ricci and Einstein
Operators}

We now investigate the square of a Dirac operator. We start recalling that the
square of the standard Dirac operator can be identified with the Hodge D'
Alembertian and that it can be separated in some interesting parts that we
called in \cite{rodoliv2007} the D'Alembertian, Ricci and Einstein operators
of $(M,%
\slg
,\mathring{D})$.

\subsection{The Square of the Dirac Operator $%
\protect\bpartial
$ Associated to $\mathring{D}$}

The square of standard Dirac operator $%
\bpartial
$ is the operator, $%
\bpartial
^{2}=%
\bpartial
\bpartial
$ $:\sec\bigwedge\nolimits^{p}T^{\ast}M\hookrightarrow\sec\mathcal{C\ell
}(M,\mathtt{g)}\rightarrow$ $\sec\bigwedge\nolimits^{p}T^{\ast}%
M\hookrightarrow\sec\mathcal{C\ell}(M,\mathtt{g)}$ given by:
\begin{equation}%
\bpartial
\text{ }^{2}=(%
\bpartial
\wedge+%
\bpartial
\lrcorner)(%
\bpartial
\wedge+%
\bpartial
\lrcorner)=(d-\delta)(d-\delta) \label{ssd}%
\end{equation}
\ 

It is quite obvious that%
\begin{equation}%
\bpartial
\text{ }^{2}=-(d\delta+\delta d), \label{ssd1}%
\end{equation}
\ and thus we recognize that $%
\bpartial
$ $^{2}\equiv\Diamond$ is the \textit{Hodge D'Alembertian }of the manifold
introduced by Eq.(\ref{705})

On the other hand, remembering the standard Dirac operator is $%
\bpartial
=\vartheta^{\alpha}\,\mathring{D}_{e_{\alpha}}$, where $\{\vartheta^{\alpha
}\}$ is the dual basis of an arbitrary basis $\{e_{\alpha}\}$ on $TU\subset
TM$ and $\mathring{D}$ is the Levi-Civita connection of the metric $%
\slg
$, we have:
\begin{align*}%
\bpartial
\text{ }^{2}  &  =(\vartheta^{\alpha}\mathring{D}_{e_{\alpha}})(\vartheta
^{\beta}\mathring{D}_{e_{\beta}})=\vartheta^{\alpha}(\vartheta^{\beta
}\mathring{D}_{e_{\alpha}}\mathring{D}_{e_{\beta}}+(\mathring{D}_{e_{\alpha}%
}\vartheta^{\beta})\mathring{D}_{e_{\beta}})\\
&  =g^{\alpha\beta}(\mathring{D}_{e_{\alpha}}\mathring{D}_{e_{\beta}%
}-\mathbf{\mathring{L}}_{\alpha\beta}^{\rho}\mathring{D}_{e_{\rho}}%
)+\vartheta^{\alpha}\wedge\vartheta^{\beta}(\mathring{D}_{e_{\alpha}}%
\mathring{D}_{e_{\beta}}-\mathbf{\mathring{L}}_{\alpha\beta}^{\rho}%
\mathring{D}_{\mathbf{e}_{\rho}}).
\end{align*}
Then defining the operators:
\begin{equation}%
\begin{tabular}
[c]{c}%
(a)\\
(b)
\end{tabular}%
\begin{array}
[c]{ccl}%
\bpartial
\cdot%
\bpartial
& = & g^{\alpha\beta}(\mathring{D}_{e_{\alpha}}\mathring{D}_{e_{\beta}%
}-\mathbf{\mathring{L}}_{\alpha\beta}^{\rho}\mathring{D}_{e_{\rho}})\\%
\bpartial
\wedge%
\bpartial
& = & \vartheta^{\alpha}\wedge\vartheta^{\beta}(\mathring{D}_{e_{\alpha}%
}\mathring{D}_{e_{\beta}}-\mathbf{\mathring{L}}_{\alpha\beta}^{\rho}%
\mathring{D}_{e_{\rho}}),
\end{array}
\label{1792}%
\end{equation}
we can write:
\begin{equation}
\Diamond=%
\bpartial
\text{ }^{2}=%
\bpartial
\cdot%
\bpartial
+%
\bpartial
\wedge%
\bpartial
\label{1796}%
\end{equation}
or,%
\begin{align}%
\bpartial
\text{ }^{2}  &  =(%
\bpartial
\lrcorner+%
\bpartial
\wedge)(%
\bpartial
\lrcorner+%
\bpartial
\wedge)\nonumber\\
&  =%
\bpartial
\lrcorner%
\bpartial
\wedge+%
\bpartial
\wedge%
\bpartial
\lrcorner\label{1796a}%
\end{align}

It is important to observe that the operators $%
\bpartial
\cdot%
\bpartial
$ and $%
\bpartial
\wedge%
\bpartial
$ do not have anything analogous in the formulation of the differential
geometry in the Cartan and Hodge bundles.

The operator $%
\bpartial
\cdot%
\bpartial
$ can also be written as:
\begin{equation}%
\bpartial
\cdot%
\bpartial
=\frac{1}{2}g^{\alpha\beta}\left[  \mathring{D}_{e_{\alpha}}\mathring
{D}_{e_{\beta}}+\mathring{D}_{e_{\beta}}\mathring{D}_{e_{\alpha}}%
-b_{\alpha\beta}^{\rho}\mathring{D}_{e_{\rho}}\right]  .
\end{equation}
Applying this operator to the 1-forms of the frame $\{\theta^{\alpha}\}$, we
get:
\begin{equation}
(%
\bpartial
\cdot%
\bpartial
)\vartheta^{\mu}=-\frac{1}{2}g^{\alpha\beta}\mathring{M}_{\!\rho}{}^{\mu}%
{}_{\alpha\beta}\theta^{\rho}, \label{1812}%
\end{equation}
where:
\begin{equation}
\mathring{M}_{\!\rho}{}^{\mu}{}_{\alpha\beta}=e_{\alpha}(\mathbf{\mathring{L}%
}_{\beta\rho}^{\mu})+e_{\beta}(\mathbf{\mathring{L}}_{\alpha\rho}^{\mu
})-\mathbf{\mathring{L}}_{\alpha\sigma}^{\mu}\mathbf{\mathring{L}}_{\beta\rho
}^{\sigma}-\mathbf{\mathring{L}}_{\beta\sigma}^{\mu}\mathbf{\mathring{L}%
}_{\alpha\rho}^{\sigma}-b_{\alpha\beta}^{\sigma}\mathbf{\mathring{L}}%
_{\sigma\rho}^{\mu}. \label{1812bis}%
\end{equation}
The proof that an object with these components is a tensor may be found in
\cite{rodoliv2007}. In particular, for every $r$-form field $\omega\in
\sec\bigwedge\nolimits^{r}T^{\ast}M$, $\omega=\frac{1}{r!}\omega_{\alpha
_{1}\ldots\alpha_{r}}\theta^{\alpha_{1}}\wedge\ldots\wedge\theta^{\alpha_{r}}%
$, we have:
\begin{equation}
(%
\bpartial
\cdot%
\bpartial
)\omega=\frac{1}{r!}g^{\alpha\beta}\mathring{D}_{\alpha}\mathring{D}_{\beta
}\omega_{\alpha_{1}\ldots\alpha_{r}}\theta^{\alpha_{1}}\wedge\ldots
\wedge\theta^{\alpha_{r}}, \label{1831}%
\end{equation}
where\/ $\mathring{D}_{\alpha}\mathring{D}_{\beta}\omega_{\alpha_{1}%
\ldots\alpha_{r}}$ are the components of the covariant derivative of\/
$\omega$, i.e., writing $\mathring{D}_{\mathbf{e}_{\beta}}\omega=\frac{1}%
{r!}\mathring{D}_{\beta}\omega_{\alpha_{1}\ldots\alpha_{r}}\theta^{\alpha_{1}%
}\wedge\ldots\wedge\theta^{\alpha_{r}}$, it is:%
\begin{equation}
\mathring{D}_{\beta}\omega_{\alpha_{1}\ldots\alpha_{r}}=e_{\beta}%
(\omega_{\alpha_{1}\ldots\alpha_{r}})-\mathbf{\mathring{L}}_{\beta\alpha_{1}%
}^{\sigma}\omega_{\sigma\alpha_{2}\ldots\alpha_{r}}-\cdots-\mathbf{\mathring
{L}}_{\beta\alpha_{r}}^{\sigma}\omega_{\alpha_{1}\ldots\alpha_{r-1}\sigma}.
\label{1832}%
\end{equation}
\hfill\medskip

\noindent In view of Eq.(\ref{1831}), we give the call the operator
$\mathring{\square}=%
\bpartial
\cdot%
\bpartial
$ the \textit{covariant} \textit{D'Alembertian}.

Note that the covariant D'Alembertian of the 1-forms $\vartheta^{\mu}$ can
also be written as:
\[
(%
\bpartial
\cdot%
\bpartial
)\vartheta^{\mu}=\mathring{g}^{\alpha\beta}\mathring{D}_{\alpha}\mathring
{D}_{\beta}\delta_{\rho}^{\mu}\vartheta^{\rho}=\frac{1}{2}\mathring{g}%
^{\alpha\beta}(\mathring{D}_{\alpha}\mathring{D}_{\beta}\delta_{\rho}^{\mu
}+\mathring{D}_{\beta}\mathring{D}_{\alpha}\delta_{\rho}^{\mu})\vartheta
^{\rho}%
\]
and therefore, taking into account the Eq.(\ref{1812}), we conclude that:
\begin{equation}
\mathring{M}_{\!\rho}{}^{\mu}{}_{\alpha\beta}=-(\mathring{D}_{\alpha}%
\mathring{D}_{\beta}\delta_{\rho}^{\mu}+\mathring{D}_{\beta}\mathring
{D}_{\alpha}\delta_{\rho}^{\mu}).
\end{equation}

By its turn, the operator $%
\bpartial
\wedge%
\bpartial
$ can also be written as:
\begin{equation}%
\bpartial
\wedge%
\bpartial
=\frac{1}{2}\vartheta^{\alpha}\wedge\vartheta^{\beta}\left[  \mathring
{D}_{\alpha}\mathring{D}_{\beta}-\mathring{D}_{\beta}\mathring{D}_{\alpha
}-c_{\alpha\beta}^{\rho}\mathring{D}_{\rho}\right]  .
\end{equation}
Applying this operator to the 1-forms of the frame $\{\vartheta^{\mu}\}$, we
get:
\begin{equation}
(%
\bpartial
\wedge%
\bpartial
)\vartheta^{\mu}=-\frac{1}{2}\mathring{R}_{\rho}{}^{\mu}{}_{\alpha\beta
}(\vartheta^{\alpha}\wedge\vartheta^{\beta})\vartheta^{\rho}%
=-\mathcal{\mathring{R}}_{\rho}^{\mu}\vartheta^{\rho},
\end{equation}
where $\mathring{R}_{\rho}{}^{\mu}{}_{\alpha\beta}$ are the components of the
curvature tensor of the connection $\mathring{D}$. Then using the second
formula in the first line of Eq.(\ref{10}) we have
\begin{equation}
\mathcal{\mathring{R}}_{\rho}^{\mu}\theta^{\rho}=\mathcal{\mathring{R}}_{\rho
}^{\mu}\llcorner\theta^{\rho}+\mathcal{\mathring{R}}_{\rho}^{\mu}\wedge
\theta^{\rho}. \label{rrr}%
\end{equation}
The second term in the r.h.s. of this equation is identically null because due
to the first Bianchi identity \ which for the particular case of the
Levi-Civita connection ($\mathcal{T}^{\mu}=0$) is $\mathcal{\mathring{R}%
}_{\rho}^{\mu}\wedge\theta^{\rho}=0$ . The first term in Eq.(\ref{rrr}) can be
written
\begin{align}
\mathcal{\mathring{R}}_{\rho}^{\mu}\llcorner\theta^{\rho}  &  =\frac{1}%
{2}\mathring{R}_{\rho}{}^{\mu}{}_{\alpha\beta}(\theta^{\alpha}\wedge
\theta^{\beta})\llcorner\theta^{\rho}\nonumber\\
&  =\frac{1}{2}\mathring{R}_{\rho}{}^{\mu}{}_{\alpha\beta}\theta^{\rho
}\lrcorner(\theta^{\alpha}\wedge\theta^{\beta})\nonumber\\
&  =-\frac{1}{2}\mathring{R}_{\rho}{}^{\mu}{}_{\alpha\beta}(\mathring{g}%
^{\rho\alpha}\theta^{\beta}-\mathring{g}^{\rho\beta}\theta^{\alpha
})\nonumber\\
&  =-\mathring{g}^{\rho\alpha}\mathring{R}_{\rho}{}^{\mu}{}_{\alpha\beta
}\theta^{\beta}=-\mathring{R}_{\beta}^{\mu}\theta^{\beta},
\end{align}
where $\mathring{R}_{\beta}^{\mu}$ are the components of the Ricci tensor of
the Levi-Civita connection $\mathring{D}$ of \texttt{ $\mathbf{g}$}. Thus we
have a really beautiful result:
\begin{equation}
(%
\bpartial
\wedge%
\bpartial
)\theta^{\mu}=\mathcal{\mathring{R}}^{\mu}, \label{ricci equation}%
\end{equation}
where $\mathcal{\mathring{R}}^{\mu}=\mathring{R}_{\beta}^{\mu}\theta^{\beta}$
are the Ricci 1-forms of the manifold. Because of this relation, we call the
operator $%
\bpartial
\wedge%
\bpartial
$ the \textit{Ricci operator\/}of the manifold associated to the Levi-Civita
connection $\mathring{D}$ of $%
\slg
$.

We can show \cite{rodoliv2007} that the Ricci operator $%
\bpartial
\wedge%
\bpartial
$ satisfies the relation\textsl{: }
\begin{equation}%
\bpartial
\wedge%
\bpartial
=\mathcal{\mathring{R}}^{\sigma}\wedge\mathbf{i}_{\sigma}+\mathcal{\mathring
{R}}^{\rho\sigma}\wedge\mathbf{i}_{\rho}\mathbf{i}_{\sigma}, \label{1918}%
\end{equation}
where the curvature $2$-forms are $\mathcal{\mathring{R}}^{\rho\sigma}%
=\frac{1}{2}\mathring{R}^{\rho\sigma}{}_{\alpha\beta}\vartheta^{\alpha}%
\wedge\vartheta^{\beta}$ and%

\begin{equation}
\mathbf{i}_{\sigma}\omega:=\vartheta_{\sigma}\lrcorner\omega.
\end{equation}
\hfill

Observe that applying the operator given by the second term in the r.h.s. of
Eq.(\ref{1918}) to the dual of the 1-forms $\vartheta^{\mu}$, we get:
\begin{align}
\mathcal{\mathring{R}}^{\rho\sigma}\wedge\mathbf{i}_{\rho}\mathbf{i}_{\sigma
}\star\vartheta^{\mu}  &  =\mathcal{\mathring{R}}_{\rho\sigma}\star
\vartheta^{\rho}\lrcorner(\vartheta^{\sigma}\lrcorner\star\vartheta^{\mu
}))\nonumber\\
&  =-\mathcal{\mathring{R}}_{\rho\sigma}\wedge\star(\vartheta^{\rho}%
\wedge\vartheta^{\sigma}\star\vartheta^{\mu})\label{440nbis}\\
&  =\star(\mathcal{\mathring{R}}_{\rho\sigma}\lrcorner(\vartheta^{\rho}%
\wedge\vartheta^{\sigma}\wedge\vartheta^{\mu})),\nonumber
\end{align}
where we have used the Eqs.(\ref{440new}). Then, recalling the definition of
the curvature forms and using the Eq.(\ref{7}), we conclude that:
\begin{equation}
\mathcal{\mathring{R}}^{\rho\sigma}\wedge(\vartheta_{\rho}\lrcorner
\vartheta_{\sigma}\lrcorner\star\vartheta^{\mu})=2\star(\mathcal{\mathring{R}%
}^{\mu}-\frac{1}{2}\mathring{R}\vartheta^{\mu})=2\star\mathcal{\mathring{G}%
}^{\mu},
\end{equation}
where $\mathring{R}$ is the scalar curvature of the manifold and $\ $the
$\mathcal{\mathring{G}}^{\mu}$ may be called the \textit{Einstein }%
$1$\textit{-form fields.}

That observation motivate us to introduce in \cite{rodoliv2007} the
\textit{Einstein operator }of the Levi-Civita connection $\mathring{D}$ of $%
\slg
$ on the manifold $M$ as the mapping $\mathring{\blacksquare}$ $:\sec
\mathcal{C\ell}(M,\mathtt{g)}\rightarrow\sec\mathcal{C\ell}(M,\mathtt{g)}$
given by:
\begin{equation}
\mathring{\blacksquare}=\frac{1}{2}\star^{-1}(\mathcal{\mathring{R}}%
^{\rho\sigma}\wedge\mathbf{i}_{\rho}\mathbf{i}_{\sigma})\star.
\end{equation}
\ 

Obviously, we have:
\begin{equation}
\mathring{\blacksquare}\theta^{\mu}=\mathcal{\mathring{G}}^{\mu}%
=\mathcal{\mathring{R}}^{\mu}-\frac{1}{2}\mathring{R}\vartheta^{\mu}.
\end{equation}
In addition, it is easy to verify that $\star^{-1}(%
\bpartial
\wedge%
\bpartial
)\star=-%
\bpartial
\wedge%
\bpartial
$ and $\star^{-1}(\mathcal{\mathring{R}}^{\sigma}\wedge\mathbf{i}_{\sigma
})\star=\mathcal{\mathring{R}}^{\sigma}\lrcorner\mathbf{j}_{\sigma}$. Thus we
can also write the Einstein operator as:
\begin{equation}
\mathring{\blacksquare}=-\frac{1}{2}(%
\bpartial
\wedge%
\bpartial
+\mathcal{\mathring{R}}^{\sigma}\lrcorner\mathbf{j}_{\sigma}),
\end{equation}
where%
\begin{equation}
\mathbf{j}_{\sigma}\mathcal{A=\vartheta}_{\sigma}\wedge\mathcal{A},
\end{equation}
for any $\mathcal{A\in}\sec%
{\displaystyle\bigwedge}
T^{\ast}M\hookrightarrow\sec\mathcal{C\ell}(M,\mathtt{g)}$.

We recall \cite{rodoliv2007} that if $\mathring{\omega}_{\rho}^{\mu}$\ are the
Levi-Civita connection 1-forms fields in an arbitrary moving frame
$\{\vartheta^{\mu}\}$\ on $(M,%
\slg
,\mathring{D})$ then:\textit{ }%
\begin{equation}%
\begin{tabular}
[c]{c}%
(a)\\
(b)
\end{tabular}%
\begin{array}
[c]{rcl}%
(%
\bpartial
\cdot%
\bpartial
)\vartheta^{\mu} & = & -(%
\bpartial
\cdot\mathring{\omega}_{\rho}^{\mu}-\mathring{\omega}_{\rho}^{\sigma}%
\cdot\mathring{\omega}_{\sigma}^{\mu})\vartheta^{\rho}\\
(%
\bpartial
\wedge%
\bpartial
)\vartheta^{\mu} & = & -(%
\bpartial
\wedge\mathring{\omega}_{\rho}^{\mu}-\mathring{\omega}_{\rho}^{\sigma}%
\wedge\mathring{\omega}_{\sigma}^{\mu})\vartheta^{\rho},
\end{array}
\label{2017}%
\end{equation}
\textit{and \ }%
\begin{equation}%
\bpartial
\text{ }^{2}\vartheta^{\mu}=-(%
\bpartial
\mathring{\omega}_{\rho}^{\mu}-\mathring{\omega}_{\rho}^{\sigma}%
\mathring{\omega}_{\sigma}^{\mu})\vartheta^{\rho}.
\end{equation}

\begin{exercise}
Show that $\vartheta_{\rho}\wedge\vartheta_{\sigma}\mathcal{\mathring{R}%
}^{\rho\sigma}=-\mathring{R},$ where $\mathring{R}$ is the curvature scalar.
\end{exercise}

\subsection{The Square of the Dirac Operator ${\mbox{\boldmath$\partial$}}$
Associated to $D$}

Consider the structure $(M,%
\slg
,D)$, where $D$ is an arbitrary Riemann-Cartan-Weyl connection and the
Clifford algebra $\mathcal{C\ell}(M,\mathtt{g)}$. Let us now compute the
square of the (general) Dirac operator ${\mbox{\boldmath$\partial$}}%
=\vartheta^{\alpha}D_{e_{\alpha}}$. As in the earlier section, we have, by one
side,
\begin{align*}
{\mbox{\boldmath$\partial$}}^{2}  &  =({\mbox{\boldmath$\partial$}}%
\lrcorner+\;{\mbox{\boldmath$\partial$}}\wedge)({\mbox{\boldmath$\partial$}}%
\lrcorner+\;{\mbox{\boldmath$\partial$}\wedge})\\
&  ={\mbox{\boldmath$\partial$}}\lrcorner{\mbox{\boldmath$\partial$}}%
\lrcorner+\;{\mbox{\boldmath$\partial$}}\lrcorner
{\mbox{\boldmath$\partial$}\wedge}+\;{\mbox{\boldmath$\partial$}\wedge
\mbox{\boldmath$\partial$}}\lrcorner+\;{\mbox{\boldmath$\partial$}\wedge
\mbox{\boldmath$\partial$}}\wedge
\end{align*}
and we write ${\mbox{\boldmath$\partial$}}\!\lrcorner
{\mbox{\boldmath$\partial$}}\lrcorner\equiv{\mbox{\boldmath$\partial$}}%
^{2}\lrcorner$, ${\mbox{\boldmath$\partial$}}\!\wedge
{\mbox{\boldmath$\partial$}}\wedge\equiv{\mbox{\boldmath$\partial$}}^{2}%
\wedge$ and
\begin{equation}
\mathcal{L}_{+}={\mbox{\boldmath$\partial$}}\lrcorner
{\mbox{\boldmath$\partial$}}\wedge+\;{\mbox{\boldmath$\partial$}}%
\wedge{\mbox{\boldmath$\partial$}}\lrcorner,
\end{equation}
so that:
\begin{equation}
{\mbox{\boldmath$\partial$}}^{2}={\mbox{\boldmath$\partial$}}^{2}%
\!\lrcorner\;+\;\mathcal{L}_{+}\;+\;{\mbox{\boldmath$\partial$}}^{2}%
\!\wedge\text{ .} \label{1071}%
\end{equation}
The operator $\mathcal{L}_{+}$ when applied to scalar functions corresponds,
for the case of a Riemann-Cartan space, to the wave operator introduced by
Rapoport \cite{rapoport} in his theory of Stochastic Mechanics. Obviously, for
the case of the standard Dirac operator, $\mathcal{L}_{+}$ reduces to the
usual Hodge D' Alembertian of the manifold, which preserve graduation of
forms. For more details see \cite{notterod}.

On the other hand, we have also:
\begin{align*}
{\mbox{\boldmath$\partial$}}^{\,2}  &  =(\vartheta^{\alpha}D_{e_{\alpha}%
})(\vartheta^{\beta}D_{\mathbf{e}_{\beta}})=\vartheta^{\alpha}(\vartheta
^{\beta}D_{e_{\alpha}}D_{e_{\beta}}+(D_{e_{\alpha}}\vartheta^{\beta
})D_{e_{\beta}})\\
&  =g^{\alpha\beta}(D_{e_{\alpha}}D_{e_{\beta}}-\mathbf{L}_{\alpha\beta}%
^{\rho}D_{e_{\rho}})+\vartheta^{\alpha}\wedge\vartheta^{\beta}(D_{e_{\alpha}%
}D_{e_{\beta}}-\mathbf{L}_{\alpha\beta}^{\rho}D_{e_{\rho}})
\end{align*}
and we can then define:
\begin{equation}%
\begin{array}
[c]{ccl}%
{\mbox{\boldmath$\partial$}}\cdot{\mbox{\boldmath$\partial$}} & = &
g^{\alpha\beta}(D_{e_{\alpha}}D_{e_{\beta}}-\mathbf{L}_{\alpha\beta}^{\rho
}D_{e_{\rho}})\\
{\mbox{\boldmath$\partial$}}\wedge{\mbox{\boldmath$\partial$}} & = &
\theta^{\alpha}\wedge\theta^{\beta}(D_{e_{\alpha}}D_{e_{\beta}}-\mathbf{L}%
_{\alpha\beta}^{\rho}D_{e_{\rho}})
\end{array}
\end{equation}
in order to have:
\begin{equation}
{\mbox{\boldmath$\partial$}}^{2}%
{=\mbox{\boldmath$\partial$}\mbox{\boldmath$\partial$}}%
={\mbox{\boldmath$\partial$}}\cdot{\mbox{\boldmath$\partial$}}%
+{\mbox{\boldmath$\partial$}}\wedge{\mbox{\boldmath$\partial$}}\text{ .}
\label{1091}%
\end{equation}
The operator ${\mbox{\boldmath$\partial$}}\cdot{\mbox{\boldmath$\partial$}}$
can also be written as:
\begin{align}
{\mbox{\boldmath$\partial$}}\cdot{\mbox{\boldmath$\partial$}}  &  =\frac{1}%
{2}\theta^{\alpha}\cdot\theta^{\beta}(D_{e_{\alpha}}D_{e_{\beta}}%
-\mathbf{L}_{\alpha\beta}^{\rho}D_{e_{\rho}})+\frac{1}{2}\theta^{\beta}%
\cdot\theta^{\alpha}(D_{e_{\beta}}D_{e_{\alpha}}-\mathbf{L}_{\beta\alpha
}^{\rho}D_{e_{\rho}})\nonumber\\
&  =\frac{1}{2}g^{\alpha\beta}[D_{e_{\alpha}}D_{e_{\beta}}+D_{e_{\beta}%
}D_{e_{\alpha}}-(\mathbf{L}_{\alpha\beta}^{\rho}+\mathbf{L}_{\beta\alpha
}^{\rho})D_{e_{\rho}}] \label{gda}%
\end{align}
or,
\begin{equation}
{\mbox{\boldmath$\partial$}}\cdot{\mbox{\boldmath$\partial$}}=\frac{1}%
{2}g^{\alpha\beta}(D_{e_{\alpha}}D_{e_{\beta}}+D_{e_{\beta}}D_{e_{\alpha}%
}-b_{\alpha\beta}^{\rho}D_{e_{\rho}})-s^{\rho}D_{e_{\rho}}, \label{gdabis}%
\end{equation}
where $s^{\rho}$ has been defined in Eq.(\ref{thes}).

By its turn, the operator ${\mbox{\boldmath$\partial$}}\wedge
{\mbox{\boldmath$\partial$}}$ can also be written as:
\begin{align*}
{\mbox{\boldmath$\partial$}}\wedge{\mbox{\boldmath$\partial$}}  &  =\frac
{1}{2}\vartheta^{\alpha}\wedge\vartheta^{\beta}(D_{e_{\alpha}}D_{e_{\beta}%
}-\mathbf{L}_{\alpha\beta}^{\rho}D_{e_{\rho}})+\frac{1}{2}\vartheta^{\beta
}\wedge\vartheta^{\alpha}(D_{e_{\beta}}D_{e_{\alpha}}-\mathbf{L}_{\beta\alpha
}^{\rho}D_{e_{\rho}})\\
&  =\frac{1}{2}\vartheta^{\alpha}\wedge\vartheta^{\beta}[D_{e_{\alpha}%
}D_{e_{\beta}}-D_{e_{\beta}}D_{e_{\alpha}}-(\mathbf{L}_{\alpha\beta}^{\rho
}-\mathbf{L}_{\beta\alpha}^{\rho})D_{e_{\rho}}]
\end{align*}
or,
\begin{equation}
{\mbox{\boldmath$\partial$}}\wedge{\mbox{\boldmath$\partial$}}=\frac{1}%
{2}\vartheta^{\alpha}\wedge\vartheta^{\beta}(D_{e_{\alpha}}D_{e_{\beta}%
}-D_{e_{\beta}}D_{\mathbf{e}_{\alpha}}-c_{\alpha\beta}^{\rho}D_{\mathbf{e}%
_{\rho}})-\mathcal{T}^{\rho}D_{\mathbf{e}_{\rho}}. \label{gricci}%
\end{equation}

\begin{remark}
For the case of a Levi-Civita connection we have similar formulas for $%
\bpartial
\cdot%
\bpartial
$ \emph{(Eq.(\ref{gda})) }and $%
\bpartial
\wedge%
\bpartial
$ \emph{(Eq.(\ref{gricci}))} with $D\mapsto\mathring{D}$, and of course,
$\mathcal{T}^{\rho}=0$, as follows directly from \emph{Eq.(\ref{1792}).}
\end{remark}

\section{Coordinate Expressions for Maxwell Equations on Lorentzian and
Riemann-Cartan Spacetimes}

\subsection{Maxwell Equations on a Lorentzian Spacetime}

We now take $(M,%
\slg
)$ as a Lorentzian manifold, i.e., $\dim M=4$ and the signature of $%
\slg
$ is $(1,3)$. We consider moreover a Lorentzian spacetime structure on $(M,%
\slg
)$, i.e., the pentuple $(M,%
\slg
,\mathring{D},\tau_{%
\slg
},\uparrow)$ and a Riemann-Cartan spacetime structure $(M,%
\slg
,D,\tau_{%
\slg
},\uparrow)$.

Now, in both spacetime structures, Maxwell equations in vacuum read:%
\begin{equation}
d\mathbf{F}=0,\hspace{0.5cm}\delta\mathbf{F}=-\mathbf{J,} \label{maxwell}%
\end{equation}
where $\mathbf{F}\in\sec%
{\displaystyle\bigwedge\nolimits^{2}}
T^{\ast}M$ is the Faraday tensor (electromagnetic field) and $\mathbf{J}%
\in\sec%
{\displaystyle\bigwedge\nolimits^{1}}
T^{\ast}M$ is the current. We observe that writing
\begin{equation}
\mathbf{F}=\frac{1}{2}F_{\mu\nu}dx^{\mu}\wedge dx^{\nu}=\frac{1}{2}F_{\mu\nu
}\theta^{\mu}\wedge\theta^{\nu}=\frac{1}{2}F_{\mu\nu}\theta^{\mu\nu},
\label{ma1}%
\end{equation}
we have using Eq.(\ref{hodge dual}) that
\begin{equation}
\star\mathbf{F}=\frac{1}{2}F_{\mu\nu}(\star\theta^{\mu\nu})=\frac{1}{2}%
^{\star}\mathbf{F}_{\rho\sigma}\vartheta^{\rho\sigma}=\frac{1}{2}(F_{\mu\nu
}\frac{1}{2}\sqrt{\left\vert \det%
\slg
\right\vert }g^{\mu\alpha}g^{\nu\beta}\epsilon_{\alpha\beta\rho\sigma
})\vartheta^{\rho\sigma}%
\end{equation}
Thus%
\begin{equation}
{}^{\star}\mathbf{F}_{\rho\sigma}=(\star\mathbf{F)}_{\rho\sigma}=\frac{1}%
{2}F_{\mu\nu}\sqrt{\left\vert \det%
\slg
\right\vert }g^{\mu\alpha}g^{\nu\beta}\epsilon_{\alpha\beta\rho\sigma}.
\label{ma3}%
\end{equation}

The homogeneous Maxwell equation $d\mathbf{F}=0$ can be writing as
$\delta\star\mathbf{F}=0$. The proof follows at once from the definition of
$\delta$ (Eq.(\ref{hodge})). Indeed, we can write%
\[
0=d\mathbf{F}=\star\star^{-1}d\star\star^{-1}\mathbf{F}=\star\delta\star
^{-1}\mathbf{F}=-\star\delta\star\mathbf{F}=0.
\]
Then $\star^{-1}\star\delta\star\mathbf{F}=0$ and we end with%
\[
\delta\star\mathbf{F}=0.
\]
(a)\textbf{ }We now express the equivalent equations $dF=0$ and $\delta\star
F=0$ in arbitrary coordinates $\{x^{\mu}\}$ covering $U\subset M$ using first
the Levi-Civita connection and noticeable formula in Eq.(\ref{d and delta}).
We have
\begin{align*}
d\mathbf{F}  &  =\theta^{\alpha}\wedge(\mathring{D}%
_{{\mbox{\boldmath$\partial$}}_{\alpha}}F)\\
&  =\frac{1}{2}\theta^{\alpha}\wedge\left[  \mathring{D}%
_{{\mbox{\boldmath$\partial$}}_{\alpha}}(F_{\mu\nu}\theta^{\mu}\wedge
\theta^{\nu})\right] \\
&  =\frac{1}{2}\theta^{\alpha}\wedge\left[  ({\partial}_{\alpha}F_{\mu\nu
})\theta^{\mu}\wedge\theta^{\nu}-F_{\mu\nu}\mathring{\Gamma}_{\alpha\rho}%
^{\mu}\theta^{\rho}\wedge\theta^{\nu}-F_{\mu\nu}\mathring{\Gamma}_{\alpha\rho
}^{\nu}\theta^{\mu}\wedge\theta^{\rho}\right] \\
&  =\frac{1}{2}\theta^{\alpha}\wedge\left[  (\mathring{D}_{\alpha}F_{\mu\nu
})\theta^{\mu}\wedge\theta^{\nu}\right] \\
&  =\frac{1}{2}D_{\alpha}F_{\mu\nu}\theta^{\alpha}\wedge\theta^{\mu}%
\wedge\theta^{\nu}\\
&  =\frac{1}{2}\left[  \frac{1}{3}\mathring{D}_{\alpha}F_{\mu\nu}%
\theta^{\alpha}\wedge\theta^{\mu}\wedge\theta^{\nu}+\frac{1}{3}\mathring
{D}_{\mu}F_{\nu\alpha}\theta^{\mu}\wedge\theta^{\nu}\wedge\theta^{\alpha
}+\frac{1}{3}\mathring{D}_{\nu}F_{\alpha\mu}\theta^{\nu}\wedge\theta^{\alpha
}\wedge\theta^{\mu}\right] \\
&  =\frac{1}{2}\left[  \frac{1}{3}\mathring{D}_{\alpha}F_{\mu\nu}%
\theta^{\alpha}\wedge\theta^{\mu}\wedge\theta^{\nu}+\frac{1}{3}\mathring
{D}_{\mu}F_{\nu\alpha}\theta^{\alpha}\wedge\theta^{\mu}\wedge\theta^{\nu
}+\frac{1}{3}\mathring{D}_{\nu}F_{\alpha\mu}\theta^{\alpha}\wedge\theta^{\mu
}\wedge\theta^{\nu}\right] \\
&  =\frac{1}{6}\left(  \mathring{D}_{\alpha}F_{\mu\nu}+\mathring{D}_{\mu
}F_{\nu\alpha}+\mathring{D}_{\nu}F_{\alpha\mu}\right)  \theta^{\alpha}%
\wedge\theta^{\mu}\wedge\theta^{\nu}.
\end{align*}

So,
\begin{equation}
d\mathbf{F}=0\Leftrightarrow\mathring{D}_{\alpha}F_{\mu\nu}+\mathring{D}_{\mu
}F_{\nu\alpha}+\mathring{D}_{\nu}F_{\alpha\mu}=0. \label{dF1}%
\end{equation}
If we calculate $d\mathbf{F}=0$ using the definition of $d$ we get:%
\begin{align}
d\mathbf{F}  &  =\frac{1}{2}({\partial}_{\alpha}F_{\mu\nu})\theta^{\alpha
}\wedge\theta^{\mu}\wedge\theta^{\nu}\label{dF2}\\
&  =\frac{1}{6}\left(  {\partial}_{\alpha}F_{\mu\nu}+{\partial}_{\mu}%
F_{\nu\alpha}+{\partial}_{\nu}F_{\alpha\mu}\right)  \theta^{\alpha}%
\wedge\theta^{\mu}\wedge\theta^{\nu},\nonumber
\end{align}
from where we get that%
\begin{equation}
d\mathbf{F}=0\Longleftrightarrow{\partial}_{\alpha}F_{\mu\nu}+{\partial}_{\mu
}F_{\nu\alpha}+{\partial}_{\nu}F_{\alpha\mu}=0\Longleftrightarrow\mathring
{D}_{\alpha}F_{\mu\nu}+\mathring{D}_{\mu}F_{\nu\alpha}+\mathring{D}_{\nu
}F_{\alpha\mu}=0. \label{dF3}%
\end{equation}
Next we calculate $\delta\star F=0$. We have
\begin{align}
\delta\star\mathbf{F}  &  =-\theta^{\alpha}\lrcorner(\mathring{D}%
_{{\mbox{\boldmath$\partial$}}_{\alpha}}\star\mathbf{F})\nonumber\\
&  =-\frac{1}{2}\theta^{\alpha}\lrcorner\left\{  \mathring{D}%
_{{\mbox{\boldmath$\partial$}}_{\alpha}}\left[  {}^{\star}F_{\mu\nu}%
\theta^{\mu}\wedge\theta^{\nu}\right]  \right\} \nonumber\\
&  =-\frac{1}{2}\theta^{\alpha}\lrcorner\left\{  ({\partial}_{\alpha}{}%
^{\star}F_{\mu\nu})\theta^{\mu}\wedge\theta^{\nu}-{}^{\star}F_{\mu\nu
}\mathring{\Gamma}_{\alpha\rho}^{\mu}\theta^{\rho}\wedge\theta^{\nu}-{}%
^{\star}F_{\mu\nu}\mathring{\Gamma}_{\alpha\rho}^{\nu}\theta^{\mu}\wedge
\theta^{\rho}\right\} \nonumber\\
&  =-\frac{1}{2}\theta^{\alpha}\lrcorner\left\{  ({\partial}_{\alpha}{}%
^{\star}F_{\mu\nu})\theta^{\mu}\wedge\theta^{\nu}-{}^{\star}F_{\rho\nu
}\mathring{\Gamma}_{\alpha\mu}^{\rho}\theta^{\mu}\wedge\theta^{\nu}-{}^{\star
}F_{\mu\rho}\mathring{\Gamma}_{\alpha\nu}^{\rho}\theta^{\mu}\wedge\theta^{\nu
}\right\} \nonumber\\
&  =-\frac{1}{2}\theta^{\alpha}\lrcorner\left\{  (\mathring{D}_{\alpha}%
{}^{\star}F_{\mu\nu})\theta^{\mu}\wedge\theta^{\nu}\right\} \nonumber\\
&  =-\frac{1}{2}\left\{  (\mathring{D}_{\alpha}{}^{\star}F_{\mu\nu}%
)g^{\alpha\mu}\theta^{\nu}-(\mathring{D}_{\alpha}{}^{\star}F_{\mu\nu
})g^{\alpha\nu}\theta^{\mu}\right\} \nonumber\\
&  =-(\mathring{D}_{\alpha}{}^{\star}F_{\mu\nu})g^{\alpha\mu}\theta^{\nu
}\nonumber\\
&  =-[\mathring{D}_{\alpha}{}(^{\star}F_{\mu\nu}g^{\alpha\mu})]\theta^{\nu
}\nonumber\\
&  =-(\mathring{D}_{\alpha}{}^{\star}F_{\,\nu}^{\alpha})]\theta^{\nu}.
\label{delf}%
\end{align}
Then we get that
\begin{equation}
\mathring{D}_{\alpha}F_{\mu\nu}+\mathring{D}_{\mu}F_{\nu\alpha}+\mathring
{D}_{\nu}F_{\alpha\mu}=0\Leftrightarrow d\mathbf{F}=0\Leftrightarrow
\delta\star\mathbf{F}=0\Longleftrightarrow\mathring{D}_{\alpha}{}^{\star
}F_{\,\nu}^{\alpha}=0. \label{dF4}%
\end{equation}

(b)\emph{ }Also, the non homogenoeous Maxwell equation $\delta\mathbf{F}=-J$
can be written using the definition of $\delta$ (Eq.(\ref{hodge})) as
$d\star\mathbf{F}=-\star\mathbf{J}$:%
\begin{align}
\delta\mathbf{F}  &  =-\mathbf{J,}\nonumber\\
(-1)^{2}\star^{-1}d\star\mathbf{F}  &  =-\mathbf{J,}\nonumber\\
\star\star^{-1}d\star\mathbf{F}  &  =-\star\mathbf{J,}\nonumber\\
d\star\mathbf{F}  &  =-\star\mathbf{J.} \label{dF5}%
\end{align}

We now express $\delta\mathbf{F=}-\mathbf{J}$ in arbitrary
coordinates\footnote{We observe that in terms of the "classical" charge and
"vector" current densities we have $\mathbf{J=}\rho\theta^{0}-j_{i}\theta^{i}%
$.} using first the Levi-Civita connection. We have following the same steps
as in Eq.(\ref{delf})
\begin{align}
\delta\mathbf{F+J}  &  =-\frac{1}{2}\theta^{\alpha}\lrcorner\left\{
\mathring{D}_{{\mbox{\boldmath$\partial$}}_{\alpha}}\left[  F_{\mu\nu}%
\theta^{\mu}\wedge\theta^{\nu}\right]  \right\}  +J_{\nu}\theta^{\nu
}\label{dF6}\\
&  =(-\mathring{D}_{\alpha}F_{\,\nu}^{\alpha}+J_{\nu})\theta^{\nu}.\nonumber
\end{align}
Then%
\begin{equation}
\delta\mathbf{F+J}=0\Leftrightarrow\mathring{D}_{\alpha}F^{\alpha\nu}=J^{\nu}.
\label{dF7}%
\end{equation}

We also observe that using the symmetry of the connection coefficients and the
antisymmetry of the $F^{\alpha\nu}$ that $\mathring{\Gamma}_{\alpha\rho}^{\nu
}F^{\alpha\rho}=-\mathring{\Gamma}_{\alpha\rho}^{\nu}F^{\alpha\rho}=0$. Also,
\[
\mathring{\Gamma}_{\alpha\rho}^{\alpha}=\partial_{\rho}\ln\sqrt{\left\vert
\det%
\slg
\right\vert }=\frac{1}{\sqrt{\left\vert \det%
\slg
\right\vert }}\partial_{\rho}\sqrt{\left\vert \det%
\slg
\right\vert },
\]
and
\begin{align*}
\mathring{D}_{\alpha}F^{\alpha\nu}  &  =\partial_{\alpha}F^{\alpha\nu
}+\mathring{\Gamma}_{\alpha\rho}^{\alpha}F^{\rho\nu}+\mathring{\Gamma}%
_{\alpha\rho}^{\nu}F^{\alpha\rho}\\
&  =\partial_{\alpha}F^{\alpha\nu}+\mathring{\Gamma}_{\alpha\rho}^{\alpha
}F^{\rho\nu}\\
&  =\partial_{\rho}F^{\rho\nu}+\frac{1}{\sqrt{\left\vert \det%
\slg
\right\vert }}\partial_{\rho}(\sqrt{\left\vert \det%
\slg
\right\vert })F^{\rho\nu}.
\end{align*}
Then%
\begin{align}
\mathring{D}_{\alpha}F^{\alpha\nu}  &  =J^{\nu},\nonumber\\
\sqrt{\left\vert \det%
\slg
\right\vert }\partial_{\rho}F^{\rho\nu}+\partial_{\rho}(\sqrt{\left\vert \det%
\slg
\right\vert })F^{\rho\nu}  &  =\sqrt{\left\vert \det%
\slg
\right\vert }J^{\nu},\nonumber\\
\partial_{\rho}(\sqrt{\left\vert \det%
\slg
\right\vert }F^{\rho\nu})  &  =\sqrt{\left\vert \det%
\slg
\right\vert }J^{\nu},\nonumber\\
\frac{1}{\sqrt{\left\vert \det%
\slg
\right\vert }}\partial_{\rho}(\sqrt{\left\vert \det%
\slg
\right\vert }F^{\rho\nu})  &  =J^{\nu}, \label{dF8}%
\end{align}
and%
\begin{equation}
\delta\mathbf{F}=0\Leftrightarrow\mathring{D}_{\alpha}F^{\alpha\nu}=J^{\nu
}\Leftrightarrow\frac{1}{\sqrt{\left\vert \det%
\slg
\right\vert }}\partial_{\rho}(\sqrt{\left\vert \det%
\slg
\right\vert }F^{\rho\nu})=J^{\nu}. \label{dF9}%
\end{equation}

\begin{exercise}
Show that in a Lorentzian spacetime Maxwell equations become Maxwell equation,
i.e.,
\begin{equation}%
\bpartial
\mathbf{F=J.} \label{meq}%
\end{equation}

\end{exercise}

\subsection{Maxwell Equations on Riemann-Cartan Spacetime}

From time to time we see papers (e.g., \cite{prasana,sabbata}) writing Maxwell
equations in a Riemann-Cartan spacetime using arbitrary coordinates and (of
course) the Riemann-Cartan connection. As we shall see such enterprises are
simple exercises, if we make use of the noticeable formulas of
Eq.(\ref{d delta bis}). Indeed, the homogeneous Maxwell equation
$d\mathbf{F}=0$ reads
\begin{equation}
d\mathbf{F}=\theta^{\alpha}\wedge(D_{{\mbox{\boldmath$\partial$}}_{\alpha}%
}\mathbf{F})-\mathcal{T}^{\alpha}\wedge(\theta_{\alpha}\lrcorner
\mathbf{F}\mathcal{)}=0
\end{equation}
or%

\begin{align*}
&  \frac{1}{6}(D_{\alpha}F_{\mu\nu}+D_{\mu}F_{\nu\alpha}+D_{\nu}F_{\alpha\mu
})\theta^{\alpha}\wedge\theta^{\mu}\wedge\theta^{\nu}\\
&  -\frac{1}{2}\frac{1}{2}T_{\rho\sigma}^{\alpha}\theta^{\rho}\wedge
\theta^{\sigma}\wedge\lbrack\theta_{\alpha}\lrcorner F_{\mu\nu}(\theta^{\mu
}\wedge\theta^{\nu})]\\
&  =\frac{1}{6}(D_{\alpha}F_{\mu\nu}+D_{\mu}F_{\nu\alpha}+D_{\nu}F_{\alpha\mu
})\theta^{\alpha}\wedge\theta^{\mu}\wedge\theta^{\nu}\\
&  -\frac{1}{2}T_{\rho\sigma}^{\alpha}F_{\mu\nu}\theta^{\rho}\wedge
\theta^{\sigma}\wedge\delta_{\alpha}^{\mu}\theta^{\nu}\\
&  =\frac{1}{6}(D_{\alpha}F_{\mu\nu}+D_{\mu}F_{\nu\alpha}+D_{\nu}F_{\alpha\mu
})\theta^{\alpha}\wedge\theta^{\mu}\wedge\theta^{\nu}\\
&  -\frac{1}{2}T_{\alpha\mu}^{\sigma}F_{\sigma\nu}\theta^{\alpha}\wedge
\theta^{\mu}\wedge\theta^{\nu}\\
&  =\frac{1}{6}(D_{\alpha}F_{\mu\nu}+D_{\mu}F_{\nu\alpha}+D_{\nu}F_{\alpha\mu
})\theta^{\alpha}\wedge\theta^{\mu}\wedge\theta^{\nu}\\
&  -\frac{1}{6}(T_{\alpha\mu}^{\sigma}F_{\sigma\nu}+T_{\mu\nu}^{\sigma
}F_{\sigma\alpha}+T_{\nu\alpha}^{\sigma}F_{\sigma\mu})\theta^{\alpha}%
\wedge\theta^{\mu}\wedge\theta^{\nu}\\
&  =\frac{1}{6}(D_{\alpha}F_{\mu\nu}+D_{\mu}F_{\nu\alpha}+D_{\nu}F_{\alpha\mu
})\theta^{\alpha}\wedge\theta^{\mu}\wedge\theta^{\nu}\\
&  +\frac{1}{6}(F_{\alpha\sigma}T_{\mu\nu}^{\sigma}+F_{\mu\sigma}T_{\nu\alpha
}^{\sigma}+F_{\nu\sigma}T_{\alpha\mu}^{\sigma})\theta^{\alpha}\wedge
\theta^{\mu}\wedge\theta^{\nu}.
\end{align*}
i.e.,%
\begin{equation}
d\mathbf{F}=0\Longleftrightarrow D_{\alpha}F_{\mu\nu}+D_{\mu}F_{\nu\alpha
}+D_{\nu}F_{\alpha\mu}+F_{\sigma\alpha}T_{\mu\nu}^{\sigma}+F_{\mu\sigma}%
T_{\nu\alpha}^{\sigma}+F_{\nu\sigma}T_{\alpha\mu}^{\sigma}=0. \label{Df10}%
\end{equation}
Also, taking into account that $d\mathbf{F}=0\Longleftrightarrow\delta
\star\mathbf{F}=0$ we have using the second noticeable formula in
Eq.(\ref{d delta bis})\emph{ }that%

\begin{equation}
\delta\star\mathbf{F=}\mathcal{-}\theta^{\alpha}\lrcorner(D_{e_{\alpha}}%
\star\mathbf{F})-\mathcal{T}^{\alpha}\lrcorner(\theta_{\alpha}\wedge
\star\mathbf{F}\mathcal{)}=0. \label{dF12}%
\end{equation}
Now,%
\begin{equation}
\theta^{\alpha}\lrcorner(D_{e_{\alpha}}\star\mathbf{F})=(D_{\alpha}{}^{\star
}F_{\,\,\nu}^{\alpha})\theta^{\nu}=(D_{\alpha}{}^{\star}F^{\alpha\nu}%
)\theta_{\nu} \label{dF13}%
\end{equation}
and
\begin{align*}
&  \mathcal{T}^{\alpha}\lrcorner(\theta_{\alpha}\wedge\star\mathbf{F}%
\mathcal{)}\\
&  =\frac{1}{4}T_{\beta\rho}^{\alpha}(\theta^{\beta}\wedge\theta^{\rho
})\lrcorner(\theta_{\alpha}\wedge{}(^{\star}F_{\mu\nu}\theta^{\mu}\wedge
\theta^{\nu})\\
&  =\frac{1}{4}T_{\beta\rho}^{\alpha}{}^{\star}F_{\mu\nu}(\theta^{\beta}%
\wedge\theta^{\rho})\lrcorner(\theta_{\alpha}\wedge\theta^{\mu}\wedge
\theta^{\nu})\\
&  =\frac{1}{4}T_{\beta\rho}^{\alpha}{}^{\star}F^{\mu\nu}\theta^{\beta
}\lrcorner\lbrack\theta^{\rho}\lrcorner(\theta_{\alpha}\wedge\theta_{\mu
}\wedge\theta_{\nu})]\\
&  =\frac{1}{4}T_{\beta\rho}^{\alpha}{}^{\star}F^{\mu\nu}\theta^{\beta
}\lrcorner(\delta_{\alpha}^{\rho}\theta_{\mu}\wedge\theta_{\nu}-\delta_{\mu
}^{\rho}\theta_{\alpha}\wedge\theta_{\nu}+\delta_{\nu}^{\rho}\theta_{\alpha
}\wedge\theta_{\mu})\\
&  =\frac{1}{4}T_{\beta\alpha}^{\alpha}{}^{\star}F^{\mu\nu}\theta^{\beta
}\lrcorner(\theta_{\mu}\wedge\theta_{\nu})-\frac{1}{4}T_{\beta\mu}^{\alpha}%
{}^{\star}F^{\mu\nu}\theta^{\beta}\lrcorner(\theta_{\alpha}\wedge\theta_{\nu
})+\frac{1}{4}T_{\beta\nu}^{\alpha}{}^{\star}F^{\mu\nu}\theta^{\beta}%
\lrcorner(\theta_{\alpha}\wedge\theta_{\mu})
\end{align*}%
\begin{align}
&  =\frac{1}{4}T_{\beta\alpha}^{\alpha}{}^{\star}F^{\mu\nu}\theta^{\beta
}\lrcorner(\theta_{\mu}\wedge\theta_{\nu})-\frac{1}{4}T_{\beta\rho}^{\mu}%
{}^{\star}F^{\rho\nu}\theta^{\beta}\lrcorner(\theta_{\mu}\wedge\theta_{\nu
})+\frac{1}{4}T_{\beta\rho}^{\mu}{}^{\star}F^{\nu\rho}\theta^{\beta}%
\lrcorner(\theta_{\mu}\wedge\theta_{\nu})\nonumber\\
&  =\frac{1}{4}(T_{\beta\alpha}^{\alpha}{}^{\star}F^{\mu\nu}-T_{\beta\rho
}^{\mu}{}^{\star}F^{\rho\nu}+T_{\beta\rho}^{\mu}{}^{\star}F^{\nu\rho}%
)\theta^{\beta}\lrcorner(\theta_{\mu}\wedge\theta_{\nu})\nonumber\\
&  =\frac{1}{4}(T_{\beta\alpha}^{\alpha}{}^{\star}F^{\mu\nu}-T_{\beta\rho
}^{\mu}{}^{\star}F^{\rho\nu}+T_{\beta\rho}^{\mu}{}^{\star}F^{\nu\rho}%
)(\delta_{\mu}^{\beta}\theta_{\nu}-\delta_{\nu}^{\beta}\theta_{\mu
})\nonumber\\
&  =\frac{1}{4}(T_{\mu\alpha}^{\alpha}{}^{\star}F^{\mu\nu}-T_{\mu\rho}^{\mu}%
{}^{\star}F^{\rho\nu}+T_{\mu\rho}^{\mu}{}^{\star}F^{\nu\rho})\theta_{\nu
}-\frac{1}{4}(T_{\mu\alpha}^{\alpha}{}^{\star}F^{\nu\mu}-T_{\mu\rho}^{\nu}%
{}^{\star}F^{\rho\mu}+T_{\mu\rho}^{\nu}{}^{\star}F^{\mu\rho})\theta_{\nu
}\nonumber\\
&  =\frac{1}{2}(T_{\mu\alpha}^{\alpha}{}^{\star}F^{\mu\nu}-T_{\mu\rho}^{\mu}%
{}^{\star}F^{\rho\nu}+T_{\mu\rho}^{\nu}{}^{\star}F^{\mu\rho})\theta_{\nu}
\label{df14}%
\end{align}

Using Eqs.(\ref{dF13}) and (\ref{df14}) in Eq.(\ref{dF12}) we get%
\begin{equation}
D_{\alpha}{}^{\star}F^{\alpha\nu}+\frac{1}{2}(T_{\mu\alpha}^{\alpha}{}^{\star
}F^{\mu\nu}-T_{\mu\rho}^{\mu}{}^{\star}F^{\rho\nu}+T_{\mu\rho}^{\nu}{}^{\star
}F^{\mu\rho})=0 \label{d15}%
\end{equation}
and we have%
\begin{equation}
d\mathbf{F}=0\Leftrightarrow\delta\star\mathbf{F=}0\Leftrightarrow D_{\alpha
}{}^{\star}F^{\alpha\nu}+\frac{1}{2}(T_{\mu\alpha}^{\alpha}{}^{\star}F^{\mu
\nu}-T_{\mu\rho}^{\mu}{}^{\star}F^{\rho\nu}+T_{\mu\rho}^{\nu}{}^{\star}%
F^{\mu\rho})=0. \label{dF16}%
\end{equation}
Finally we express the non homogenous Maxwell equation $\delta\mathbf{F}%
=-\mathbf{J}$ in arbitrary coordinates using the Riemann-Cartan connection. We
have
\begin{align}
\delta\mathbf{F}  &  =\mathcal{-}\theta^{\alpha}\lrcorner(D_{e_{\alpha}%
}\mathbf{F})-\mathcal{T}^{\alpha}\lrcorner(\theta_{\alpha}\wedge
\mathbf{F}\mathcal{)}\nonumber\\
&  =-[D_{\alpha}{}F^{\alpha\nu}+\frac{1}{2}(T_{\mu\alpha}^{\alpha}{}^{\star
}F^{\mu\nu}-T_{\mu\rho}^{\mu}{}^{\star}F^{\rho\nu}+T_{\mu\rho}^{\nu}{}^{\star
}F^{\mu\rho})]\theta_{\nu}=-J^{\nu}\theta_{\nu}, \label{dF18}%
\end{align}
i.e.,%
\begin{equation}
D_{\alpha}{}F^{\alpha\nu}+\frac{1}{2}(T_{\mu\alpha}^{\alpha}{}^{\star}%
F^{\mu\nu}-T_{\mu\rho}^{\mu}{}^{\star}F^{\rho\nu}+T_{\mu\rho}^{\nu}{}^{\star
}F^{\mu\rho})=J^{\nu}. \label{d19}%
\end{equation}

\begin{exercise}
Show \emph{(}use \emph{Eq.(\ref{d delta bis}))} that in a Riemann-Cartan
spacetime Maxwell equations become Maxwell equation, i.e.,%
\begin{equation}
{\mbox{\boldmath$\partial$}}\mathbf{F}=\mathbf{J}+\mathcal{T}^{\mathbf{a}%
}\lrcorner(\theta_{\mathbf{a}}\wedge\mathbf{F})-\mathcal{T}^{\mathbf{a}}%
\wedge(\theta_{\mathbf{a}}\lrcorner\mathbf{F}). \label{MERC}%
\end{equation}

\end{exercise}

\section{Bianchi Identities}

We rewrite Cartan's structure equations\textit{ for an arbitrary
Riemann-Cartan structure }$(M,%
\slg
,D,\tau_{%
\slg
})$ where $\dim M=n$ and $%
\slg
$ is a metric of signature $(p,q)$, with $p+q=n$ using an arbitrary cotetrad
$\{\theta^{\mathbf{a}}\}$ as%

\begin{equation}%
\begin{array}
[c]{l}%
\mathcal{T}^{\mathbf{a}}=d\theta^{\mathbf{a}}+\omega_{\mathbf{b}}^{\mathbf{a}%
}\wedge\theta^{\mathbf{b}}=\mathbf{D}\theta^{\mathbf{a}},\\
\mathcal{R}_{\mathbf{b}}^{\mathbf{a}}=d\omega_{\mathbf{b}}^{\mathbf{a}}%
+\omega_{\mathbf{c}}^{\mathbf{a}}\wedge\omega_{\mathbf{b}}^{\mathbf{c}}%
\end{array}
\label{Cartan strc}%
\end{equation}
where
\begin{align}
&  \omega_{\mathbf{b}}^{\mathbf{a}}=\omega_{\mathbf{cb}}^{\mathbf{a}}%
\theta^{\mathbf{c}},\nonumber\\
&  \mathcal{T}^{\mathbf{a}}=\frac{1}{2}T_{\mathbf{bc}}^{\mathbf{a}}%
\theta^{\mathbf{b}}\wedge\theta^{\mathbf{c}}\\
&  \mathcal{R}_{\mathbf{b}}^{\mathbf{a}}=\frac{1}{2}R_{\mathbf{b}}%
{}^{\mathbf{a}}{}_{\!\mathbf{cd}}\theta^{\mathbf{c}}\wedge\theta^{\mathbf{d}}.
\label{cart bis}%
\end{align}
Since the $\mathcal{T}^{\mathbf{a}}$ and the $\mathcal{R}_{\mathbf{b}%
}^{\mathbf{a}}$ are \textit{index form fields} we can apply to those objects
the exterior covariant differential (Eq.(\ref{559new4})). We get
\begin{align}
\mathbf{D}\mathcal{T}^{\mathbf{a}}  &  =d\mathcal{T}^{\mathbf{a}}%
+\omega_{\mathbf{b}}^{\mathbf{a}}\wedge\mathcal{T}^{\mathbf{b}}=d^{2}%
\theta^{\mathbf{a}}+d(\omega_{\mathbf{b}}^{\mathbf{a}}\wedge\theta
^{\mathbf{b}})+\omega_{\mathbf{b}}^{\mathbf{a}}\wedge\mathcal{T}^{\mathbf{b}%
}\nonumber\\
&  =d\omega_{\mathbf{b}}^{\mathbf{a}}\wedge\theta^{\mathbf{b}}-\omega
_{\mathbf{b}}^{\mathbf{a}}\wedge d\theta^{\mathbf{b}}+\omega_{\mathbf{b}%
}^{\mathbf{a}}\wedge\mathcal{T}^{\mathbf{b}}\nonumber\\
&  =d\omega_{\mathbf{b}}^{\mathbf{a}}\wedge\theta^{\mathbf{b}}-\omega
_{\mathbf{b}}^{\mathbf{a}}\wedge(\mathcal{T}^{\mathbf{b}}-\omega_{\mathbf{c}%
}^{\mathbf{b}}\wedge\theta^{\mathbf{c}})+\omega_{\mathbf{b}}^{\mathbf{a}%
}\wedge\mathcal{T}^{\mathbf{b}}\nonumber\\
&  =(d\omega_{\mathbf{b}}^{\mathbf{a}}+\omega_{\mathbf{c}}^{\mathbf{a}}%
\wedge\omega_{\mathbf{b}}^{\mathbf{c}})\wedge\theta^{\mathbf{b}}\nonumber\\
&  =\mathcal{R}_{\mathbf{b}}^{\mathbf{a}}\wedge\theta^{\mathbf{b}} \label{b1}%
\end{align}
Also,%
\begin{align}
\mathbf{D}\mathcal{R}_{\mathbf{b}}^{\mathbf{a}}  &  =d\mathcal{R}_{\mathbf{b}%
}^{\mathbf{a}}+\omega_{\mathbf{c}}^{\mathbf{a}}\wedge\mathcal{R}_{\mathbf{b}%
}^{\mathbf{c}}-\omega_{\mathbf{b}}^{\mathbf{c}}\wedge\mathcal{R}_{\mathbf{c}%
}^{\mathbf{a}}\nonumber\\
&  =d^{2}\omega_{\mathbf{b}}^{\mathbf{a}}+d\omega_{\mathbf{c}}^{\mathbf{a}%
}\wedge\omega_{\mathbf{b}}^{\mathbf{c}}-d\omega_{\mathbf{b}}^{\mathbf{c}%
}\wedge\omega_{\mathbf{c}}^{\mathbf{a}}-\mathcal{R}_{\mathbf{c}}^{\mathbf{a}%
}\wedge\omega_{\mathbf{b}}^{\mathbf{c}}+\mathcal{R}_{\mathbf{b}}^{\mathbf{c}%
}\wedge\omega_{\mathbf{c}}^{\mathbf{a}}\nonumber\\
&  =d\omega_{\mathbf{c}}^{\mathbf{a}}\wedge\omega_{\mathbf{b}}^{\mathbf{c}%
}-\mathcal{(}d\omega_{\mathbf{c}}^{\mathbf{a}}+\omega_{\mathbf{d}}%
^{\mathbf{a}}\wedge\omega_{\mathbf{c}}^{\mathbf{d}}\mathcal{)}\wedge
\omega_{\mathbf{b}}^{\mathbf{c}}-d\omega_{\mathbf{c}}^{\mathbf{a}}\wedge
\omega_{\mathbf{b}}^{\mathbf{c}}+(d\omega_{\mathbf{b}}^{\mathbf{c}}%
+\omega_{\mathbf{d}}^{\mathbf{c}}\wedge\omega_{\mathbf{b}}^{\mathbf{d}}%
)\wedge\omega_{\mathbf{c}}^{\mathbf{a}}\nonumber\\
&  =-\omega_{\mathbf{d}}^{\mathbf{a}}\wedge\omega_{\mathbf{c}}^{\mathbf{d}%
}\wedge\omega_{\mathbf{b}}^{\mathbf{c}}+\omega_{\mathbf{d}}^{\mathbf{c}}%
\wedge\omega_{\mathbf{b}}^{\mathbf{d}}\wedge\omega_{\mathbf{c}}^{\mathbf{a}%
}\nonumber\\
&  =-\omega_{\mathbf{d}}^{\mathbf{a}}\wedge\omega_{\mathbf{c}}^{\mathbf{d}%
}\wedge\omega_{\mathbf{b}}^{\mathbf{c}}+\omega_{\mathbf{c}}^{\mathbf{d}}%
\wedge\omega_{\mathbf{b}}^{\mathbf{c}}\wedge\omega_{\mathbf{d}}^{\mathbf{a}%
}\nonumber\\
&  =-\omega_{\mathbf{d}}^{\mathbf{a}}\wedge\omega_{\mathbf{c}}^{\mathbf{d}%
}\wedge\omega_{\mathbf{b}}^{\mathbf{c}}+\omega_{\mathbf{d}}^{\mathbf{a}}%
\wedge\omega_{\mathbf{c}}^{\mathbf{d}}\wedge\omega_{\mathbf{b}}^{\mathbf{c}%
}=0. \label{b2}%
\end{align}
So, we have the general Bianchi identities which are valid for any one of the
\ metrical compatible structures\footnote{For non metrical compatible
structures we have more general equations than the Cartan structure equations
and thus more general identities, see \cite{rodoliv2007}.} classified in
Section 2,%
\begin{align}
\mathbf{D}\mathcal{T}^{\mathbf{a}}  &  =\mathcal{R}_{\mathbf{b}}^{\mathbf{a}%
}\wedge\theta^{\mathbf{b}},\nonumber\\
\mathbf{D}\mathcal{R}_{\mathbf{b}}^{\mathbf{a}}  &  =0. \label{bianchi}%
\end{align}

\subsection{Coordinate Expressions of the First Bianchi Identity}

Taking advantage of the calculations we done for the coordinate expressions of
Maxwell equations we can write in a while:%

\begin{align}
\mathbf{D}\mathcal{T}^{\mathbf{a}}  &  =d\mathcal{T}^{\mathbf{a}}%
+\omega_{\mathbf{b}}^{\mathbf{a}}\wedge\mathcal{T}^{\mathbf{b}}\nonumber\\
&  =\frac{1}{3!}\left(  \partial_{\mu}T_{\alpha\beta}^{\mathbf{a}}+\omega
_{\mu\mathbf{b}}^{\mathbf{a}}T_{\alpha\beta}^{\mathbf{b}}+\partial_{\alpha
}T_{\beta\mu}^{\mathbf{a}}+\omega_{\alpha\mathbf{b}}^{\mathbf{a}}T_{\beta\mu
}^{\mathbf{b}}+\partial_{\beta}T_{\mu\alpha}^{\mathbf{a}}+\omega
_{\beta\mathbf{b}}^{\mathbf{a}}T_{\mu\alpha}^{\mathbf{b}}\right)  \theta^{\mu
}\wedge\theta^{\alpha}\wedge\theta^{\beta}. \label{bi1}%
\end{align}
Now,
\begin{equation}
\partial_{\mu}T_{\alpha\beta}^{\mathbf{a}}=(\partial_{\mu}q_{\rho}%
^{\mathbf{a}})T_{\alpha\beta}^{\rho}+q_{\rho}^{\mathbf{a}}\partial_{\mu
}T_{\alpha\beta}^{\rho}, \label{bi2}%
\end{equation}
and using the freshman identity (Eq.(\ref{freshman})) we can write
\begin{equation}
\omega_{\mu\mathbf{b}}^{\mathbf{a}}T_{\alpha\beta}^{\mathbf{b}}=\omega
_{\mu\mathbf{b}}^{\mathbf{a}}q_{\rho}^{\mathbf{b}}T_{\alpha\beta
}^{\mathbf{\rho}}=L_{\mu\mathbf{b}}^{\mathbf{a}}q_{\rho}^{\mathbf{b}}%
T_{\alpha\beta}^{\rho}-(\partial_{\mu}q_{\rho}^{\mathbf{a}})T_{\alpha\beta
}^{\rho}. \label{bi3}%
\end{equation}
So,%

\begin{align}
&  \partial_{\mu}T_{\alpha\beta}^{\mathbf{a}}+\omega_{\mu\mathbf{b}%
}^{\mathbf{a}}T_{\alpha\beta}^{\mathbf{b}}\nonumber\\
&  =q_{\rho}^{\mathbf{a}}\partial_{\mu}T_{\alpha\beta}^{\rho}+\Gamma
_{\mu\mathbf{b}}^{\mathbf{a}}q_{\rho}^{\mathbf{b}}T_{\alpha\beta}^{\rho
}\nonumber\\
&  =q_{\rho}^{\mathbf{a}}(D_{\mu}T_{\alpha\beta}^{\rho}+\Gamma_{\mu\alpha
}^{\kappa}T_{\kappa\beta}^{\rho}+\Gamma_{\mu\beta}^{\kappa}T_{\alpha\kappa
}^{\rho}). \label{bi4}%
\end{align}
Now, recalling that $T_{\mu\alpha}^{\kappa}=\Gamma_{\mu\alpha}^{\kappa}%
-\Gamma_{\alpha\mu}^{\kappa}$ we can write%
\begin{align}
&  q_{\rho}^{\mathbf{a}}(\Gamma_{\mu\alpha}^{\kappa}T_{\kappa\beta}^{\rho
}+\Gamma_{\mu\beta}^{\kappa}T_{\alpha\kappa}^{\rho})\theta^{\mu}\wedge
\theta^{\alpha}\wedge\theta^{\beta}\label{bi5}\\
&  =q_{\rho}^{\mathbf{a}}T_{\mu\alpha}^{\kappa}T_{\kappa\beta}^{\rho}%
\theta^{\mu}\wedge\theta^{\alpha}\wedge\theta^{\beta}.\nonumber
\end{align}
Using these formulas we can write%
\begin{align}
&  \mathbf{D}\mathcal{T}^{\mathbf{a}}\nonumber\\
&  =\frac{1}{3!}q_{\rho}^{\mathbf{a}}\left\{  D_{\mu}T_{\alpha\beta}^{\rho
}+D_{\alpha}T_{\beta\mu}^{\rho}+D_{\beta}T_{\mu\alpha}^{\rho}+T_{\mu\alpha
}^{\kappa}T_{\kappa\beta}^{\rho}+T_{\alpha\beta}^{\kappa}T_{\kappa\mu}^{\rho
}+T_{\beta\mu}^{\kappa}T_{\kappa\alpha}^{\rho}\right\}  \theta^{\mu}%
\wedge\theta^{\alpha}\wedge\theta^{\beta}. \label{bi6}%
\end{align}
Now, the coordinate representation of \ $\mathcal{R}_{\mathbf{b}}^{\mathbf{a}%
}\wedge\theta^{\mathbf{b}}$ is:%
\begin{equation}
\mathcal{R}_{\mathbf{b}}^{\mathbf{a}}\wedge\theta^{\mathbf{b}}=\frac{1}%
{3!}q_{\rho}^{\mathbf{a}}(R_{\mu}{}^{\rho}{}_{\alpha\beta}+R_{\alpha}{}^{\rho
}{}_{\beta\mu}+R_{\beta}{}^{\rho}{}_{\mu\alpha})\theta^{\mu}\wedge
\theta^{\alpha}\wedge\theta^{\beta}, \label{bi7}%
\end{equation}
and thus the coordinate expression of the first Bianchi identity is:%
\begin{equation}
D_{\mu}T_{\alpha\beta}^{\rho}+D_{\alpha}T_{\beta\mu}^{\rho}+D_{\beta}%
T_{\mu\alpha}^{\rho}=(R_{\mu}{}^{\rho}{}_{\alpha\beta}+R_{\alpha}{}^{\rho}%
{}_{\!\beta\mu}+R_{\beta}{}^{\rho}{}_{\mu\alpha})-(T_{\mu\alpha}^{\kappa
}T_{\kappa\beta}^{\rho}+T_{\alpha\beta}^{\kappa}T_{\kappa\mu}^{\rho}%
+T_{\beta\mu}^{\kappa}T_{\kappa\alpha}^{\rho}), \label{bi8}%
\end{equation}
which we can write as%
\begin{equation}%
{\displaystyle\sum\limits_{(\mu\alpha\beta)}}
R_{\mu}{}^{\rho}{}_{\alpha\beta}=%
{\displaystyle\sum\limits_{(\mu\alpha\beta)}}
\left(  D_{\mu}T_{\alpha\beta}^{\rho}-T_{\mu\beta}^{\kappa}T_{\kappa\alpha
}^{\rho}\right)  ,
\end{equation}
with $%
{\displaystyle\sum\limits_{(\mu\alpha\beta)}}
$ denoting as usual the sum over cyclic permutation of the indices $(\mu
\alpha\beta)$. For the particular case of a Levi-Civita connection
$\mathring{D}$ since the $T_{\alpha\beta}^{\rho}=0$ we have the standard form
of the first Bianchi identity in classical Riemannian geometry, i.e.,
\begin{equation}
R_{\mu}{}^{\rho}{}_{\alpha\beta}+R_{\alpha}{}^{\rho}{}_{\beta\mu}+R_{\beta}%
{}^{\rho}{}_{\mu\alpha}=0. \label{cbi}%
\end{equation}

If we now recall the steps that lead us to Eq.(\ref{dF16}) we can write for
the torsion $2$-form fields $\mathcal{T}^{\mathbf{a}}$,%

\begin{align}
d\mathcal{T}^{\mathbf{a}}  &  =\star\star^{-1}d\star\star^{-1}\mathcal{T}%
^{\mathbf{a}}\nonumber\\
&  =(-1)^{n-2}\star\delta\star^{-1}\mathcal{T}^{\mathbf{a}}=(-1)^{n-2}%
(-1)^{n-2}\mathrm{sgn}%
\slg
\star\delta\star\mathcal{T}^{\mathbf{a}}\nonumber\\
&  =(-1)^{n-2}\star^{-1}\delta\star\mathcal{T}^{\mathbf{a}}. \label{bi9}%
\end{align}
with $\mathrm{sgn}%
\slg
=\det%
\slg
/\left\vert \det%
\slg
\right\vert $. Then we can write the first Bianchi identity as%
\begin{equation}
\delta\star\mathcal{T}^{\mathbf{a}}=(-1)^{n-2}\star\lbrack\mathcal{R}%
_{\mathbf{b}}^{\mathbf{a}}\wedge\theta^{\mathbf{b}}-\omega_{\mathbf{b}%
}^{\mathbf{a}}\wedge\mathcal{T}^{\mathbf{b}}], \label{bi10}%
\end{equation}
and taking into account that
\begin{align}
\star(\mathcal{R}_{\mathbf{b}}^{\mathbf{a}}\wedge\theta^{\mathbf{b}})  &
=\star(\theta^{\mathbf{b}}\wedge\mathcal{R}_{\mathbf{b}}^{\mathbf{a}}%
)=\theta^{\mathbf{b}}\lrcorner\star\mathcal{R}_{\mathbf{b}}^{\mathbf{a}%
},\nonumber\\
\star(\omega_{\mathbf{b}}^{\mathbf{a}}\wedge\mathcal{T}^{\mathbf{b}})  &
=\omega_{\mathbf{b}}^{\mathbf{a}}\lrcorner\star\mathcal{T}^{\mathbf{b}},
\label{bi11}%
\end{align}
we end with\
\begin{equation}
\delta\star\mathcal{T}^{\mathbf{a}}=(-1)^{n-2}(\theta^{\mathbf{b}}%
\lrcorner\star\mathcal{R}_{\mathbf{b}}^{\mathbf{a}}-\omega_{\mathbf{b}%
}^{\mathbf{a}}\lrcorner\star\mathcal{T}^{\mathbf{b}}). \label{bi12}%
\end{equation}
This is the first Bianchi identity written in terms of duals. To calculate its
coordinate expression, we recall the steps that lead us to Eq.(\ref{dF16}) and
write directly for the torsion $2$-form fields $\mathcal{T}^{\mathbf{a}}$%

\begin{align}
&  \delta\star\mathcal{T}^{\mathbf{a}}\nonumber\\
&  =-(D_{\alpha}{}^{\star}T^{\mathbf{a}\alpha\nu}+\frac{1}{2}(T_{\mu\alpha
}^{\alpha}{}^{\star}T^{\mathbf{a}\mu\nu}-T_{\mu\rho}^{\mu}{}^{\star
}T^{\mathbf{a}\rho\nu}+T_{\mu\rho}^{\nu}{}^{\star}T^{\mathbf{a}\mu\rho
}))\theta_{\nu}. \label{bi13}%
\end{align}

Also, writing%
\begin{equation}
\star\mathcal{R}_{\mathbf{b}}^{\mathbf{a}}=\frac{1}{2}{}^{\ast}%
R_{\mathbf{b\;cd}}^{\;\mathbf{a}}\theta^{\mathbf{c}}\wedge\theta^{\mathbf{d}},
\end{equation}
we have:%

\begin{align}
\star(\mathcal{R}_{\mathbf{b}}^{\mathbf{a}}\wedge\theta^{\mathbf{b}})  &
=\theta^{\mathbf{b}}\lrcorner\star\mathcal{R}_{\mathbf{b}}^{\mathbf{a}%
}\nonumber\\
&  =\frac{1}{2}\theta^{\mathbf{b}}\lrcorner{}(^{\star}R_{\mathbf{b\;cd}%
}^{\;\mathbf{a}}\theta^{\mathbf{c}}\wedge\theta^{\mathbf{d}})\nonumber\\
&  ={}^{\star}R_{\mathbf{b\;cd}}^{\;\mathbf{a}}\eta^{\mathbf{bc}}%
\theta^{\mathbf{d}}\nonumber\\
&  ={}^{\star}R_{\mathbf{\;\;\,cd}}^{\mathbf{ca}}\theta^{\mathbf{d}}={}%
^{\star}R_{\mathbf{\;\;\,c}}^{\mathbf{ca\;d}}\theta_{\mathbf{d}}={}^{\star
}R_{\mathbf{\;\;\,c}}^{\mathbf{ca\;d}}q_{\mathbf{d}}^{\nu}\theta_{\nu}.
\label{bi14}%
\end{align}
On the other hand we can also write:%
\begin{align*}
\star(\mathcal{R}_{\mathbf{b}}^{\mathbf{a}}\wedge\theta^{\mathbf{b}})  &
=\theta^{\mathbf{b}}\lrcorner\star\mathcal{R}_{\mathbf{b}}^{\mathbf{a}}\\
&  =\frac{1}{2}\vartheta^{\mathbf{b}}\lrcorner{}(\frac{1}{(n-2)!}%
R_{\mathbf{b}}^{\;\mathbf{akl}}\epsilon_{\mathbf{klmn}}\theta^{\mathbf{m}%
}\wedge\theta^{\mathbf{n}})\\
&  =\frac{1}{2}\frac{1}{(n-2)!}(R_{\mathbf{b}}^{\;\mathbf{akl}}\epsilon
_{\mathbf{klmn}}\eta^{\mathbf{bm}}\wedge\theta^{\mathbf{n}}-R_{\mathbf{b}%
}^{\;\mathbf{akl}}\epsilon_{\mathbf{klmn}}\eta^{\mathbf{bn}}\wedge
\theta^{\mathbf{m}})\\
&  =\frac{1}{(n-2)!}R_{\mathbf{b}}^{\;\mathbf{akl}}\epsilon_{\mathbf{klmn}%
}\eta^{\mathbf{bm}}\theta^{\mathbf{n}}=\frac{1}{(n-2)!}R^{\mathbf{makl}%
}\epsilon_{\mathbf{klmn}}\theta^{\mathbf{n}}\\
&  =\frac{1}{(n-2)!}R_{\mathbf{m}}^{\;\,\mathbf{akl}}\epsilon_{\mathbf{kl}%
}^{\;\;\,\mathbf{mn}}\theta_{\mathbf{n}}\\
&  =\frac{1}{(n-2)!}R_{\mathbf{m}}^{\;\,\mathbf{akl}}\epsilon_{\mathbf{kl}%
}^{\;\;\,\mathbf{mn}}q_{\mathbf{n}}^{\nu}\theta_{\nu}.
\end{align*}
from where we get in agreement with Eq.(\ref{hodge dual}) the formula%
\begin{equation}
^{\star}R_{\mathbf{\;\;\,cd}}^{\mathbf{ca\;}}=\frac{1}{(n-2)!}R^{\mathbf{makl}%
}\epsilon_{\mathbf{mkld}}, \label{bi15}%
\end{equation}
which shows explicitly that \ $^{\star}R_{\mathbf{\;\;\,cd}}^{\mathbf{ca\;}}$
are not the components of the \textit{Ricci} tensor.

Moreover,%
\begin{align}
&  \omega_{\mathbf{b}}^{\mathbf{a}}\lrcorner\star\mathcal{T}^{\mathbf{b}%
}\label{bi16}\\
&  =\frac{1}{2}\omega_{\alpha\mathbf{b}}^{\mathbf{a}}\theta^{\alpha}%
\lrcorner(^{\star}T^{\mathbf{b\mu\nu}}\theta_{\mu}\wedge\theta_{\nu
})\nonumber\\
&  =^{\star}\!T^{\mathbf{b\mu\nu}}\omega_{\alpha\mathbf{b}}^{\mathbf{a}}%
\theta_{\nu}.\nonumber
\end{align}
Collecting the above formulas we end with%
\begin{equation}
D_{\alpha}{}^{\star}T^{\mathbf{a}\alpha\nu}+\frac{1}{2}(T_{\mu\alpha}^{\alpha
}{}^{\star}T^{\mathbf{a}\mu\nu}-T_{\mu\rho}^{\mu}{}^{\star}T^{\mathbf{a}%
\rho\nu}+T_{\mu\rho}^{\nu}{}^{\star}T^{\mathbf{a}\mu\rho})=(-1)^{n-1}(^{\star
}R_{\mathbf{\;\;\,c}}^{\mathbf{ca\;d}}q_{\mathbf{d}}^{\nu}-\omega
_{\alpha\mathbf{b}}^{\mathbf{a}{}\star}T^{\mathbf{b\alpha\nu}}), \label{bi17}%
\end{equation}
which is another expression for the first Bianchi identity written in terms of duals.

\begin{remark}
Consider, e.g., the term $D_{\alpha}{}^{\star}T^{\mathbf{a}\alpha\nu}$ in the
above equation and write
\begin{equation}
D_{\alpha}{}^{\star}T^{\mathbf{a}\alpha\nu}=D_{\alpha}(q_{\rho}^{\mathbf{a}}%
{}^{\star}T^{\rho\alpha\nu}). \label{bi18}%
\end{equation}
We now show that
\begin{equation}
D_{\alpha}(q_{\rho}^{\mathbf{a}}{}^{\star}T^{\rho\alpha\nu})\neq q_{\rho
}^{\mathbf{a}}D_{\alpha}{}^{\star}T^{\rho\alpha\nu}. \label{bi19}%
\end{equation}
Indeed, recall that we already found that \
\begin{equation}
(D_{\alpha}{}^{\star}T^{\mathbf{a}\alpha\nu})\theta_{\nu}=-\mathbf{\delta
\star}\mathcal{T}^{\mathbf{a}}+\frac{1}{2}(T_{\mu\alpha}^{\alpha}{}^{\star
}T^{\mathbf{a}\mu\nu}-T_{\mu\rho}^{\mu}{}^{\star}T^{\mathbf{a}\rho\nu}%
+T_{\mu\rho}^{\nu}{}^{\star}T^{\mathbf{a}\mu\rho})\theta_{\nu}, \label{bi20'}%
\end{equation}
and taking into account the second formula in \emph{Eq.(\ref{d delta bis})} we
can write
\begin{equation}
\theta^{\alpha}\lrcorner(D_{\mathbf{\partial}_{\alpha}}\star\mathcal{T}%
^{\mathbf{a}})=-\mathbf{\delta\star}\mathcal{T}^{\mathbf{a}}+\frac{1}%
{2}(T_{\mu\alpha}^{\alpha}{}^{\star}T^{\mathbf{a}\mu\nu}-T_{\mu\rho}^{\mu}%
{}^{\star}T^{\mathbf{a}\rho\nu}+T_{\mu\rho}^{\nu}{}^{\star}T^{\mathbf{a}%
\mu\rho})\theta_{\nu}. \label{bi21}%
\end{equation}
Now, writing $\star\mathcal{T}^{\mathbf{a}}=\frac{1}{2}q_{\rho}^{\mathbf{a}}$
$^{{}\star}T^{\rho\mu\nu}\theta_{\mu}\wedge\theta_{\nu}$ and get
\begin{align}
&  \theta^{\alpha}\lrcorner(D_{\mathbf{\partial}_{\alpha}}\star\mathcal{T}%
^{\mathbf{a}})\nonumber\\
&  =\frac{1}{2}\theta^{\alpha}\lrcorner\lbrack D_{\mathbf{\partial}_{\alpha}%
}(q_{\rho}^{\mathbf{a}}{}^{\star}T^{\mathbf{\rho}\mu\nu}\theta_{\mu}%
\wedge\theta_{\nu})]\nonumber\\
&  =\frac{1}{2}\vartheta^{\alpha}\lrcorner\lbrack\partial_{\alpha}(q_{\rho
}^{\mathbf{a}}{}^{\star}T^{\mathbf{\rho}\mu\nu})\theta_{\mu}\wedge\theta_{\nu
}+q_{\rho}^{\mathbf{a}}{}^{\star}T^{\mathbf{\rho}\mu\nu}D_{\mathbf{\partial
}_{\alpha}}(\theta_{\mu}\wedge\theta_{\nu})]\nonumber\\
&  =\frac{1}{2}\vartheta^{\alpha}\lrcorner\lbrack(\partial_{\alpha}q_{\rho
}^{\mathbf{a}}){}^{\star}T^{\mathbf{\rho}\mu\nu}\theta_{\mu}\wedge\theta_{\nu
}+q_{\rho}^{\mathbf{a}}\partial_{\alpha}({}^{\star}T^{\mathbf{\rho}\mu\nu
})\theta_{\mu}\wedge\theta_{\nu}+q_{\rho}^{\mathbf{a}}{}^{\star}%
T^{\mathbf{\rho}\mu\nu}D_{\mathbf{\partial}_{\alpha}}(\theta_{\mu}\wedge
\theta_{\nu})]\nonumber\\
&  =\frac{1}{2}\vartheta^{\alpha}\lrcorner\lbrack(\partial_{\alpha}q_{\rho
}^{\mathbf{a}}){}^{\star}T^{\mathbf{\rho}\mu\nu}\theta_{\mu}\wedge\theta_{\nu
}+q_{\rho}^{\mathbf{a}}D_{\alpha}({}^{\star}T^{\mathbf{\rho}\mu\nu}%
)\theta_{\mu}\wedge\theta_{\nu}]\nonumber\\
&  =(\partial_{\alpha}q_{\rho}^{\mathbf{a}}){}^{\star}T^{\mathbf{\rho}\mu\nu
}\delta_{\mu}^{\alpha}\theta_{\nu}+q_{\rho}^{\mathbf{a}}D_{\alpha}({}^{\star
}T^{\mathbf{\rho}\mu\nu})\delta_{\mu}^{\alpha}\theta_{\nu}. \label{bi22}%
\end{align}
Comparing the \emph{Eq.(\ref{bi20})} with \emph{Eq.(\ref{bi21})} using
\emph{Eq.(\ref{bi22})} we get%
\begin{equation}
D_{\alpha}{}^{\star}T^{\mathbf{a}\alpha\nu}\theta_{\nu}=D_{\alpha}(q_{\rho
}^{\mathbf{a}}{}^{\star}T^{\mathbf{a}\alpha\nu}\theta_{\nu})=(\partial
_{\alpha}q_{\rho}^{\mathbf{a}}){}^{\star}T^{\mathbf{\rho}\mu\nu}+q_{\rho
}^{\mathbf{a}}D_{\alpha}({}^{\star}T^{\mathbf{\rho}\mu\nu}), \label{bi23}%
\end{equation}
thus proving our statement and showing the danger of applying a so called
"tetrad postulate" asserting without due care on the meaning of the symbols
that \textquotedblleft\ the covariant derivative of the tetrad is zero, and
thus using \textquotedblleft$D_{\alpha}q_{\rho}^{\mathbf{a}}=0$%
\textquotedblright.\textquotedblright
\end{remark}

\begin{exercise}
Show that the coordinate expression of the second Bianchi identity
$\mathbf{D}\mathcal{R}_{\mathbf{b}}^{\mathbf{a}}=0$ is%
\begin{equation}%
{\displaystyle\sum\limits_{(\mu\nu\rho)}}
D_{\mu}R_{\beta\;\nu\rho}^{\;\alpha}=%
{\displaystyle\sum\limits_{(\mu\nu\rho)}}
T_{\nu\mu}^{\alpha}R_{\beta\;\alpha\rho}^{\;\alpha}. \label{bi24}%
\end{equation}

\end{exercise}

\begin{exercise}
Calculate $\star\mathcal{R}_{\mathbf{b}}^{\mathbf{a}}\wedge\theta^{\mathbf{b}%
}$ in an orthonormal basis.
\end{exercise}

\noindent\textbf{Solution: }First we recall the $\star\mathcal{R}_{\mathbf{b}%
}^{\mathbf{a}}\wedge\theta^{\mathbf{b}}=\theta^{\mathbf{b}}\wedge
\star\mathcal{R}_{\mathbf{b}}^{\mathbf{a}}$ and then use the formula in the
third line of Eq.(\ref{440new}) to write:%
\begin{align}
\theta^{\mathbf{b}}\wedge\star\mathcal{R}_{\mathbf{b}}^{\mathbf{a}}  &
=-\star(\theta^{\mathbf{b}}\lrcorner\mathcal{R}_{\mathbf{b}}^{\mathbf{a}%
})\nonumber\\
&  =-\star\left[  \frac{1}{2}\theta^{\mathbf{b}}\lrcorner(R_{\mathbf{b\;cd}%
}^{\;\mathbf{a}}\theta^{\mathbf{c}}\wedge\theta^{\mathbf{d}})\right]
\nonumber\\
&  =-\star\lbrack R_{\mathbf{b\;cd}}^{\;\mathbf{a}}\eta^{\mathbf{bc}}%
\theta^{\mathbf{d}}]\nonumber\\
&  =-\star\lbrack R_{\;\;\mathbf{cd}}^{\mathbf{ca}}\theta^{\mathbf{d}}%
]=-\star\lbrack R_{\;\;\mathbf{dc}}^{\mathbf{ac}}\theta^{\mathbf{d}%
}]\nonumber\\
&  =-\star\lbrack R_{\mathbf{d}}^{\mathbf{a}}\theta^{\mathbf{d}}%
]=-\star\mathcal{R}^{\mathbf{a}} \label{bi25}%
\end{align}
Of course, if the connection is the Levi-Civita one we get%
\begin{equation}
\theta^{\mathbf{b}}\wedge\star\mathcal{\mathring{R}}_{\mathbf{b}}^{\mathbf{a}%
}=-\star(\theta^{\mathbf{b}}\lrcorner\mathcal{\mathring{R}}_{\mathbf{b}%
}^{\mathbf{a}})=-\star\mathring{R}_{\mathbf{b}}^{\mathbf{a}}\theta
^{\mathbf{b}}=-\star\mathcal{\mathring{R}}^{\mathbf{a}}. \label{bi29}%
\end{equation}

\section{A Remark on Evans 101$^{th}$ Paper on "ECE Theory"}

Eq. (\ref{bi17}) or its equivalent Eq.(\ref{bi23}) is to be compared with a
wrong one derived by Evans from where he now claims that \ the
Einstein-Hilbert (gravitational) theory which uses\ in its formulation the
Levi-Civita connection $\mathring{D}$ is incompatible with the first Bianchi
identity. Evans conclusion follows because he thinks to have derived
\textquotedblleft from first principles\textquotedblright\ that
\begin{equation}%
\begin{tabular}
[c]{l}%
$\mathbf{D\star}\mathcal{T}^{\mathbf{a}}=\star\mathcal{R}_{\mathbf{b}%
}^{\mathbf{a}}\wedge\theta^{\mathbf{b}},$%
\end{tabular}
\ \ \ \ \ \ \ \label{Evans 24}%
\end{equation}
an equation that \textit{if} true implies as we just see from Eq.(\ref{bi25})
that for the Levi-Civita connection for which the $\mathcal{T}^{\mathbf{a}}=0$
the Ricci tensor of the connection $\mathring{D}$ is null.

We show below that Eq.(\ref{Evans 24}) is a false one in two different ways,
firstly by deriving the correct equation for $\mathbf{D\star}\mathcal{T}%
^{\mathbf{a}}$ and secondly by showing explicit counterexamples for some
trivial structures.

Before doing that let us show that we can derive from the first Bianchi
identity that%
\begin{equation}
\mathring{R}_{\mathbf{a\;cd}}^{\;\mathbf{a}}=0, \label{e3bis}%
\end{equation}
an equation that eventually may lead Evans in believing that for a Levi-Civita
connection the first Bianchi identity implies that the Ricci tensor is null.
As we know, for a Levi-Civita connection the first Bianchi identity gives
(with $\mathcal{R}_{\mathbf{b}}^{\mathbf{a}}\longmapsto\mathcal{\mathring{R}%
}_{\mathbf{b}}^{\mathbf{a}})$:
\begin{equation}
\mathcal{\mathring{R}}_{\mathbf{b}}^{\mathbf{a}}\wedge\theta^{\mathbf{b}}=0.
\label{e1}%
\end{equation}

Contracting this equation with $\theta_{\mathbf{a}}$ we get%
\begin{align*}
\theta_{\mathbf{a}}\lrcorner(\mathcal{\mathring{R}}_{\mathbf{b}}^{\mathbf{a}%
}\wedge\theta^{\mathbf{b}})  &  =\theta_{\mathbf{a}}\lrcorner(\theta
^{\mathbf{b}}\wedge\mathcal{\mathring{R}}_{\mathbf{b}}^{\mathbf{a}})\\
&  =\delta_{\mathbf{a}}^{\mathbf{b}}\mathcal{\mathring{R}}_{\mathbf{b}%
}^{\mathbf{a}}-\theta^{\mathbf{b}}\wedge(\theta_{\mathbf{a}}\lrcorner
\mathcal{\mathring{R}}_{\mathbf{b}}^{\mathbf{a}})\\
&  =\mathcal{\mathring{R}}_{\mathbf{a}}^{\mathbf{a}}-\frac{1}{2}%
\theta^{\mathbf{b}}\wedge\lbrack\theta_{\mathbf{a}}\lrcorner(\mathring
{R}_{\mathbf{b\;cd}}^{\;\mathbf{a}}\theta^{\mathbf{c}}\wedge\theta
^{\mathbf{d}})]\\
&  =\mathcal{\mathring{R}}_{\mathbf{a}}^{\mathbf{a}}-\mathring{R}%
_{\mathbf{b\;ad}}^{\;\mathbf{a}}\theta^{\mathbf{b}}\wedge\theta^{\mathbf{d}}%
\end{align*}
Now, the second term in this last equation is null because according to the
Eq.(\ref{ricci}),\ $-R_{\mathbf{b\;ad}}^{\;\mathbf{a}}=R_{\mathbf{b\;da}%
}^{\;\mathbf{a}}=R_{\mathbf{bd}}^{\;}$ are the components of the Ricci tensor,
which is a symmetric tensor for the Levi-Civita connection. For the first term
we get
\begin{equation}
\mathring{R}_{\mathbf{a\;cd}}^{\;\mathbf{a}}\theta^{\mathbf{c}}\wedge
\theta^{\mathbf{d}}=0, \label{e2}%
\end{equation}
which implies that as we stated above that%
\begin{equation}
\mathring{R}_{\mathbf{a\;cd}}^{\;\mathbf{a}}=0. \label{e3}%
\end{equation}
But according to Eq.(\ref{ricci}) the$\ \mathring{R}_{\mathbf{a\;cd}%
}^{\;\mathbf{a}}$ are not the components of the Ricci tensor, and so there is
not any contradiction. \ As an additional verification recall that the
standard form of the first Bianchi identity in Riemannian geometry is%
\begin{equation}
\mathring{R}_{\mathbf{b}}{}^{\mathbf{a}}{}_{\!\mathbf{cd}}+\mathring
{R}_{\mathbf{c}}{}^{\mathbf{a}}{}_{\!\mathbf{db}}+\mathring{R}_{\mathbf{d}}%
{}^{\mathbf{a}}{}_{\!\mathbf{bc}}.=0
\end{equation}
Making $\mathbf{b}=\mathbf{a}$ we get%
\begin{align}
&  \mathring{R}_{\mathbf{a}}{}^{\mathbf{a}}{}_{\!\mathbf{cd}}+\mathring
{R}_{\mathbf{c}}{}^{\mathbf{a}}{}_{\!\mathbf{da}}+\mathring{R}_{\mathbf{d}}%
{}^{\mathbf{a}}{}_{\!\mathbf{ac}}\nonumber\\
&  =\mathring{R}_{\mathbf{a}}{}^{\mathbf{a}}{}_{\!\mathbf{cd}}-\mathring
{R}_{\mathbf{c}}{}^{\mathbf{a}}{}_{\!\mathbf{ad}}+\mathring{R}_{\mathbf{d}}%
{}^{\mathbf{a}}{}_{\!\mathbf{ac}}\nonumber\\
&  =\mathring{R}_{\mathbf{a}}{}^{\mathbf{a}}{}_{\!\mathbf{cd}}+\mathring
{R}_{\mathbf{cd}}-\mathring{R}_{\mathbf{dc}}\nonumber\\
&  =\mathring{R}_{\mathbf{a}}{}^{\mathbf{a}}{}_{\!\mathbf{cd}}=0. \label{e4}%
\end{align}

\begin{center}

\end{center}

\section{Direct Calculation of $\mathbf{D}\star\mathcal{T}^{\mathbf{a}}$}

We now present \ using results of Clifford bundle formalism, recalled above
(for details, see, e.g., \cite{rodoliv2007}) a calculation of $\mathbf{D}%
\star\mathcal{T}^{\mathbf{a}}$.

We start from Cartan first structure equation
\begin{equation}
\mathcal{T}^{\mathbf{a}}=d\theta^{\mathbf{a}}+\omega_{\mathbf{b}}^{\mathbf{a}%
}\wedge\theta^{\mathbf{b}}.
\end{equation}
By definition
\begin{equation}
\mathbf{D}\star\mathcal{T}^{\mathbf{a}}=d\star\mathcal{T}^{\mathbf{a}}%
+\omega_{\mathbf{b}}^{\mathbf{a}}\wedge\star\mathcal{T}^{\mathbf{b}}.
\label{du1}%
\end{equation}

Now, if we recall Eq.(\ref{545}), since the $\mathcal{T}^{\mathbf{a}}\in\sec%
{\displaystyle\bigwedge\nolimits^{2}}
T^{\ast}M\hookrightarrow\sec\mathcal{C\ell(}M,\mathtt{g)}$ we can write
\begin{equation}
d\star\mathcal{T}^{\mathbf{a}}=\star\delta\mathcal{T}^{\mathbf{a}}.
\label{du2}%
\end{equation}

We next calculate $\delta\mathcal{T}^{\mathbf{a}}$. We have:%
\begin{align}
\delta\mathcal{T}^{\mathbf{a}}  &  =\delta\left(  d\theta^{\mathbf{a}}%
+\omega_{\mathbf{b}}^{\mathbf{a}}\wedge\theta^{\mathbf{b}}\right) \nonumber\\
&  =\delta d\theta^{\mathbf{a}}+d\delta\theta^{\mathbf{a}}-d\delta
\theta^{\mathbf{a}}+\delta(\omega_{\mathbf{b}}^{\mathbf{a}}\wedge
\theta^{\mathbf{b}})\text{.} \label{du3}%
\end{align}

Next we recall the definition of the Hodge D'Alembertian which, recalling
Eq.(\ref{ssd}) permit us to write the first two terms in Eq.(\ref{du3}) as the
negative of the square of the standard Dirac operator (associated with the
Levi-Civita connection)\footnote{Be patient, the Riemann-Cartan connection
will appear in due time.}. We then get:%
\begin{align}
\delta\mathcal{T}^{\mathbf{a}}  &  =-%
\bpartial
\text{ }^{2}\theta^{\mathbf{a}}-d\delta\theta^{\mathbf{a}}+\delta
(\omega_{\mathbf{b}}^{\mathbf{a}}\wedge\theta^{\mathbf{b}})\nonumber\\
&  \overset{\text{Eq.(\ref{1796})}}{=}-\mathring{\square}\theta^{\mathbf{a}}-(%
\bpartial
\wedge%
\bpartial
)\theta^{\mathbf{a}}-d\delta\theta^{\mathbf{a}}+\delta(\omega_{\mathbf{b}%
}^{\mathbf{a}}\wedge\theta^{\mathbf{b}})\nonumber\\
&  \overset{\text{Eq.(\ref{ricci equation})}}{=}-\mathring{\square}%
\theta^{\mathbf{a}}-\mathcal{\mathring{R}}^{\mathbf{a}}-d\delta\theta
^{\mathbf{a}}+\delta(\omega_{\mathbf{b}}^{\mathbf{a}}\wedge\theta^{\mathbf{b}%
})\nonumber\\
&  =-\mathring{\square}\theta^{\mathbf{a}}-\mathcal{R}^{\mathbf{a}%
}+\mathcal{J}^{\mathbf{a}}-d\delta\theta^{\mathbf{a}}+\delta(\omega
_{\mathbf{b}}^{\mathbf{a}}\wedge\theta^{\mathbf{b}}) \label{du4}%
\end{align}
where we have used Eq(\ref{1174}) to write
\begin{equation}
\mathcal{R}^{\mathbf{a}}=R_{\mathbf{b}}^{\mathbf{a}}\theta^{\mathbf{b}%
}=(\mathring{R}_{\mathbf{b}}^{\mathbf{a}}+J_{\mathbf{b}}^{\mathbf{a}}%
)\theta^{\mathbf{b}}. \label{du5}%
\end{equation}
So, we have%
\[
d\star\mathcal{T}^{\mathbf{a}}=-\star\mathring{\square}\theta^{\mathbf{a}%
}-\star\mathcal{R}^{\mathbf{a}}+\star\mathcal{J}^{\mathbf{a}}-\star
d\delta\theta^{\mathbf{a}}+\star\delta(\omega_{\mathbf{b}}^{\mathbf{a}}%
\wedge\theta^{\mathbf{b}})
\]
and finally%
\begin{equation}
D\star\mathcal{T}^{\mathbf{a}}=-\star\mathring{\square}\theta^{\mathbf{a}%
}-\star\mathcal{R}^{\mathbf{a}}+\star\mathcal{J}^{\mathbf{a}}-\star
d\delta\theta^{\mathbf{a}}+\star\delta(\omega_{\mathbf{b}}^{\mathbf{a}}%
\wedge\theta^{\mathbf{b}})+\omega_{\mathbf{b}}^{\mathbf{a}}\wedge
\star\mathcal{T}^{\mathbf{b}} \label{du6}%
\end{equation}
or equivalently recalling Eq.(\ref{440new})
\begin{equation}
D\star\mathcal{T}^{\mathbf{a}}=-\star\mathring{\square}\theta^{\mathbf{a}%
}-\star\mathcal{R}^{\mathbf{a}}+\star\mathcal{J}^{\mathbf{a}}-\star
d\delta\theta^{\mathbf{a}}+\star\delta(\omega_{\mathbf{b}}^{\mathbf{a}}%
\wedge\theta^{\mathbf{b}})-\star(\omega_{\mathbf{b}}^{\mathbf{a}}%
\lrcorner\mathcal{T}^{\mathbf{b}}) \label{du7b}%
\end{equation}

\begin{remark}
\emph{Eq.(\ref{du7b})} does not implies that $D\star\mathcal{T}^{\mathbf{a}%
}=\star\mathcal{R}_{\mathbf{b}}^{\mathbf{a}}\wedge\theta^{\mathbf{b}}$ because
taking into account \emph{Eq.(\ref{bi25})}
\begin{equation}
\star\mathcal{R}_{\mathbf{b}}^{\mathbf{a}}\wedge\theta^{\mathbf{b}}%
=-\star\mathcal{R}^{\mathbf{a}}\neq D\star\mathcal{T}^{\mathbf{a}}%
=-\star\mathring{\square}\theta^{\mathbf{a}}-\star\mathcal{R}^{\mathbf{a}%
}+\star\mathcal{J}^{\mathbf{a}}-\star d\delta\theta^{\mathbf{a}}+\star
\delta(\omega_{\mathbf{b}}^{\mathbf{a}}\wedge\theta^{\mathbf{b}})-\star
(\omega_{\mathbf{b}}^{\mathbf{a}}\lrcorner\mathcal{T}^{\mathbf{b)}}
\label{du7bb}%
\end{equation}
in general.
\end{remark}

So, for a Levi-Civita connection we have that $D\star\mathcal{T}^{\mathbf{a}%
}=0$ and then Eq.(\ref{du6}) implies
\begin{equation}
D\star\mathcal{T}^{\mathbf{a}}=0\Leftrightarrow-\mathring{\square}%
\theta^{\mathbf{a}}-\mathcal{\mathring{R}}^{\mathbf{a}}-d\delta\theta
^{\mathbf{a}}+\delta(\mathring{\omega}_{\mathbf{b}}^{\mathbf{a}}\wedge
\theta^{\mathbf{b}})=0
\end{equation}
or since $\mathring{\omega}_{\mathbf{b}}^{\mathbf{a}}\wedge\theta^{\mathbf{b}%
}=-d\theta^{\mathbf{b}}$ for a Levi-Civita connection,%
\begin{equation}
D\star\mathcal{T}^{\mathbf{a}}=0\Leftrightarrow-\mathring{\square}%
\theta^{\mathbf{a}}-\mathcal{\mathring{R}}^{\mathbf{a}}-d\delta\theta
^{\mathbf{a}}-\delta d\theta^{\mathbf{a}}=0 \label{du8}%
\end{equation}
or yet
\begin{equation}
-\mathring{\square}\theta^{\mathbf{a}}-\mathcal{\mathring{R}}^{\mathbf{a}}=-%
\bpartial
\hspace{0.01cm}^{2}\,\theta^{\mathbf{a}}=d\delta\theta^{\mathbf{a}}+\delta
d\theta^{\mathbf{a}}, \label{du9}%
\end{equation}
an identity that we already mentioned above (Eq.(\ref{ssd1})).

\subsection{Einstein Equations}

The reader can easily verify that Einstein equations in the Clifford bundle
formalism is written as:%
\begin{equation}
\mathcal{\mathring{R}}^{\mathbf{a}}-\frac{1}{2}\mathring{R}\theta^{\mathbf{a}%
}=%
\slT
^{\mathbf{a}}, \label{E1}%
\end{equation}
where $\mathring{R}$ is the scalar curvature and $%
\slT
^{\mathbf{a}}=-T_{\mathbf{b}}^{\mathbf{a}}\theta^{\mathbf{b}}$ are the
energy-momentum $1$-form fields. Comparing \ Eq.(\ref{du8}) with
Eq.(\ref{E1}). We immediately get the "wave equation" for the cotetrad
fields:
\begin{equation}%
\slT
^{\mathbf{a}}=-\frac{1}{2}\mathring{R}\theta^{\mathbf{a}}-\mathring{\square
}\theta^{\mathbf{a}}-d\delta\theta^{\mathbf{a}}-\delta d\theta^{\mathbf{a}},
\label{E2}%
\end{equation}
which does not implies that the Ricci tensor is null.

\begin{remark}
We see from Eq.(\ref{E2}) that a Ricci flat spacetime is characterized by the
equality of the Hodge and covariant D' Alembertians acting on the coterad
fields, i.e.,%
\begin{equation}
\mathring{\square}\theta^{\mathbf{a}}=\Diamond\theta^{\mathbf{a}}, \label{E3}%
\end{equation}
a non trivial result.
\end{remark}

\begin{exercise}
Using \emph{Eq.(\ref{1831})} and \emph{Eq.(\ref{1832} ) write } $\mathring
{\square}\theta^{\mathbf{a}}$ in terms of the connection coefficients of the
Riemann-Cartan connection.
\end{exercise}

\section{Two Counterexamples to Evans (Wrong) Equation \textquotedblleft%
$\mathbf{D}\star\mathcal{T}^{\mathbf{a}}=\star\mathcal{R}_{\mathbf{b}%
}^{\mathbf{a}}\wedge\theta^{\mathbf{b}}$\textquotedblright}

\subsection{The Riemannian Geometry of $S^{2}$}

Consider the well known Riemannian structure on the unit radius sphere
\cite{goschu} $\{S^{2},%
\slg
,\mathring{D}\}$. Let \ $\{x^{i}\}$, $x^{1}=\vartheta$ , $x^{2}=\varphi$,
$0<\vartheta<\pi$, $0<\varphi<2\pi$, be spherical coordinates covering
$U=\{S^{2}-l\}$, where $l$ is the curve joining the north and south poles.

A coordinate basis for $TU$ is then $\{{\mbox{\boldmath$\partial$}}_{\mu}\}$
and its dual basis is $\{\theta^{\mu}=dx^{\mu}\}$. The Riemannian metric $%
\slg
\in\sec T_{0}^{2}M$ is given by
\begin{equation}%
\slg
=d\vartheta\otimes d\vartheta+\sin^{2}\vartheta d\varphi\otimes d\varphi
\label{ce1}%
\end{equation}
and the metric \texttt{g}$\in\sec T_{2}^{0}M$ of the cotangent space is%
\begin{equation}
\mathtt{g}={\mbox{\boldmath$\partial$}}_{1}\otimes{\mbox{\boldmath$\partial$}}%
_{1}+\frac{1}{\sin^{2}\vartheta}{\mbox{\boldmath$\partial$}}_{2}%
\otimes{\mbox{\boldmath$\partial$}}_{2}. \label{ce2}%
\end{equation}
An orthonormal basis for $TU$ is then $\{\mathbf{e}_{\mathbf{a}}\}$ with%
\begin{equation}
\mathbf{e}_{\mathbf{1}}={\mbox{\boldmath$\partial$}}_{1},\mathbf{e}%
_{\mathbf{2}}=\frac{1}{\sin\vartheta}{\mbox{\boldmath$\partial$}}_{2},
\label{ce3}%
\end{equation}
with dual basis $\{\theta^{\mathbf{a}}\}$ given by%
\begin{equation}
\theta^{\mathbf{1}}=d\vartheta,\theta^{\mathbf{2}}=\sin\vartheta d\varphi.
\label{ce4}%
\end{equation}
The structure coefficients of the orthonormal basis are
\begin{equation}
\lbrack\mathbf{e}_{\mathbf{i}},\mathbf{e}_{\mathbf{j}}]=c_{\mathbf{ij}%
}^{\mathbf{k}}\mathbf{e}_{\mathbf{k}} \label{ce4a}%
\end{equation}
and can be evaluated, e.g., by calculating $d\theta^{\mathbf{i}}=-\frac{1}%
{2}c_{\mathbf{jk}}^{\mathbf{i}}\theta^{\mathbf{j}}\wedge\theta^{\mathbf{k}}$.
We get immediately that the only non null coefficients are%
\begin{equation}
c_{\mathbf{12}}^{\mathbf{2}}=-c_{\mathbf{21}}^{\mathbf{2}}=-\cot\theta.
\label{ce6}%
\end{equation}
To calculate the connection $1$-form $\omega_{\mathbf{d}}^{\mathbf{c}}$ we use
Eq.(\ref{conna}), i.e.,
\[
\omega^{\mathbf{cd}}=\frac{1}{2}(-c_{\mathbf{jk}}^{\mathbf{c}}\eta
^{\mathbf{dj}}+c_{\mathbf{jk}}^{\mathbf{d}}\eta^{\mathbf{cj}}-\eta
^{\mathbf{ca}}\eta_{\mathbf{bk}}\eta^{\mathbf{dj}}c_{\mathbf{ja}}^{\mathbf{b}%
})\theta^{\mathbf{k}}.
\]
Then,

\begin{equation}
\omega^{\mathbf{21}}=\frac{1}{2}(-c_{\mathbf{12}}^{\mathbf{2}}\eta
^{\mathbf{11}}-\eta^{\mathbf{22}}\eta_{\mathbf{22}}\eta^{\mathbf{11}%
}c_{\mathbf{12}}^{\mathbf{2}})\theta^{\mathbf{2}}=\cot\vartheta\theta
^{\mathbf{2}}.
\end{equation}
Then%

\begin{align}
\omega^{\mathbf{21}}  &  =-\omega^{\mathbf{12}}=\cot\vartheta\theta
^{\mathbf{2}},\nonumber\\
\omega_{\mathbf{1}}^{\mathbf{2}}  &  =-\omega_{\mathbf{2}}^{\mathbf{1}}%
=\cot\vartheta\theta^{\mathbf{2}},
\end{align}%
\begin{equation}
\mathring{\omega}_{\mathbf{21}}^{\mathbf{2}}=\cot\vartheta\text{, }%
\mathring{\omega}_{\mathbf{11}}^{\mathbf{2}}=0. \label{ce7}%
\end{equation}
Now, from Cartan' s second structure equation we have%
\begin{align}
\mathcal{\mathring{R}}_{\mathbf{2}}^{\mathbf{1}}  &  =d\mathring{\omega
}_{\mathbf{2}}^{\mathbf{1}}+\mathring{\omega}_{\mathbf{1}}^{\mathbf{1}}%
\wedge\mathring{\omega}_{\mathbf{1}}^{\mathbf{1}}+\mathring{\omega}%
_{2}^{\mathbf{1}}\wedge\mathring{\omega}_{\mathbf{2}}^{\mathbf{2}}%
=d\mathring{\omega}_{\mathbf{2}}^{\mathbf{1}}\label{ce9}\\
&  =\theta^{\mathbf{1}}\wedge\theta^{\mathbf{2}}\nonumber
\end{align}
and\footnote{Observe that with our definition of the Ricci tensor it results
that $\mathring{R}=\mathring{R}_{\mathbf{1}}^{\mathbf{1}}+\mathring
{R}_{\mathbf{2}}^{\mathbf{2}}=-1.$}
\begin{equation}
\mathring{R}_{\mathbf{2\;12}}^{\;\mathbf{1}}=-\mathring{R}_{\mathbf{2\;21}%
}^{\;\mathbf{1}}=-\mathring{R}_{\mathbf{1\;12}}^{\;\mathbf{2}}=\mathring
{R}_{\mathbf{1\;21}}^{\;\mathbf{2}}=\frac{1}{2}. \label{ce10}%
\end{equation}

Now, let us calculate $\star\mathcal{R}_{\mathbf{2}}^{\mathbf{1}}\in\sec%
{\displaystyle\bigwedge\nolimits^{0}}
T^{\ast}M$. We have
\begin{align}
\star\mathcal{R}_{\mathbf{2}}^{\mathbf{1}}  &  =\widetilde{\mathcal{R}%
_{\mathbf{2}}^{\mathbf{1}}}\lrcorner\tau_{%
\slg
}=-(\theta^{\mathbf{1}}\wedge\theta^{\mathbf{2}})\lrcorner(\theta^{\mathbf{1}%
}\wedge\theta^{\mathbf{2}})=-\theta^{\mathbf{1}}\theta^{\mathbf{2}}%
\theta^{\mathbf{1}}\theta^{\mathbf{2}}\nonumber\\
&  =\left(  \theta^{\mathbf{1}}\right)  ^{2}\left(  \theta^{\mathbf{2}%
}\right)  ^{2}=1 \label{ce11'}%
\end{align}
and
\begin{equation}
\star\mathcal{R}_{\mathbf{a}}^{\mathbf{1}}\wedge\theta^{\mathbf{a}%
}=\mathcal{R}_{\mathbf{2}}^{\mathbf{1}}\wedge\theta^{\mathbf{2}}%
=\theta^{\mathbf{2}}\neq0. \label{ce12}%
\end{equation}
Now, Evans equation implies that $\star\mathcal{R}_{\mathbf{a}}^{\mathbf{1}%
}\wedge\theta^{\mathbf{1}}=0$ for a Levi-Civita connection and thus as
promised we exhibit a counterexample to his wrong equation.

\begin{remark}
We recall that the first Bianchi identity for $(S^{2},%
\slg
,\mathring{D})$, i.e., $\mathbf{D}\mathcal{T}^{\mathbf{a}}=\mathcal{R}%
_{\mathbf{b}}^{\mathbf{a}}\wedge\theta^{\mathbf{b}}=0$ which translate in the
orthonormal basis used above in $\mathring{R}_{\mathbf{b}}{}^{\mathbf{a}}%
{}_{\!\mathbf{cd}}+\mathring{R}_{\mathbf{c}}{}^{\mathbf{a}}{}_{\!\mathbf{db}%
}+\mathring{R}_{\mathbf{d}}{}^{\mathbf{a}}{}_{\!\mathbf{bc}}.=0$ is rigorously
valid. Indeed, we have
\begin{align}
\mathring{R}_{\mathbf{2}}{}^{\mathbf{1}}{}_{\!\mathbf{12}}+\mathring
{R}_{\mathbf{1}}{}^{\mathbf{1}}{}_{\!\mathbf{21}}+\mathring{R}_{\mathbf{2}}%
{}^{\mathbf{1}}{}_{\!\mathbf{21}}  &  =\mathring{R}_{\mathbf{2}}{}%
^{\mathbf{1}}{}_{\!\mathbf{12}}-\mathring{R}_{\mathbf{2}}{}^{\mathbf{1}}%
{}_{\!\mathbf{12}}=0,\nonumber\\
\mathring{R}_{\mathbf{1}}{}^{\mathbf{2}}{}_{\!\mathbf{12}}+\mathring
{R}_{\mathbf{1}}{}^{\mathbf{2}}{}_{\!\mathbf{21}}+\mathring{R}_{\mathbf{2}}%
{}^{\mathbf{2}}{}_{\!\mathbf{21}}  &  =\mathring{R}_{\mathbf{1}}{}%
^{\mathbf{2}}{}_{\!\mathbf{12}}-\mathring{R}_{\mathbf{1}}{}^{\mathbf{2}}%
{}_{\!\mathbf{12}}=0. \label{ce13}%
\end{align}
\ 
\end{remark}

\subsection{The Teleparallel Geometry of $(\mathring{S}^{2},%
\slg
,D)$}

\begin{figure}
[h]
\begin{center}
\includegraphics[
natheight=1.222800in,
natwidth=2.270100in,
height=3.4921in,
width=6.448in
]%
{../Documents and Settings/Waldir/Meus 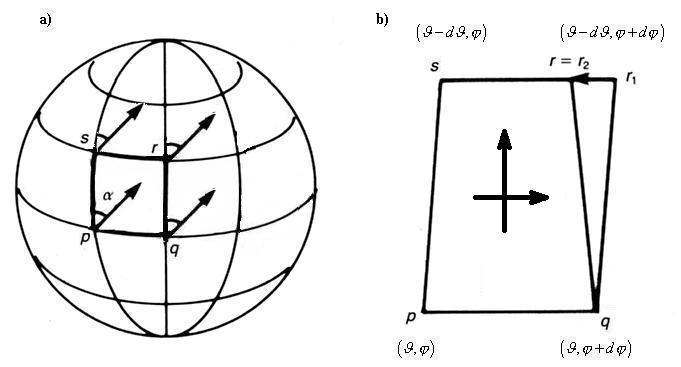}%
\caption{Geometrical Characterization of the Nunes Connection.}%
\label{fignunes}%
\end{center}
\end{figure}

Consider the manifold $\mathring{S}^{2}$ $=\{S^{2}\backslash$\textrm{north
pole}$\}\subset\mathbb{R}^{3}$, it is an sphere excluding the north pole. Let
\texttt{ }$%
\slg
\in\sec T_{2}^{0}\mathring{S}^{2}$ be the standard Riemann metric field for
$\mathring{S}^{2}$ (Eq.(\ref{ce1})). \ Now, consider besides the Levi-Civita
connection another one, $D$, here called the Nunes (or navigator
\cite{nakahara}) connection\footnote{See some historical detials in
\cite{rodoliv2007}.}. It is defined by the following parallel transport rule:
a vector is parallel transported along a curve, if at any $x\in\mathring
{S}^{2}$ the angle between the vector and the vector tangent to the latitude
line passing through that point is constant during the transport (see Figure
\ref{fignunes})

\hspace{-8cm}%

As before $(x^{1},x^{2})=(\vartheta,\varphi)$ $0<\vartheta<\pi$,
$0<\varphi<2\pi$, denote the standard spherical coordinates of a $\mathring
{S}^{2}$ of unitary radius, which covers $U=\{\mathring{S}^{2}-l\}$, where $l$
is the curve joining the north and south poles.

Now, it is obvious from what has been said above that our connection is
characterized by
\begin{equation}
D_{\mathbf{e}_{\mathbf{j}}}\mathbf{e}_{\mathbf{i}}=0. \label{6x}%
\end{equation}

Then taking into account the definition of the curvature tensor we have%

\begin{equation}
\mathbf{R(e}_{\mathbf{k}},\theta^{\mathbf{a}},\mathbf{e}_{\mathbf{i}%
},\mathbf{e}_{\mathbf{j}}\mathbf{)}=\theta^{\mathbf{a}}\left(  \left[
D_{\mathbf{e}_{\mathbf{i}}}D_{\mathbf{e}_{\mathbf{j}}}-D_{\mathbf{e}%
_{\mathbf{j}}}D_{\mathbf{e}_{\mathbf{i}}}-D_{[\mathbf{e}_{\mathbf{i}%
},\mathbf{e}_{\mathbf{j}}]}^{c}\right]  \mathbf{e}_{\mathbf{k}}\right)  =0.
\end{equation}

Also, taking into account the definition of the torsion operation we have%
\begin{align}
\mathbf{\tau}(\mathbf{e}_{\mathbf{i}},\mathbf{e}_{\mathbf{j}})  &
=T_{\mathbf{ij}}^{\mathbf{k}}\mathbf{e}_{\mathbf{k}}=D_{\mathbf{e}%
_{\mathbf{j}}}\mathbf{e}_{\mathbf{i}}-D_{\mathbf{e}_{\mathbf{i}}}%
\mathbf{e}_{\mathbf{j}}-[\mathbf{e}_{\mathbf{i}},\mathbf{e}_{\mathbf{j}%
}]\nonumber\\
&  =[\mathbf{e}_{\mathbf{i}},\mathbf{e}_{\mathbf{j}}]=c_{\mathbf{ij}%
}^{\mathbf{k}}\mathbf{e}_{\mathbf{k}},
\end{align}%
\begin{equation}
T_{\mathbf{21}}^{\mathbf{2}}=-T_{\mathbf{12}}^{\mathbf{2}}=\cot\vartheta
\text{, }T_{\mathbf{21}}^{\mathbf{1}}=-T_{\mathbf{12}}^{\mathbf{1}}=0.
\label{tp1}%
\end{equation}

It follows that the unique non null torsion $2$-form is:
\[
\mathcal{T}^{\mathbf{2}}=-\cot\vartheta\theta^{\mathbf{1}}\wedge
\theta^{\mathbf{2}}.
\]

If you still need more details, concerning this last result, consider Figure
\ref{fignunes}(b) which shows the standard parametrization of the points
$p,q,r,s$ in terms of the spherical coordinates introduced above
\cite{nakahara}. According to the geometrical meaning of torsion, we determine
its value at a given point by calculating the difference between the
(infinitesimal)\footnote{This wording, of course, means that this vectors are
identified as elements of the appropriate tangent spaces.} vectors $pr_{1}$
and $pr_{2}$ determined as follows. If we transport the vector $pq$ along $ps$
we get the vector $\vec{v}=sr_{1}$ such that $\left\vert
\slg
(\vec{v},\vec{v})\right\vert ^{\frac{1}{2}}=\sin\vartheta\triangle\varphi$. On
the other hand, if we transport the vector $ps$ along $pr$ we get the vector
$qr_{2}=qr$. Let $\vec{w}=sr$. Then,%

\begin{equation}
\left\vert
\slg
(\vec{w},\vec{w})\right\vert ^{\frac{1}{2}}=\sin(\vartheta-\triangle
\vartheta)\triangle\varphi\simeq\sin\vartheta\triangle\varphi-\cos
\vartheta\triangle\vartheta\triangle\varphi,
\end{equation}
Also,%

\begin{equation}
\vec{u}=r_{1}r_{2}=-u(\frac{1}{\sin\vartheta}{\mbox{\boldmath$\partial$}}%
_{2})\text{, \ \ }u=\left\vert
\slg
(\vec{u},\vec{u})\right\vert =\cos\vartheta\triangle\vartheta\triangle\varphi.
\end{equation}
Then, the connection $D$ of the structure $(\mathring{S}^{2},%
\slg
,D)$ has a non null torsion tensor $\Theta$. \ Indeed, the component of
$\vec{u}=r_{1}r_{2}$ in the direction ${\mbox{\boldmath$\partial$}}_{2}$ is
precisely $T_{\vartheta\varphi}^{\varphi}\triangle\vartheta\triangle\varphi$.
So, we get (recalling that $D{}_{{\mbox{\boldmath$\partial$}}_{j}%
}{\mbox{\boldmath$\partial$}}_{i}=\Gamma_{ji}^{k}{\mbox{\boldmath$\partial$}}%
_{k})$
\begin{equation}
T_{\vartheta\varphi}^{\varphi}=\left(  \Gamma_{\vartheta\varphi}^{\varphi
}-\Gamma_{\varphi\vartheta}^{\varphi}\right)  =-\cot\vartheta.
\end{equation}

\begin{exercise}
Show that $D$ is metrical compatible, i.e., $D%
\slg
=0$.
\end{exercise}

\noindent\textbf{Solution:}%
\begin{align}
0  &  =\text{ }D_{\mathbf{e}_{\mathbf{c}}}%
\slg
(\mathbf{e}_{\mathbf{i}},\mathbf{e}_{\mathbf{j}})=(D_{\mathbf{e}_{\mathbf{c}}}%
\slg
)(\mathbf{e}_{\mathbf{i}},\mathbf{e}_{\mathbf{j}})+%
\slg
(D_{\mathbf{e}_{\mathbf{c}}}\mathbf{e}_{\mathbf{i}},\mathbf{e}_{\mathbf{j}})+%
\slg
(\mathbf{e}_{\mathbf{i}},D_{\mathbf{e}_{\mathbf{c}}}\mathbf{e}_{\mathbf{j}%
})\nonumber\\
&  =(D_{\mathbf{e}_{\mathbf{c}}}%
\slg
)(\mathbf{e}_{\mathbf{i}},\mathbf{e}_{\mathbf{j}})
\end{align}

\begin{remark}
Our counterexamples that involve the parallel transport rules defined by a
Levi-Civita connection and a teleparallel connection in $\mathring{S}^{2}$
show clearly that we cannot \textit{mislead} the Riemann curvature tensor of a
connection defined in a \ given manifold with the fact that the manifold may
be bend as a surface in an Euclidean manifold where it is embedded. Neglecting
this fact may generate a lot of wishful thinking.
\end{remark}

\section{Conclusions}

In this paper after recalling the main definitions and a collection of tricks
of the trade concerning the calculus of differential forms on the Cartan,
Hodge and Clifford bundles over a Riemannian or Riemann-Cartan space or a
Lorentzian or Riemann-Cartan spacetime we solved with details several
exercises involving different grades of difficult and which we believe, may be
of some utility for pedestrians and even for experts on the subject. In
particular we found using technology of the Clifford bundle formalism the
correct equation for $\mathbf{D}\star\mathcal{T}^{\mathbf{a}}$. We show that
the result found in \cite{evans101}, namely \textquotedblleft$\mathbf{D}%
\star\mathcal{T}^{\mathbf{a}}=\star\mathcal{R}_{\mathbf{b}}^{\mathbf{a}}%
\wedge\mathcal{T}^{\mathbf{b}}$\textquotedblright\ is wrong since it
contradicts the right formula we found. Besides that, the wrong formula is
also contradicted by two simple counterexamples that we exhibited in Section
15 . The last sentence before the conclusions is a crucial remark, which each
one seeking truth must always keep in mind: do not confuse the Riemann
curvature tensor\footnote{The remark applies also to the torsion of a
connection.} of a connection defined in a given manifold with the fact that
the manifold may be bend as a surface in an Euclidean manifold where it is embedded.

We end the paper with a necessary explanation. An attentive reader may ask:
Why write a bigger paper as the present one to show wrong a result not yet
published in a scientific journal? The justification is that Dr. Evans
maintains a site on his (so called) \textquotedblleft ECE
theory\textquotedblright\ which is read by thousand of people that\ thus are
being continually mislead, thinking that its author is creating a new
Mathematics and a new Physics. Besides that, due to the low Mathematical level
of many referees, Dr. Evans from time to time succeed in\ publishing his
papers in SCI journals, as the recent ones., \cite{e1,e2}. In the past we
already showed that several published papers by Dr. Evans and colleagues
contain serious flaws (see, e.g., \cite{carod,rodqui}) and recently some other
authors spent time writing papers to correct Mr. Evans claims (see,
e.g.,\cite{B, bha,B3,h,ho,t'hooft}) It is our hope that our effort and of the
ones by those authors just quoted serve to counterbalance Dr. Evans
\ influence on a general public\footnote{And we hope also on many scientists,
see a partial list in \cite{e-free, e-free1}!} which being anxious for
novelties may be eventually mislead by people that claim among other things to
\textit{know }\cite{e-free,e-free1,e1,e2} how to project devices to withdraw
energy from the vacuum.

\begin{acknowledgement}
The author is grateful to Prof. E. A. Notte-Cuello, for have checking all
calculations and discovered several misprints, that are now corrected.
Moreover, the author will be grateful to any one which point any misprints or
eventual errors.
\end{acknowledgement}

\end{document}